\documentclass[pre,aps,twocolumn,showpacs]{revtex4-1}

\usepackage{graphicx}
\usepackage{amsfonts}
\usepackage{amsmath}
\usepackage{mathtools}
\usepackage{float}
\usepackage{epsfig}
\usepackage{color}
\usepackage{gensymb}
\usepackage{multirow}
\usepackage[mathscr]{euscript}

\newcommand{\e}[1]{\ensuremath{\times 10^{#1}}}
\newcommand{\ee}[1]{\ensuremath{10^{#1}}}
\newcommand{\T}{\ensuremath{T_E}}

\begin{document}

\title{Ground states of stealthy hyperuniform potentials: I. Entropically favored configurations}

\author{G. Zhang}

%\email{gezhang@princeton.edu}

\affiliation{\emph{Department of Chemistry}, \emph{Princeton University},
Princeton, New Jersey 08544, USA}

\author{F. H. Stillinger}

%\email{fhs@princeton.edu}

\affiliation{\emph{Department of Chemistry}, \emph{Princeton University},
Princeton, New Jersey 08544, USA}

\author{S. Torquato}

\email{torquato@electron.princeton.edu}

\affiliation{\emph{Department of Chemistry, Department of Physics,
Princeton Institute for the Science and Technology of
Materials, and Program in Applied and Computational Mathematics}, \emph{Princeton University},
Princeton, New Jersey 08544, USA}

\pacs{}

\begin{abstract}
Systems of particles interacting with ``stealthy'' pair potentials have been shown to possess infinitely degenerate disordered hyperuniform classical ground states with novel physical properties. Previous attempts to sample the infinitely degenerate ground states used energy minimization techniques, introducing algorithmic dependence that is artificial in nature. Recently, an ensemble theory of stealthy hyperuniform ground states was formulated to predict the structure and thermodynamics that was shown to be in excellent agreement with corresponding computer simulation results in the canonical ensemble (in the zero-temperature limit). In this paper, we provide details and justifications of the simulation procedure, which involves performing molecular dynamics simulations at sufficiently low temperatures and minimizing the energy of the snapshots for both the high-density disordered regime, where the theory applies, as well as lower densities. We also use numerical simulations to extend our study to the lower-density regime. We report results for the pair correlation functions, structure factors, and Voronoi cell statistics. 
In the high-density regime, we verify
the theoretical ansatz that stealthy disordered ground states behave like ``pseudo'' disordered equilibrium hard-sphere
systems in Fourier space. 
The pair statistics obey certain exact integral conditions with very high accuracy.
These results show that as the density decreases from the high-density limit, the disordered ground states in the canonical ensemble are characterized by an increasing
degree of short-range order
and eventually the system undergoes
a phase transition to crystalline ground states.
In the crystalline regime (low densities), there exist aperiodic structures that are part of the ground-state manifold, but yet are not entropically favored.
We also provide numerical evidence suggesting that different forms of stealthy pair potentials produce the same ground-state ensemble in the zero-temperature limit. Our techniques may be applied to sample the zero-temperature limit of the canonical ensemble of other potentials with highly degenerate ground states.
\end{abstract}

\maketitle

\section{Introduction}

There has been long-standing interest in the phase behavior of many-particle systems in $d$-dimensional Euclidean spaces $\mathbb{R}^d$ in which the particles interact with soft, bounded pair potentials 
\cite{stillinger1976phase, lowen1994melting, guenza1997local, likos2001soft, likos2001criterion, mladek2006formation, 
mladek2008multiple, krekelberg2009generalized, batten2011novel, zhao2012analysis, capone2012telechelic, torquato2015ensemble}. 
Considerable attention has been devoted to the determination of the classical ground states (global energy minima) of such interactions 
\cite{guenza1997local, mladek2006formation, capone2012telechelic, torquato2015ensemble}.
While typical interactions lead to unique classical ground states, certain special pair potentials are characterized by degenerate classical ground states---a phenomenon that has attracted recent attention
\cite{torquato2015ensemble, fan1991constraints, uche2004constraints, suto2005crystalline, suto2006from, uche2006collective, batten2008classical, batten2009interactions, batten2011inherent, zachary2011anomalous, martis2013exotic}.

One family of such pair interactions are the ``stealthy potentials'' because their ground states correspond to configurations that completely suppress single scattering for a range of wave numbers.
The Fourier transforms of these potentials are bounded and non-negative and have compact support \cite{torquato2015ensemble}, and hence they have corresponding direct-space potentials that are bounded and long ranged.
Because of their special construction in Fourier space, finding the ground states of stealthy potentials is equivalent to constraining the structure factor to be zero for wave vectors $\bf k$ contained within
the support of the Fourier transformed potential \cite{torquato2015ensemble}, as will be summarized in Sec.~\ref{detail}.
In the case when the constrained wave vectors lie in the radial interval $0 < |\mathbf k| \le K$, the stealthy ground states fall within the class of hyperuniform states of matter \cite{torquato2003local} and can be tuned to have 
varying degrees of disorder.
Disordered hyperuniform systems in general are of current interest because they are characterized by an anomalously large suppression of long-wavelength density fluctuations
and can exist as equilibrium or nonequilibrium states, either classically or quantum mechanically 
\cite{donev2005unexpected, zachary2011hyperuniform, jiao2011maximally, chen2014equilibrium, berthier2011suppressed,
kurita2011incompressibility,dreyfus2015diagnosing,lesanovsky2014out,jack2014hyperuniformity, de2015toward, 
degl2015thz, yu2015bloch, jiao2014avian, torquato2008point}.
Moreover, because disordered hyperuniform states of matter have characteristics that lie between a crystal and a liquid \cite{torquato2015ensemble},
they are endowed with novel physical properties \cite{batten2008classical, batten2009interactions, florescu2009designer, florescu2013optical, man2013isotropic, man2013photonic, 
haberko2013direct,laurin2014hollow,hejna2013nearly,xie2013hyperuniformity,brahim2015small}.

When a dimensionless parameter $\chi$, inversely proportional to the number density $\rho$ and proportional to $K^d$ (size of the constrained region) is sufficiently small,
the hyperuniform ground states are infinitely degenerate and counterintuitively disordered (i.e., isotropic without any Bragg peaks) \cite{torquato2015ensemble}.
However, when $\chi$ is large enough ($\rho$ is sufficiently small), there is a phase transition to a regime in which the ground states are crystalline or highly ordered 
\cite{fan1991constraints, uche2004constraints, suto2005crystalline, batten2009interactions}.
For each spatial dimension $d$, there is a special value of $\chi$, $\chi^*_{\mbox{max}}$, at which the ground state is unique \cite{unique}.
The unique ground state is the dual (reciprocal lattice) of the densest Bravais lattice packing in each dimension \cite{torquato2015ensemble}.
In two and higher dimensions, as soon as $\chi$ drops below $\chi^*_{\mbox{max}}$, the set of the ground states become uncountably infinite and gradually includes progressively less ordered structures \cite{torquato2015ensemble}.
Similarly to stealthy potentials, a family of two-, three-, and four-body potentials that lead to disordered ground states has also been defined in Fourier space and studied \cite{uche2006collective, batten2008classical, zachary2011anomalous}.

Due to the complexity of the problem, almost all previous investigations of the ground states employed computer simulations. 
Such numerical studies were carried out in one, two and three dimensions \cite{fan1991constraints, uche2004constraints, uche2006collective, batten2008classical, batten2009interactions}. 
The ground states were sampled by minimization of potential energy at fixed densities starting from random initial conditions in a $d$-dimensional cubic simulation box under periodic boundary conditions.
A few optimization techniques were employed to find the global energy minima with very high precision \cite{uche2004constraints, uche2006collective}.
%Previous researchers found that the ground states in one dimension contain only integer lattices when $0.5 < \chi < 1=\chi^*_{\mbox{max}}$ but also contain disordered configurations when $\chi < 0.5$ \cite{fan1991constraints}. 
%The ground states in two dimensions are crystalline when $0.78<\chi<0.91=\chi^*_{\mbox{max}}$, ``wavy crystalline'' when $0.58<\chi<0.78$, and disordered when $\chi<0.58$ \cite{uche2004constraints}. 
%The ground states in three dimensions are crystalline when $0.50<\chi<0.98=\chi^*_{\mbox{max}}$ and disordered when $\chi<0.50$ \cite{uche2006collective}.
%The one-dimensional results can also be proved analytically \cite{fan1991constraints}.

Generally, a numerically obtained ground-state configuration depends on the number of particles $N$ within the fundamental cell, initial particle configuration, shape of the fundamental cell, and particular optimization technique used \cite{torquato2015ensemble}. 
Adding to the complexity of the problem is that the disordered ground states are highly degenerate with a configurational dimensionality that depends on the density, 
and there are an infinite number of distinct ways to sample this complex ground-state manifold, each with its own probability measure. 
These nontrivial aspects had made the task of formulating a statistical-mechanical theory of stealthy degenerate ground states a daunting one. 
Recently, we have formulated such an ensemble theory that yields analytical predictions of the structural characteristics and other properties of stealthy degenerate ground states \cite{torquato2015ensemble}. 
A number of exact results for the thermodynamic and structural properties of these ground states were derived that applied to {\it general} ensembles.
We then specialized our results to the canonical ensemble (in the zero-temperature limit) by exploiting an ansatz 
that stealthy disordered ground states (for sufficiently small $\chi$) behave remarkably like ``pseudo" disordered equilibrium hard-sphere systems in Fourier space. 
Our theoretical predictions for the pair correlation function $g_2(r)$ and structure factor $S(k)$ of these entropically favored disordered ground states 
were shown to be in agreement with corresponding computer simulations across the first three space dimensions.
We also made predictions for the corresponding excited states for sufficiently small temperatures that were in agreement with simulations.

Because the focus of that previous investigation was the development of ensemble theories, few simulation details were presented about how the canonical ensemble was sampled to produce stealthy disordered ground states. 
One aim of the present paper is to provide a comprehensive description of the numerical procedure that we used to produce the simulation results in Ref.~\onlinecite{torquato2015ensemble}. 
Moreover, here we also extend those results by applying the simulation procedure to study numerically the ground states in the canonical ensemble for all allowable values of $\chi$ 
and thus investigate the entire phase diagram for the entropically favored states across the first three space dimensions.
In the second paper of this series, we will study the exotic aperiodic ``wavy phases'' identified in previous numerical
work \cite{uche2004constraints} (or ``stacked-slider phases,'' as called in the sequel to this paper \cite{zhang2015ground2}), a special part of the ground-state manifold. An analytical model will enable an even more detailed study of this phase.

As a justification of sampling the canonical ensemble instead of minimizing energy, we also demonstrate here how a variety of different optimization techniques affect the ground states that are sampled, which was not previously investigated \cite{uche2004constraints, uche2006collective, batten2008classical}.
This investigation reveals that the pair statistics of the ground-state configurations indeed generally depend on the algorithm.
Moreover, we show here that the energy minimization results depend on the initial conditions as well.
We also provide the reason why the simulations in Ref.~\onlinecite{torquato2015ensemble} and this paper employ noncubic, possibly deforming, simulation boxes for $d \ge 2$.
Because almost all previous numerical simulations were performed using some specific form of stealthy potentials, we show here that different forms of stealthy potentials produce identical pair correlation functions, suggesting that the specific choice of the potential form does not affect the ensemble being sampled.

Among our major findings, we show that energy minimizations starting from random initial conditions may lead to clustering of particles, 
the degree of which depends on the algorithm for a finite range of $\chi$ below 1/2 across the first three space dimensions.
When minimizing the energy starting from configurations equilibrated at some temperature $T_E$, the ground-state configurations discovered depend on $T_E$.
However, the algorithm dependence diminishes in the $T_E \to 0$ limit.
We also demonstrate that the pair statistics [$g_2(r)$ and $S(k)$] in this limit do not depend on the particular form of the stealthy potential.
The similarity between the structure factor in this limit and the pair correlation function of an equilibrium hard-sphere system in direct space \cite{torquato2015ensemble} 
is valid for $\chi$ up to some dimension-dependent values between 0.25 and 0.33 in the first three space dimensions.
Beyond this range of $\chi$, the hard-sphere analogy in Fourier space undergoes modification.
As $\chi$ increases further (to the value of about 0.4 in two dimensions, for example), the first peak in the structure factor diminishes while second peak in the structure factor grows and engulfs the first peak.
Our simulated pair statistics obey certain exact integral conditions in Ref.~\onlinecite{torquato2015ensemble} with very high accuracy, indicating the high fidelity of the numerical results.
In the infinite-system-size limit, at $\chi=0.5$, the entropically favored ground states undergo a transition from disordered states to crystalline states. 
Depending on the dimension, this phase transition can occur when aperiodic structures still are part of the ground state manifold, 
demonstrating that crystalline (ordered) structures can have a higher entropy than disordered structures.

The rest of the paper is organized as follows: In Sec.~\ref{detail}, we briefly summarize the numerical collective-coordinate procedure 
and other details of the simulation that we employ in the present paper with justifications. 
In Sec.~\ref{dependence}, we study the dependence of the results on a variety of energy minimization algorithms, initial conditions, and the forms of the stealthy potentials.
In Sec.~\ref{ensemble}, we provide pair correlation function, structure factor, Voronoi cell-volume distribution, and configuration snapshots of the stealthy hyperuniform ground states obtained from the canonical ensemble in the zero-temperature limit.
We provide concluding remarks and discussion in Sec.~\ref{conclusion}, including suggestions for sampling the canonical ensemble in the zero-temperature limit of other potentials with degenerate disordered ground states.

\section{Mathematical Relations and Simulation Procedure}
\label{detail}

As detailed in Sec.~II of Ref.~\onlinecite{torquato2015ensemble}, we simulate point processes in periodic fundamental cells (i.e. simulation boxes) with a pairwise additive potential $v(\mathbf r)$ such that its Fourier transform exists.
Under nearest image convention, the total potential energy can be calculated by summing over all pairs of particles:
\begin{equation}
\Phi({\bf r}^N) = \sum_{i<j} v(\mathbf r_{ij}),
\end{equation} 
where $N$ is the number of particles, $\mathbf r^N \equiv \mathbf r_1, \mathbf r_2, ..., \mathbf r_N$ is the locations of the particles in $d$-dimensional Euclidean space, and $\mathbf r_{ij} = \mathbf r_i -\mathbf r_j $.
Instead of summing over all pairwise contributions in the real space, the potential energy can also be represented in Fourier space:
\begin{equation}
\Phi({\bf r}^N) =\frac{1}{2v_F} \left[\sum_{\bf k} {\tilde v}({\bf k})|{\tilde n}({\bf k})|^2-N \sum_{\bf k} {\tilde v}({\bf k})\right],
\label{pot_f}
\end{equation}
where $v_F$ is the volume of the fundamental cell, 
${\tilde v}({\mathbf k})= \int_{v_F} v(\mathbf r) \exp(-i \mathbf k \cdot \mathbf r) d\mathbf r$ is the Fourier transform of the pair potential, 
${\tilde n}({\bf k }) = \sum_{j=1}^{N} \exp(-i{\bf k \cdot r}_j)$ is the complex collective density variable [with ${\tilde n}(\mathbf k=0)=N$], 
and both summations are over all reciprocal lattice vector $\mathbf k$'s appropriate to the fundamental cell. 
For every $\mathbf k \neq \mathbf 0$, ${\tilde n}({\mathbf k })$ is related to the structure factor, $S(\mathbf k)$, via
\begin{equation}
S(\mathbf k) = \frac{|{\tilde n}({\mathbf k })|^2}{N}.
\end{equation}
Given a ${\tilde v}({\mathbf k})$, the corresponding real-space pair potential is
\begin{equation}
v( \mathbf r) = \frac{1}{v_F} \sum_{\bf k} {\tilde v}({\mathbf k}) \exp(i \bf k \cdot r).
\end{equation} 
In a finite-sized system, the real-space pair potential has the same periodicity as the fundamental cell. 
Therefore, in the infinite-volume limit, the cell periodicity disappears.

A family of ``stealthy'' potentials, which completely suppress single scattering for all wave vectors within a specific cutoff in their ground states, 
are defined as \cite{fan1991constraints, uche2004constraints, uche2006collective, batten2008classical, batten2009interactions, batten2011inherent}:
\begin{equation}
{\tilde v}({\mathbf k})=
\begin{cases}
\displaystyle V(k),& \text{if $|\mathbf k| \le K$,} \\
0,& \text{otherwise,}
\end{cases}
\label{stealthy}
\end{equation}
where $V(k)$ is a positive isotropic function and $K$ is a constant. 
%Changing $K$ only changes the length scale of the potential. 
%While changing the function form of $V(k)$ could theoretically change the energy landscape, we will demonstrate it does not change the statistics of the entropically favored ground states.
In this paper we always take $K=1$, which sets the length scale. We will also use $V(k)=1$ unless otherwise specified.
In the infinite-system-size limit, the isotropic ${\tilde v}({\mathbf k})$ correspond to an isotropic real-space pair potential $v(\bf r)$ \cite{torquato2015ensemble}. 
However, for finite systems, the corresponding $v(\bf r)$ is anisotropic. 
In Appendix~\ref{finite}, we compare the infinite-system-size limit $v(\bf r)$ with the finite-size $v(\bf r)$'s in different-shaped simulation boxes and select the simulation box shape to be used in this paper based on which $v(\bf r)$ is closest to the infinite-size-limit $v(\bf r)$.

From Eqs.~\eqref{pot_f}~and~\eqref{stealthy}, one can see that a configuration is a stealthy ground state if ${\tilde n}(\mathbf k )=0$ for all $\mathbf k$ points such that $0< |\mathbf k| \le K$.
Therefore, finding a ground state of a stealthy potential is equivalent to constraining ${\tilde n}(\mathbf k )=0$ for all of those $\mathbf k$ points.
However, in a simulation, one does not need to check all of the constraints. 
As detailed in Ref.~\onlinecite{torquato2015ensemble}, if there are $(2M+1)$ $\mathbf k$ points within the constrained radius, only $M$ of them are independent and needed to be constrained to zero. 
Equation~\eqref{pot_f} can be simplified as \cite{factor}:
\begin{equation}
\Phi({\bf r}^N) =\frac{1}{v_F} \sum_{\bf k} {\tilde v}({\bf k})|{\tilde n}({\bf k})|^2 +\Phi_0,
\label{pot_f2}
\end{equation}
where the sum is over all {\it independent} constraints, and 
\begin{equation}
\Phi_0=[N(N-1) - 2N \sum_{\bf k} {\tilde v}({\bf k})]/(2v_F)
\end{equation}
is a constant independent of the particle positions $\mathbf r^N$.
We now introduce a parameter
\begin{equation}
\chi=\frac{M}{d(N-1)},
\label{chi}
\end{equation}
which determines the degree to which the ground states are constrained, and therefore the degeneracy and disorder of the ground states \cite{uche2004constraints}.
Note that the constraints depend on $K$ and the fundamental cell but are independent of the specific shape of $\tilde v(\mathbf k)$ as long as $\tilde v(\mathbf k)>0$ for all $0<|\mathbf k| \le K$.
Therefore, changing $\tilde v(\mathbf k)$ does not change the set of the ground states. However, there is no proof that changing $\tilde v(\mathbf k)$ does not change the relative sampling weights of the ground states.

In this paper we study various systems with different $\chi$'s and $N$'s. One numerical complication is that these numbers cannot be chosen arbitrarily, since $M=\chi d(N-1)$ must be an integer consistent with the specific shape of the simulation box. (For example, a list of the allowed $M$ values for a two-dimensional square box is given in Table II of Ref.~\onlinecite{uche2004constraints}.) This constraint is especially hard to meet when simulating multiple systems at the same $\chi$ value across dimensions. In fact, both $\chi$ and $N$ in Table~\ref{Np1} (see Appendix \ref{number}) had to be chosen carefully to meet this constraint.

Taking the gradient of Eq.~(\ref{pot_f2}) yields the forces on particles:
\begin{equation}
{\bf F}_j=-{\bf \bigtriangledown}_j \Phi({\bf r}^N)=\frac{2}{v_F}\sum_{\bf k} {\bf k}\, {\tilde v}({\bf k})\,\mbox{Im}[
{\tilde n}({\bf k})\exp(i {\bf k\cdot r}_j)] ,
\label{force}
\end{equation}
where the sum is also over all independent constraints. This equation enables us to perform both energy minimizations and molecular dynamics (MD) simulations.
In an energy minimization, a derivative-based algorithm is used. 
The first term on the right side of Eq.~\eqref{pot_f2} is provided to the algorithm as the objective function and the negative of the force in Eq.~\eqref{force} is provided as the derivative.
In order to minimize energy, we have tried different algorithms including the MINOP algorithm \cite{dennis1979two}, 
 the steepest descent algorithm allowing large steps \cite{gslMultiMin}, 
the low-storage BFGS (L-BFGS) algorithm \cite{nocedal1980updating, liu1989limited, nlopt}, 
the Polak-Ribiere conjugate gradient algorithm \cite{grippo1997globally, gslMultiMin},
and our ``local gradient descent'' algorithm described in Appendix~\ref{LocalGradientDescent}.
When $\chi < 0.5$, the objective function always ends up being very close to zero (the minimum). 
The maximum ending objective function for different algorithms varies from as high as $\ee{-7}$ for a conjugate gradient algorithm to $\ee{-17}$ for the local gradient descent and steepest descent algorithms 
to $\ee{-20}$ for the L-BFGS algorithm, and to as low as $\ee{-25}$ for the MINOP algorithm.
From our practical point of view, all of these algorithms are precise enough, since an error of $\ee{-7}$ or lower is indiscernible from any results presented below.
Because the L-BFGS algorithm is the fastest, we will use it unless otherwise specified.

The energy minimizations, if started from random initial configurations, will sample an algorithm-dependent, nonequilibrium ensemble.
To sample the canonical ensemble at a given equilibrium temperature $T_E$ we use MD simulations.
One important parameter in MD simulations is the integration time step. 
Since the optimal choice of the time step depends on the temperature, 
and the latter varies across several orders of magnitude in this paper, 
we desire a systematic way to determine the optimal time step.
Starting from an energy minimized configuration and a very small time step (0.01 in dimensionless units), 
we repeat the following steps $\ee{4}$ times to equilibrate the system and find a suitable time step:
\begin{itemize}
\item Assign a random velocity from Boltzmann distribution at $T_E$ to each particle.
\item Calculate the total (kinetic and potential) energy of the system $E_1$.
\item Evolve the system 1500 time steps using the velocity Verlet algorithm \cite{frenkel2001understanding}.
\item Calculate the total energy of the system $E_2$.
\item If $|\ln \frac{E_1}{E_2}| > 1\e{-5}$, then the time step is too large and errors will build up quickly. Therefore, we decrease the time step by $5\%$. On the other hand, if $|\ln \frac{E_1}{E_2}| < 4\e{-6}$, there is still some room to increase the time step. Since increasing the time step increases the efficiency of MD simulations, we increase the time step by $5\%$.
\end{itemize}
After the system is equilibrated and the time step is chosen, we perform constant temperature MD simulations with particle velocity resetting \cite{andersen1980molecular}. A randomly chosen particle is assigned a random velocity, drawn from Maxwell-Boltzmann distribution, every 100 steps. 
We take a sample configuration every 3000 time steps until we have sampled 20 000 configurations unless otherwise specified.
This amounts to an implementation of the generation of configurations in the canonical ensemble.

The above MD procedure works well for $\chi<0.5$. However, two new features arise when it is applied to $\chi \ge 0.5$ in all dimensions. 
First, the potential energy surface develops local minima and energy barriers that can trap the system if $T_E$ is too small.
We address this problem by using simulated annealing, employing a thermodynamic cooling schedule \cite{nourani1998comparison} which starts at $T=2\e{-3}$ and ends at $\ee{-6}$.
Note that, by adopting a cooling schedule, we concede that we may only take one sample at the end of each MD trajectory, whereas a fixed-temperature MD trajectory produces multiple samples.

The second new feature is that the entropically favored ground states are crystalline for $\chi \ge 0.5$. Unlike disordered structures, a crystalline structure has long-range order and may not ``fit'' in simulation boxes with certain shapes.
To overcome the second problem, we simulate an isothermal-isobaric ensemble with a deformable simulation box. 
Every 20 MD time steps, 10 Monte Carlo trial moves to deform the simulation box are attempted. 
The pressure is calculated from Eq.~(41) of Ref.~\onlinecite{torquato2015ensemble}.

We employed the Wang-Landau Monte Carlo \cite{wang2001efficient} to attempt to determine the entropically favored ground states for $\chi>0.5$ in two and three dimensions. The Wang-Landau Monte Carlo is used to calculate the microcanonical entropy $\mathscr S(\Phi)$ as a function of the potential energy $\Phi$. We limit our simulations to the energy range $3\e{-10} < \Phi-\Phi_0 < \ee{-9}$ (in dimensionless units), where $\Phi_0$ is the ground state energy, by rejecting any trial move that violates this energy tolerance. This energy range is evenly divided into 1000 bins. Starting from a perfect crystal structure 
%of $7^3$ particles in 3D or $20^2$ particles in 2D 
in a simulation box shaped like a fundamental cell, small perturbations are introduced so the energy is within the range.
After that, 60 stages of Monte Carlo simulations are performed, each stage containing $3\e{7}$ trial moves. The ``modification factor'' in Ref. \cite{wang2001efficient} is $f=\exp[5/(n+10)]$, where $n$ is the number of stages.

\section{Dependence on energy minimization algorithm, MD temperature, 
and ${{ \tilde {\MakeLowercase v}}(\mathbf \MakeLowercase k)}$} 
\label{dependence}

In this section, we present numerical simulation results demonstrating that:
\begin{itemize}
\item Energy minimizations starting from Poisson initial configurations using different algorithms can yield ground states with different pair correlation functions.
\item Energy minimizations starting from MD snapshots at different temperatures can yield ground states with different pair correlation functions.
\item For configurations obtained by minimizing energy starting from MD snapshots at sufficiently small temperature, pair correlation functions do not depend on the minimization algorithm and the form of the stealthy potential.
\end{itemize}
These results motivate the reason why we ultimately study and report results in Sec. \ref{ensemble} in the canonical ensemble in the zero-temperature limit.
For concreteness and visual clarity, we present results here in two dimensions.
However, we have verified that all of the conclusions here also apply to one and three dimensions. 

\begin{figure}[H]
\begin{center}
\includegraphics[width=0.45\textwidth]{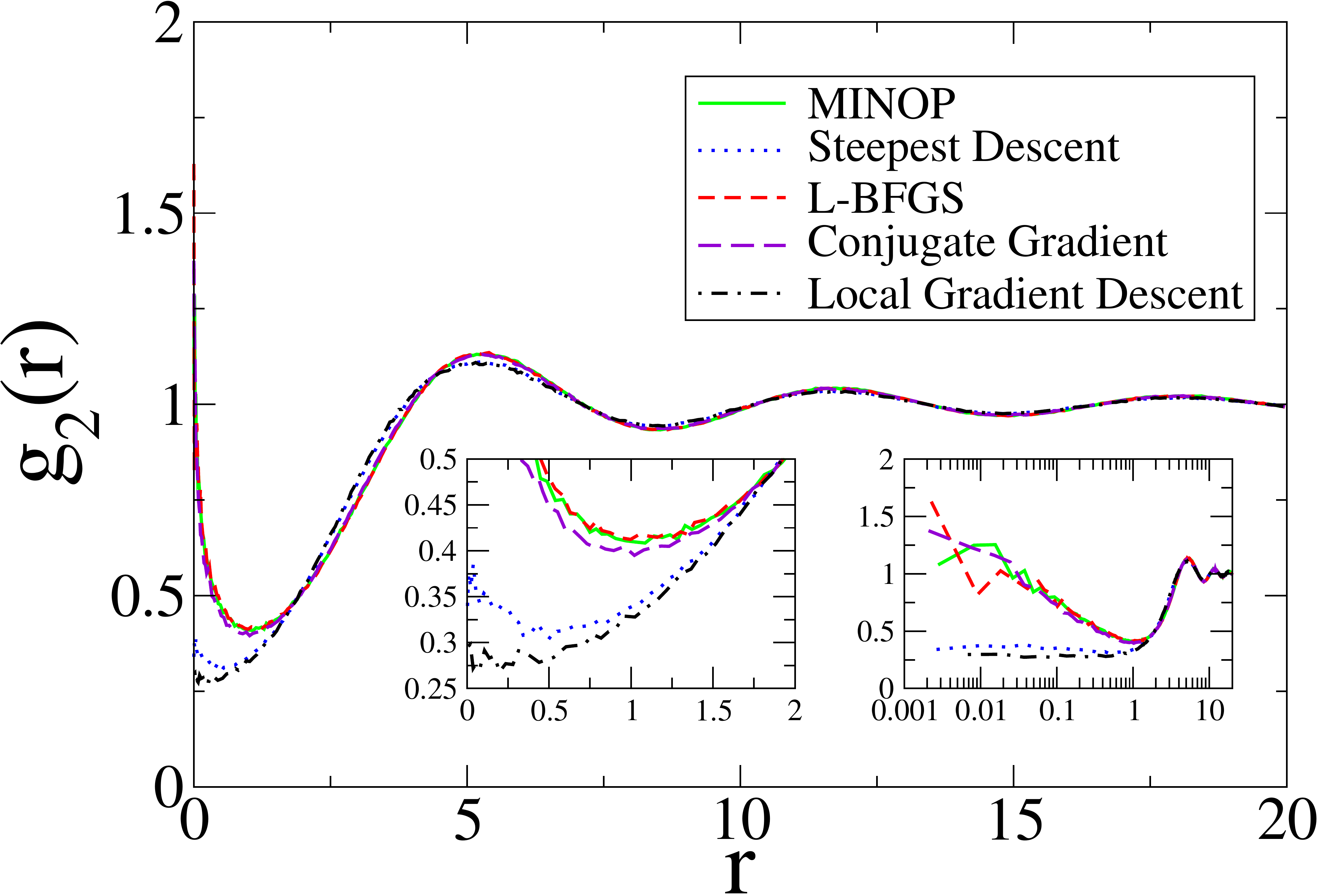}
\end{center}
\caption{(Color online) Pair correlation function as obtained from different optimization algorithms (as described in the legend) starting from Poisson initial configurations in two dimensions at $\chi=0.2$. Each curve is averaged over 20 000 configurations of 136 particles each. The left inset zooms in near the origin, showing the differences between the five algorithms more clearly. The right inset uses a semilogarithmic scale to show $g_2(r) \propto \log(r)$ near the origin.}
\label{OptimizationAlgorithm}
\end{figure}

\begin{figure}[H]
\begin{center}
\includegraphics[width=0.45\textwidth]{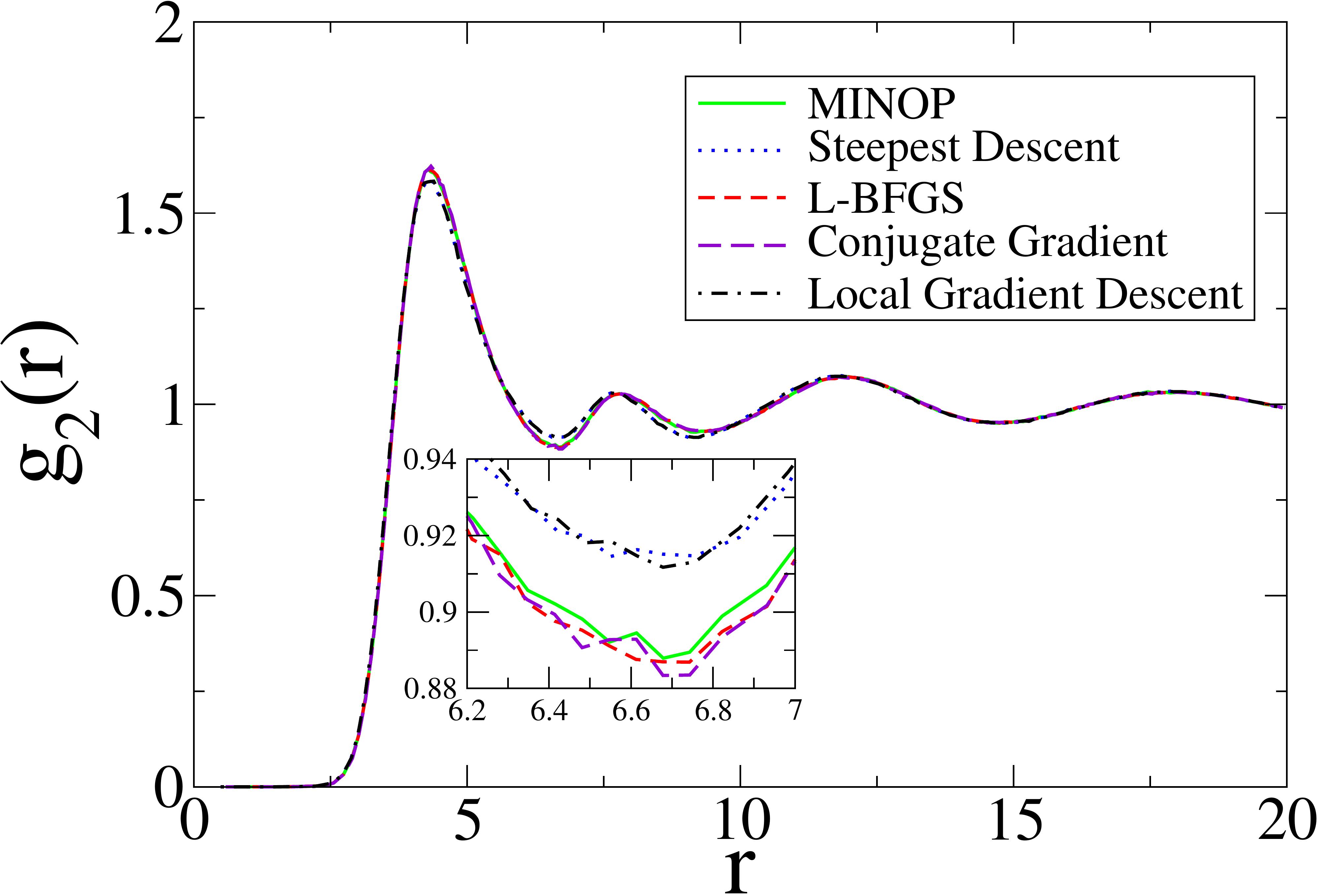}
\end{center}
\caption{(Color online) As in Fig.~\ref{OptimizationAlgorithm}, except that $\chi=0.4$ and each curve is averaged over 20 000 configurations of 151 particles each. The inset zooms in near the first well, showing the differences between the five algorithms more clearly.}
\label{OptimizationAlgorithm2}
\end{figure}

We performed energy minimizations starting from Poisson initial configurations (i.e., $T_E \to \infty$ state at fixed density) using each of the five numerical algorithms mentioned in Sec.~\ref{detail} at $\chi=0.2$ and $\chi=0.4$.
The results are shown in Figs.~\ref{OptimizationAlgorithm}~and~\ref{OptimizationAlgorithm2}.
At $\chi=0.2$, the pair correlation functions produced by the MINOP algorithm and the L-BFGS algorithm are almost identical. 
However, the pair correlation function produced by the conjugate gradient algorithm noticeably differs. 
The steepest descent algorithm and our local gradient descent algorithm produce a significantly different pair correlation function with a much weaker peak at $r=0$.
The pair correlation functions produced by some algorithms appear to have $g_2(r) \propto \log(r)$ divergence near the origin.
Since this divergence means particles have a tendency to form clusters, we call it a ``clustering effect.''
At $\chi=0.4$, the clustering effect disappears, but the pair statistics produced by different algorithms still differs.
The fact that different optimization algorithms produce different pair statistics means that they sample the ground-state manifold with different weights. 
In other words, different optimization algorithms are sampling different ground-state ensembles. 

\begin{figure}[H]
\begin{center}
\includegraphics[width=0.45\textwidth]{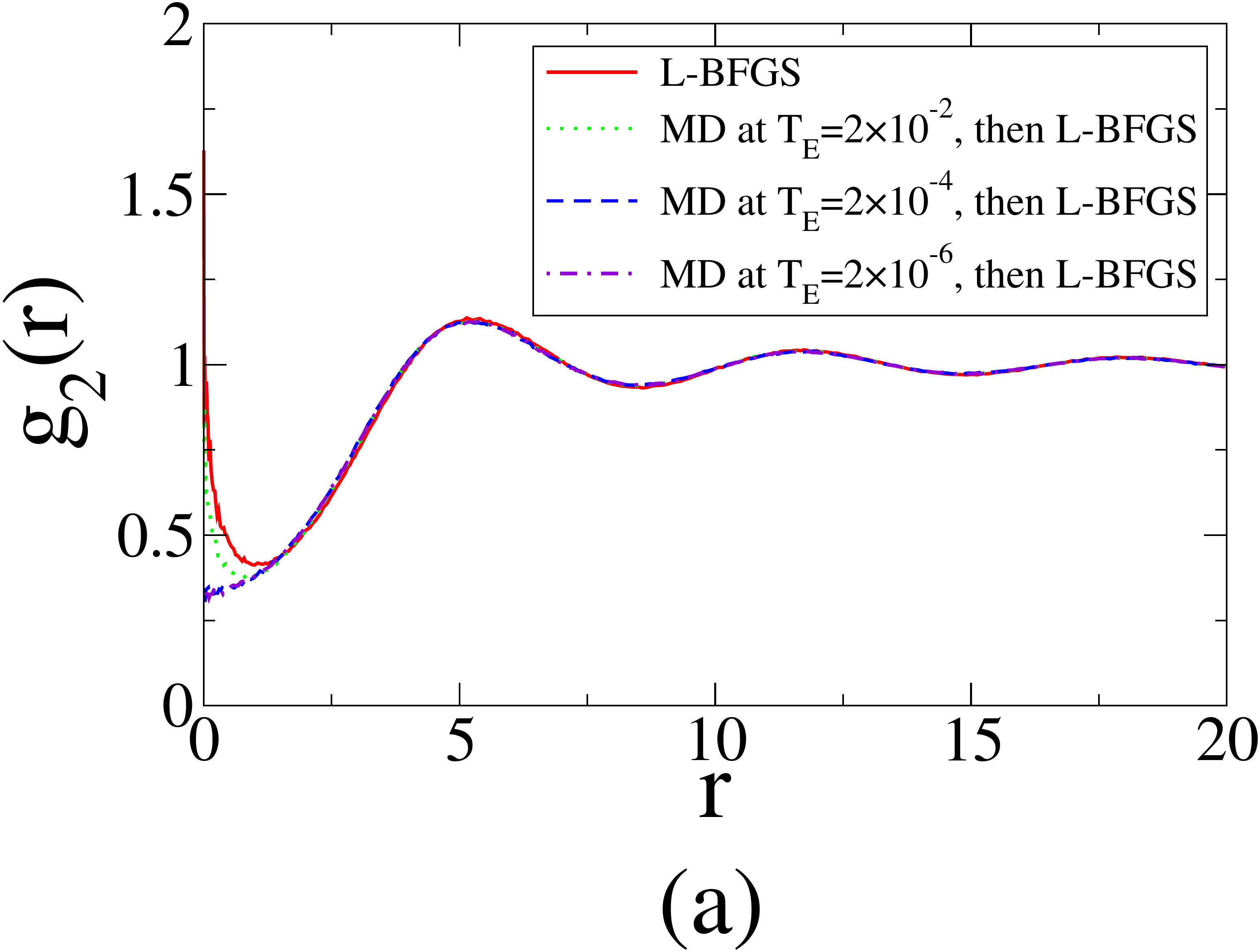}
\includegraphics[width=0.45\textwidth]{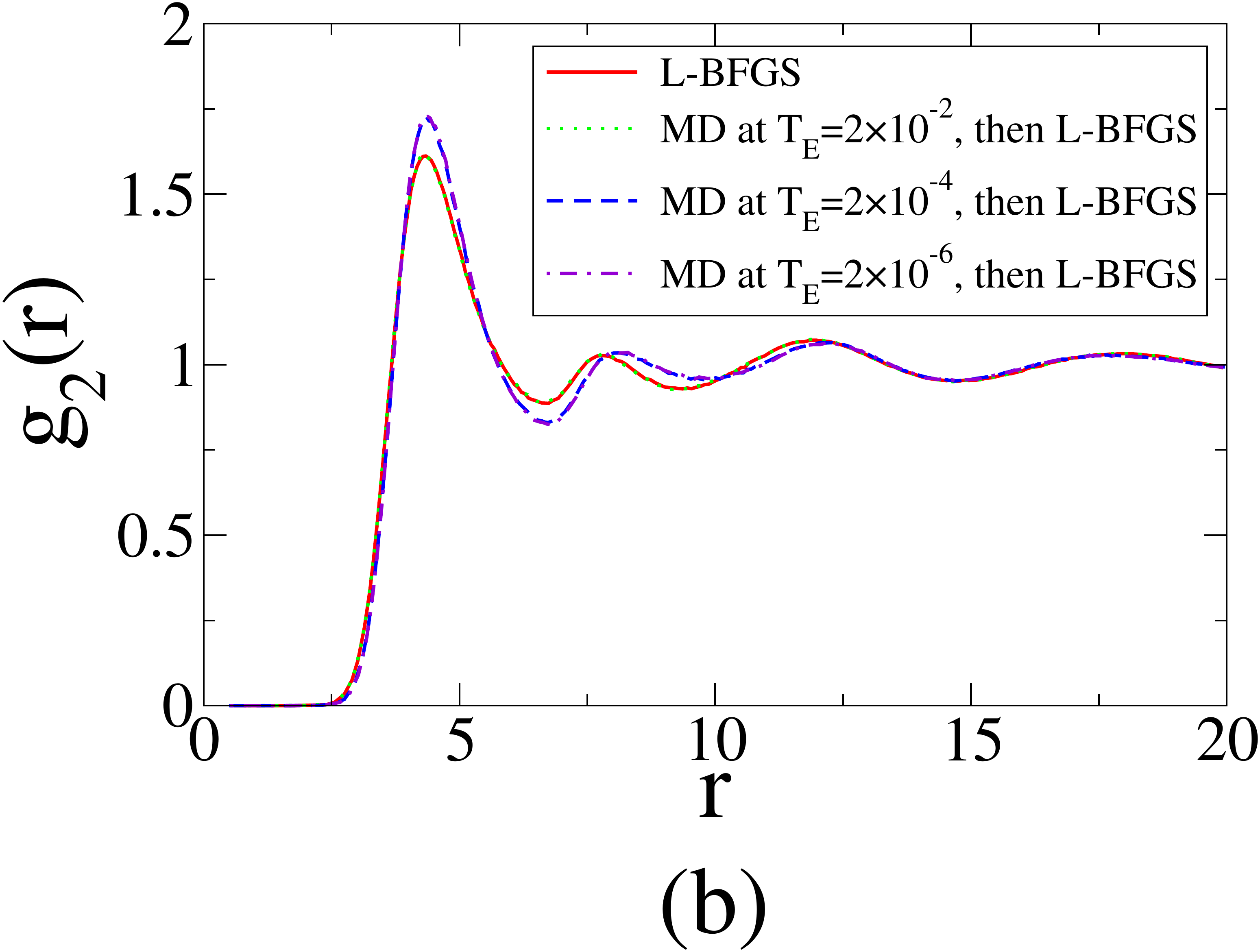}
\end{center}
\caption{(Color online) Pair correlation function produced by L-BFGS algorithm starting from snapshots of MD at different equilibration temperatures $\T$, (a) $\chi=0.2$ and (b) $\chi=0.4$. Each curve is averaged over 20 000 configurations of 136 particles each or 151 particles each.}
\label{T_E}
\end{figure}

In order to avoid the complexity caused by the details of various optimization algorithms, we turn our interest to the canonical ensemble in the $T \to 0$ limit. 
To sample this ensemble, we perform MD simulations at sufficiently small temperature $T_E$, periodically take ``snapshots,'' and then use a minimization algorithm to bring each snapshot to a ground state.
To determine a ``sufficiently small'' $T_E$, we calculated the pair correlation functions at various $T_E$'s and present them in Fig.~\ref{T_E}.
The energy minimization result starting from $T_E \to \infty$ initial configurations clearly display the ``clustering effect'' at $\chi=0.2$.
When $T_E$ goes to zero, the ``clustering effect'' also diminishes. 
At $\chi=0.4$, particles develop hard cores [$g_2(0)=0$], therefore there is no clustering even if $\T$ is large or infinite.
However, the peak height of $g_2(r)$ becomes dependent on $\T$ at this $\chi$ value. 
For both $\chi$ values, the pair correlation functions of the two lowest $\T$'s are almost identical, verifying that the $\T \to 0$ limit exists. These results show that $T_E=2\e{-6}$ is sufficiently small in two dimensions. Similarly, we have found that $T_E=2\e{-4}$ and $T_E=1\e{-6}$ are sufficiently small in one and three dimensions, respectively.
These temperatures are used in generating all of the results presented in Sec.~\ref{ensemble1}.

\begin{figure}[H]
\begin{center}

\includegraphics[width=0.45\textwidth]{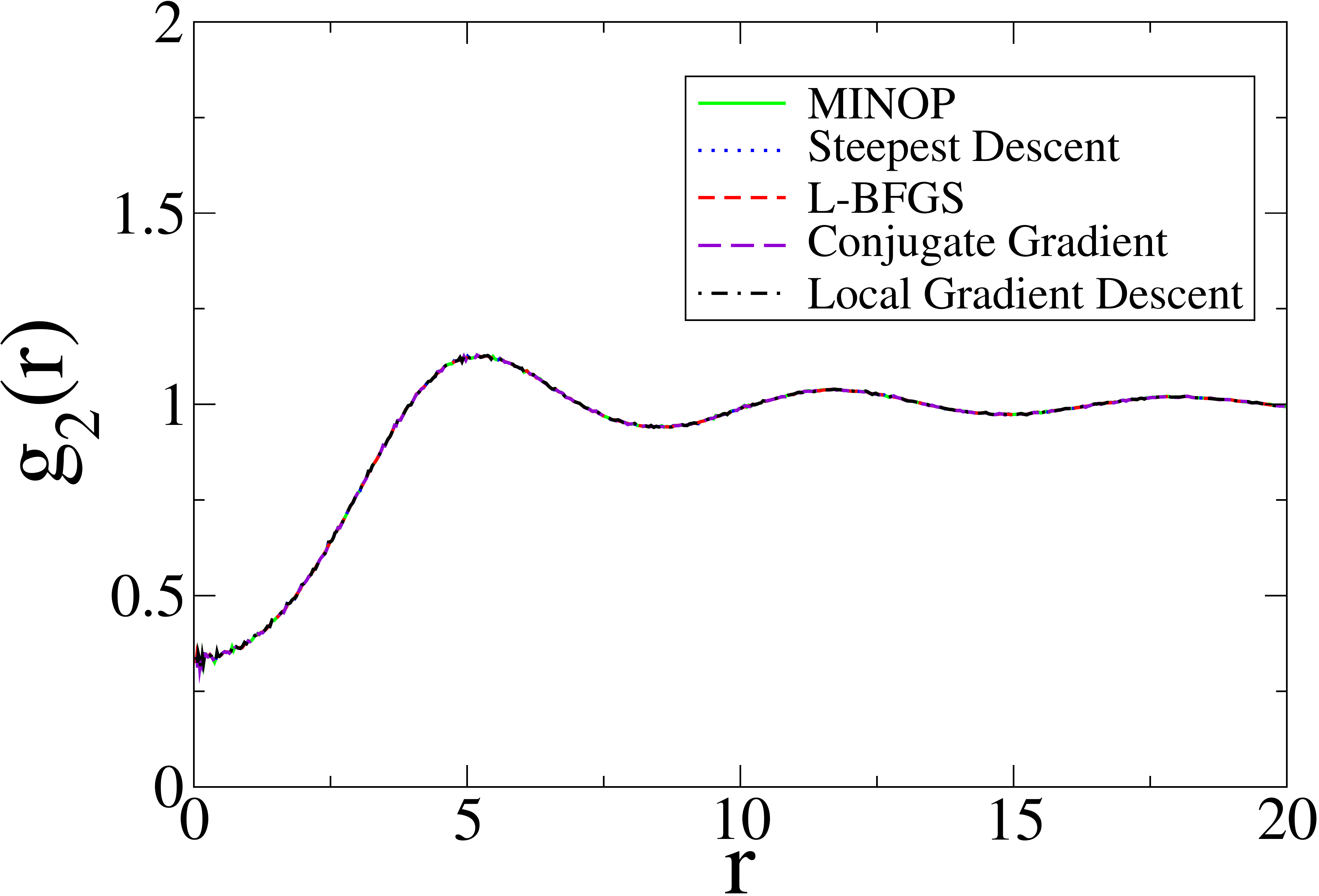}
\end{center}
\caption{(Color online) Pair correlation function produced by the five different algorithms starting from snapshots of MD at equilibration temperature $\T=2\e{-6}$ at $\chi=0.2$. Each curve is averaged over 20 000 configurations of 136 particles each.}
\label{alg}
\end{figure}

The energy minimization result starting from Poisson initial configurations differs for different algorithms, but the canonical ensemble in the $T \to 0$ limit should not depend on any particular algorithm.
After finding that $T_E=2\e{-6}$ is sufficiently small, we confirm the disappearing of algorithmic dependence by calculating the pair correlation function produced by different energy minimization algorithms starting from MD snapshots at $\T=2\e{-6}$.
Figure~\ref{alg} shows the results. The curves for all algorithms almost coincide.

\begin{figure}[H]
\begin{center}
\includegraphics[width=0.45\textwidth]{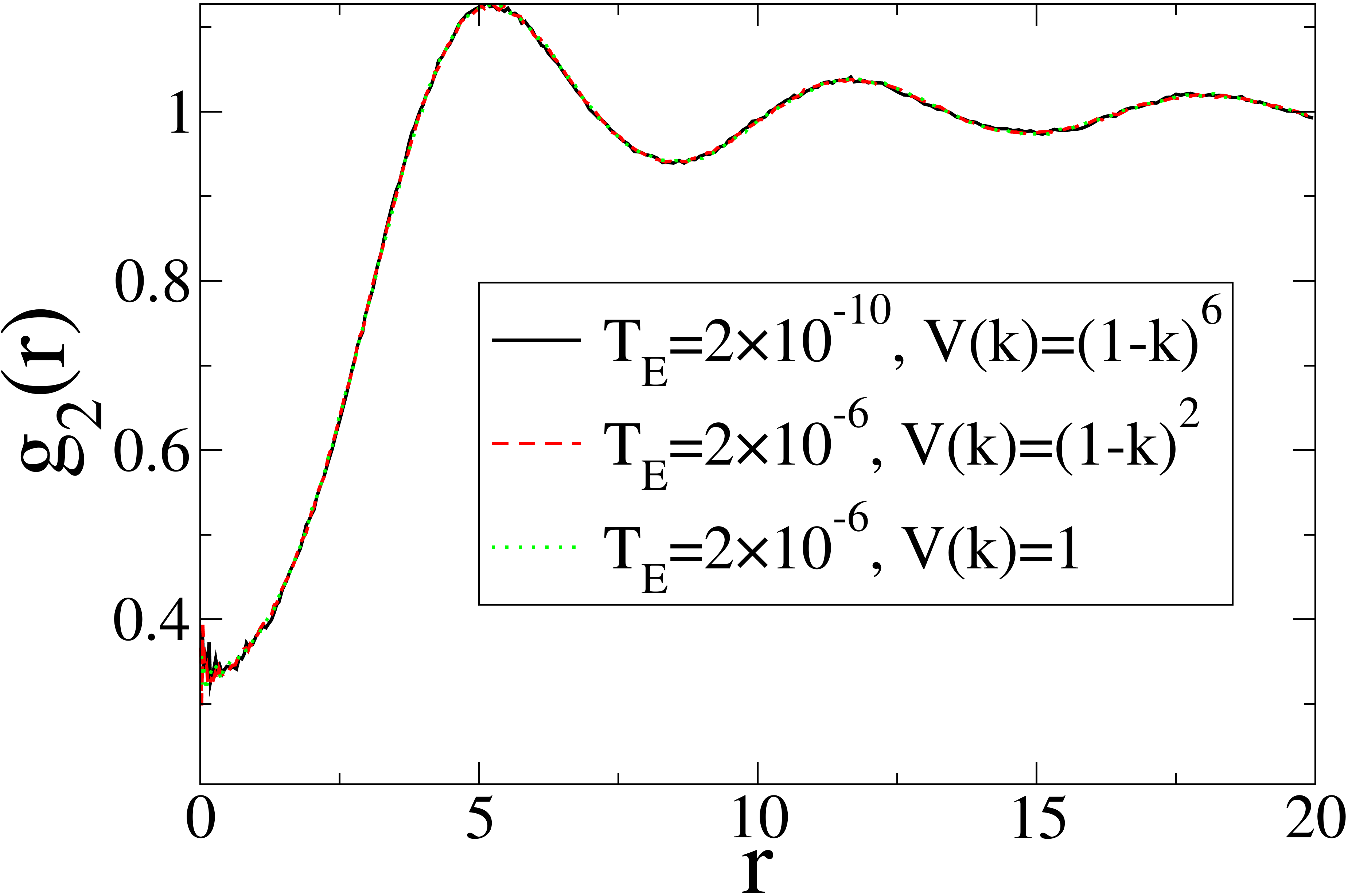}
\end{center}
\caption{(Color online) Pair correlation function produced by different potentials starting from snapshots of MD at sufficiently low temperature at $\chi=0.2$. Each curve is averaged over 20 000 configurations of 136 particles each.}
\label{ComparePot}
\end{figure}

Last, the function $V(k)$ in Eq.~\eqref{stealthy} can have different forms. This paper mainly use $V(k)=1$ but we also want to know if the results obtained using this form
are equivalent to those generated using other positive isotropic forms of $V(k)$ as well.
In principle, stealthy potentials of any form should have the same set of ground-state
configurations, but the form of the stealthy potential could theoretically affect the curvature of the potential energy surface near each ground-state configurations and thus also affect their relative weights.
Figure~\eqref{ComparePot} shows the pair correlation function produced by different $V(k)$'s. The pair correlation functions for $V(k)=1$ and $V(k)=(1-k)^2$ at $\T=2\e{-6}$ are almost identical.
For $V(k)=(1-k)^6$, we initially tried $\T=2\e{-6}$ but found that the ``clustering effect'' is still noticeable. 
We further lowered the temperature to $\T=2\e{-10}$ to completely suppress the ``clustering effect'' to produce a pair correlation function identical to that of $V(k)=1$ and $V(k)=(1-k)^2$ potentials.
This result suggests that the functional form of $V(k)$ does not produce noticeable differences in the ground-state ensembles in the $T \to 0$ limit of the canonical ensemble.

\section{Canonical ensemble in the $T\to 0$ limit}
\label{ensemble}

We will show here that the entropically favored ground states in the canonical ensemble in the $T\to 0$ limit for the first three space dimensions differ markedly below and above $\chi = 0.5$. 
For $\chi < 0.5$, the entropically favored ground states are disordered while for $\chi \ge 0.5$ the entropically favored ground states are crystalline.
Therefore, we will characterize them differently. For $\chi < 0.5$, we will report the pair correlation function, structure factor, and Voronoi cell statistics.
For sufficiently small $\chi$, we will show that the simulation results agree well with theory \cite{torquato2015ensemble}.
For $\chi \ge 0.5$, we will report the crystal structures. 
The numbers of particles in all of the systems reported in this section are collected in Appendix~\ref{number}.

\subsection{$\chi<0.5$ region}
\label{ensemble1}

\begin{figure}[H]
\begin{center}
\includegraphics[width=0.98\textwidth]{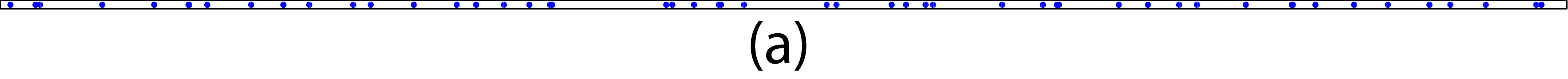}
\includegraphics[width=0.98\textwidth]{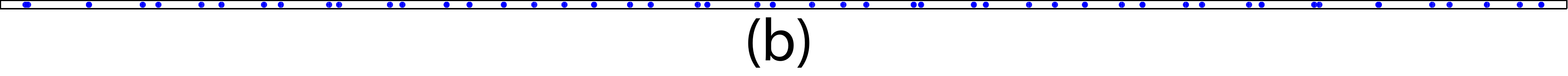}
\end{center}
\caption{(Color online) Representative one-dimensional entropically favored stealthy ground states at (a) $\chi=0.1$ and (b) $\chi=0.4$.}
\label{conf_1d}
\end{figure}

\begin{figure}[H]
\begin{center}
\includegraphics[width=0.45\textwidth]{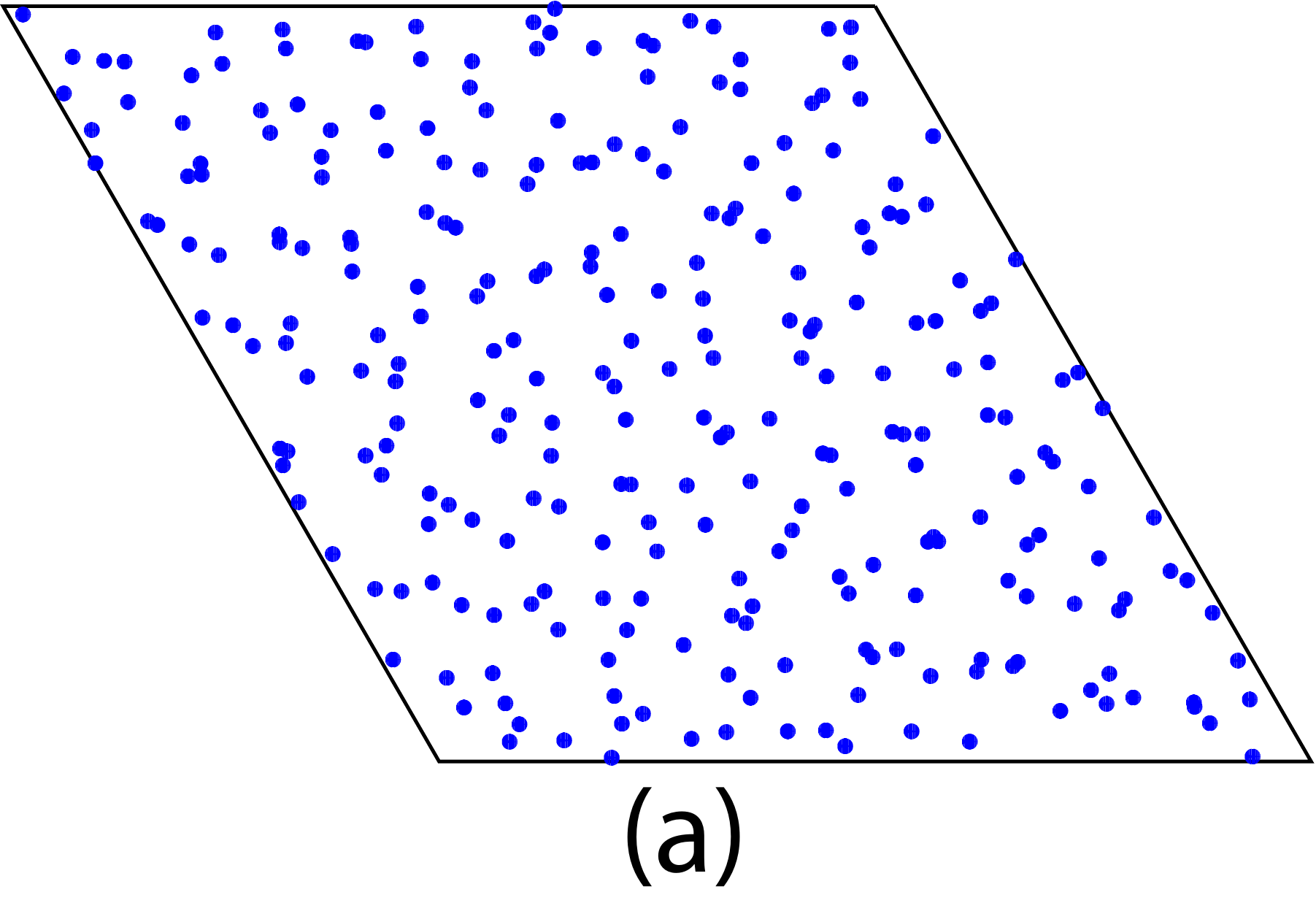}
\includegraphics[width=0.45\textwidth]{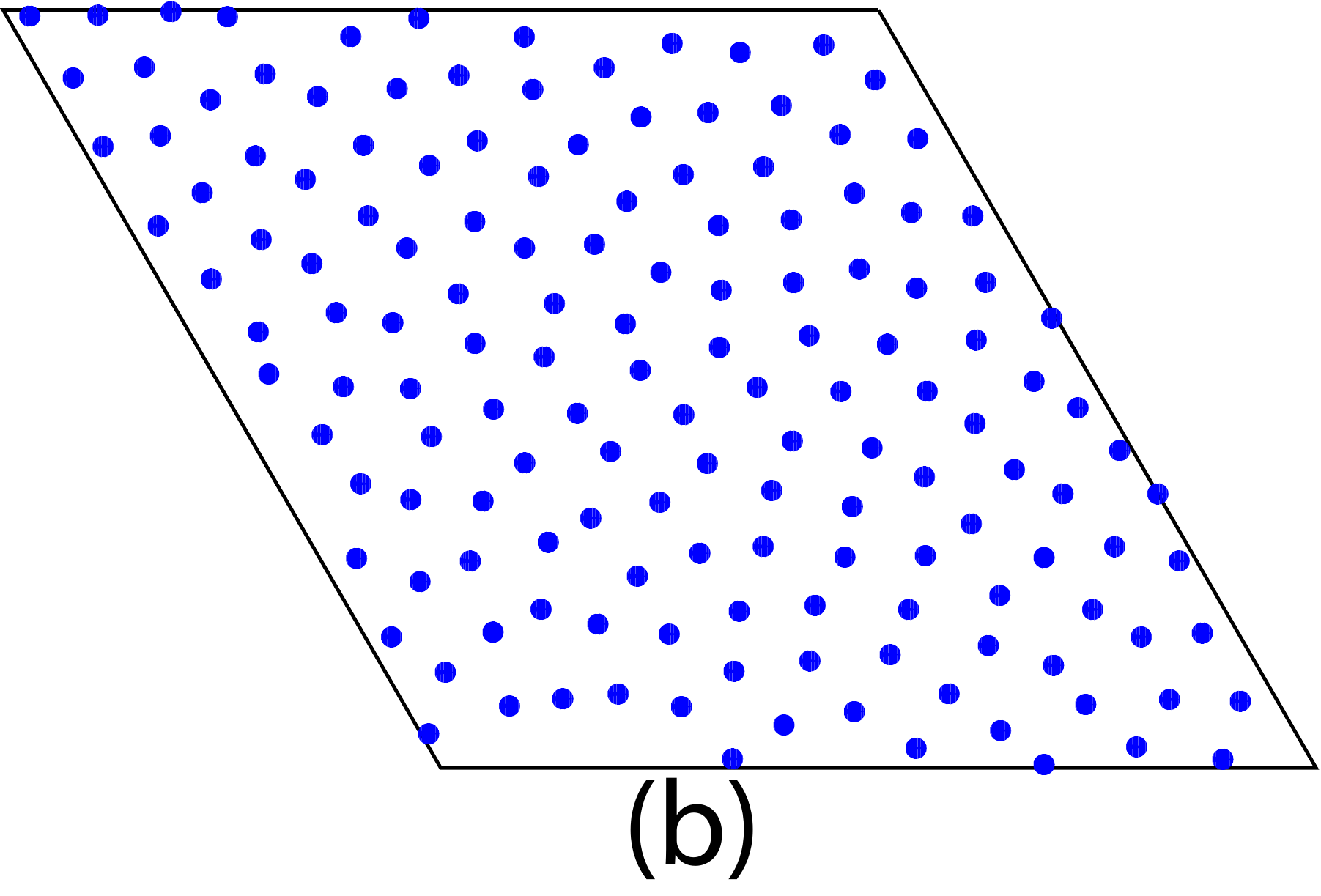}
\end{center}
\caption{(Color online) Representative two-dimensional entropically favored stealthy ground states at (a) $\chi=0.1$ and (b) $\chi=0.4$.}
\label{conf_2d}
\end{figure}

\begin{figure}[H]
\begin{center}
\includegraphics[width=0.45\textwidth, trim=280 100 280 280, clip]{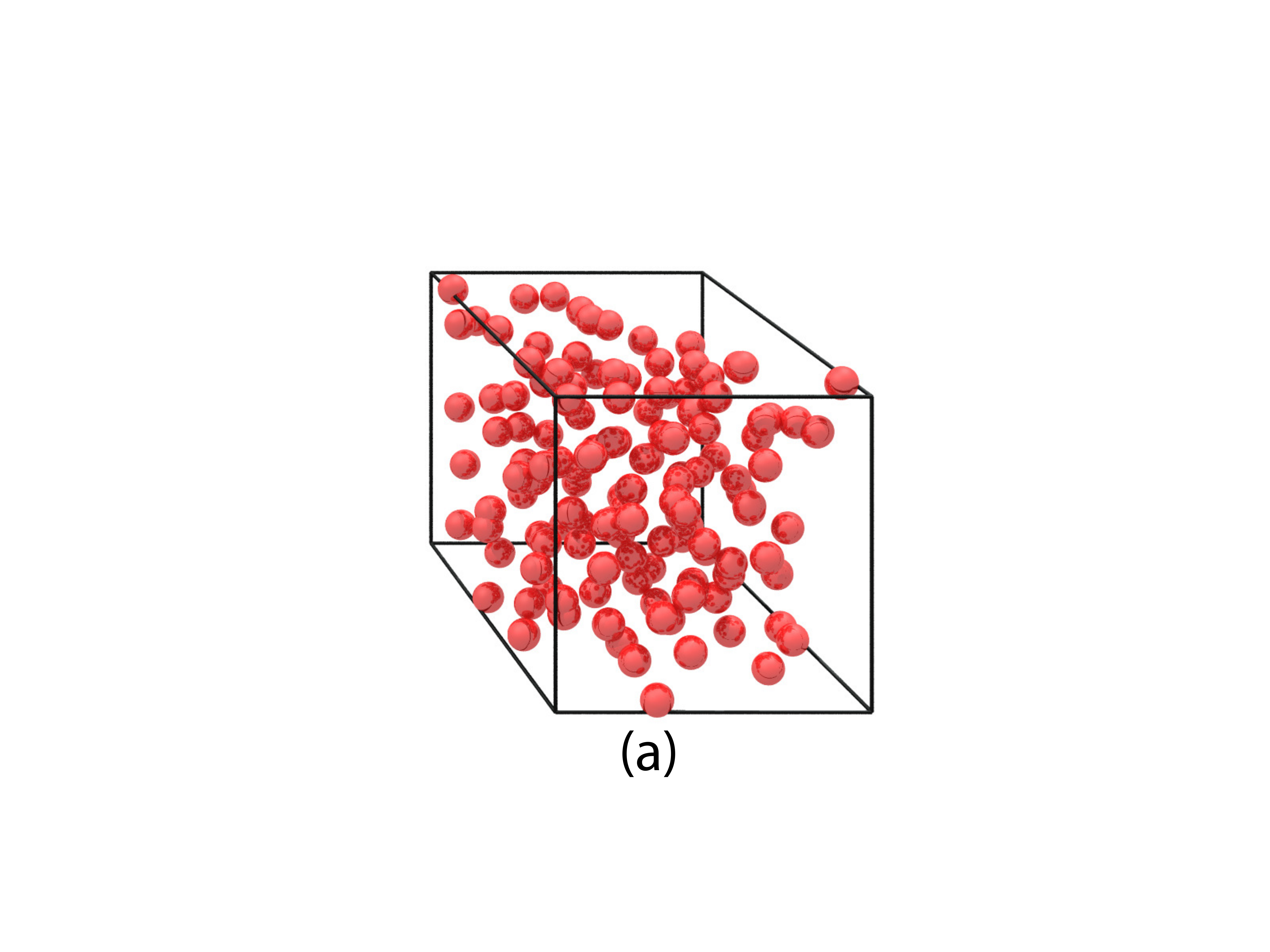}
\includegraphics[width=0.45\textwidth, trim=280 100 280 280, clip]{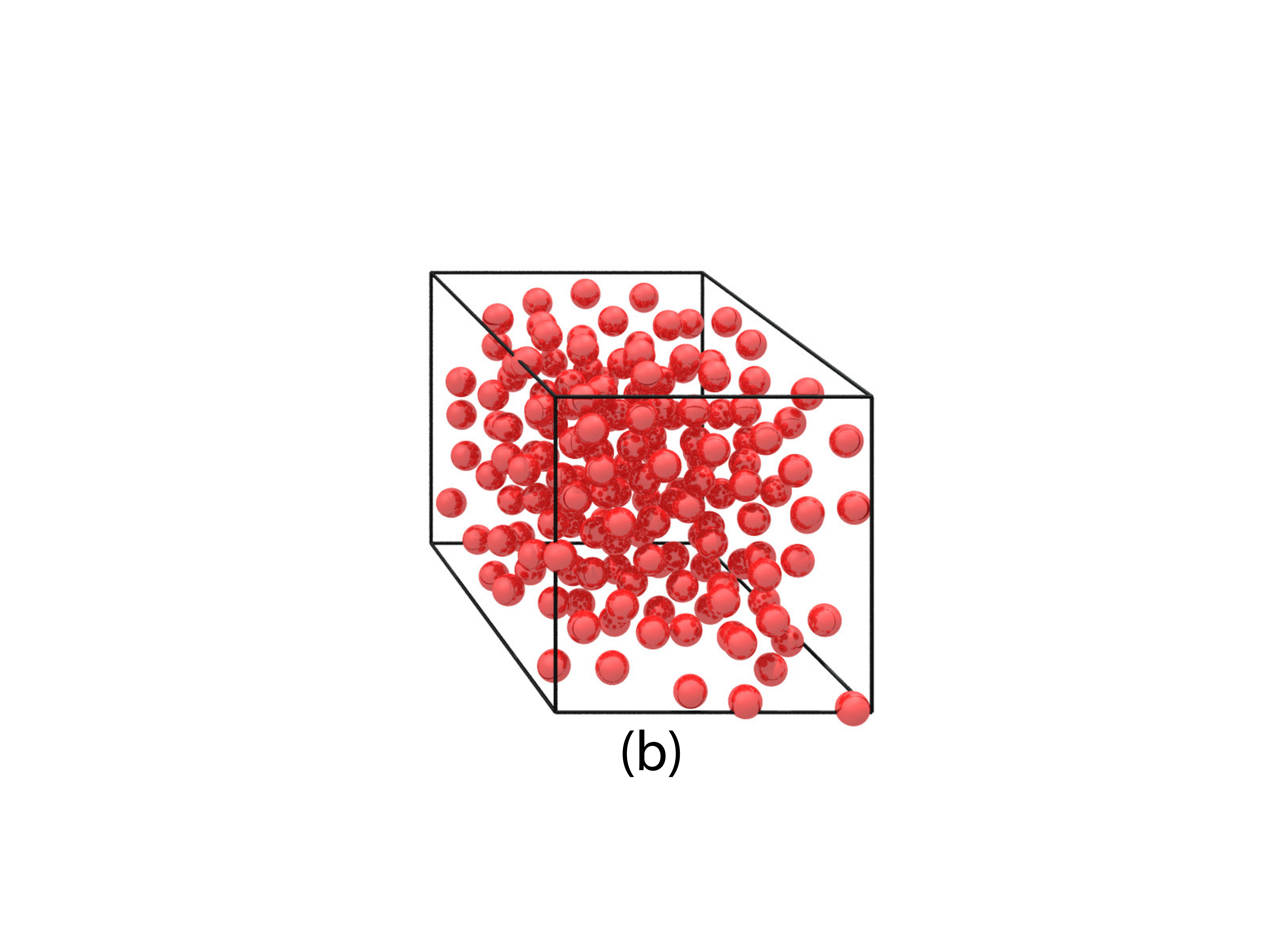}
\end{center}
\caption{(Color online) Representative three-dimensional entropically favored stealthy ground states at (a) $\chi=0.1$ and (b) $\chi=0.4$.}
\label{conf_3d}
\end{figure}

Representative entropically favored stealthy ground states in the first three space dimensions at $\chi=0.1$ and $\chi=0.4$ are shown in Figs.~\ref{conf_1d}-\ref{conf_3d}. 
As $\chi$ increases from 0.1 to 0.4, the stealthiness increases, accompanied with a visually perceptible increase in short-range order. 
This trend in short-range order is consistent with previous studies \cite{uche2004constraints, uche2006collective, batten2008classical}.

\onecolumngrid


\begin{figure}[H]
\begin{center}
\includegraphics[width=0.3\textwidth]{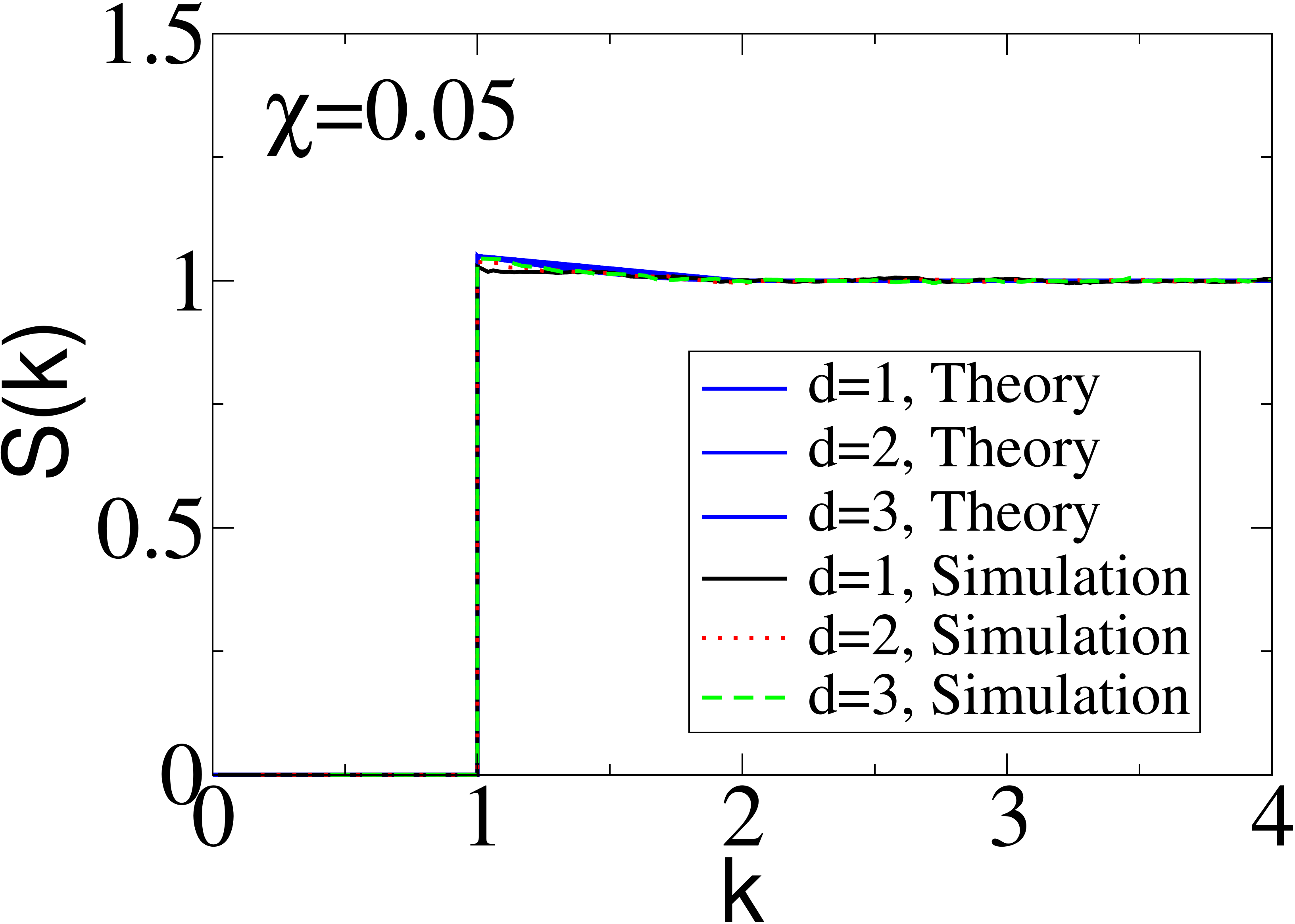}
\includegraphics[width=0.3\textwidth]{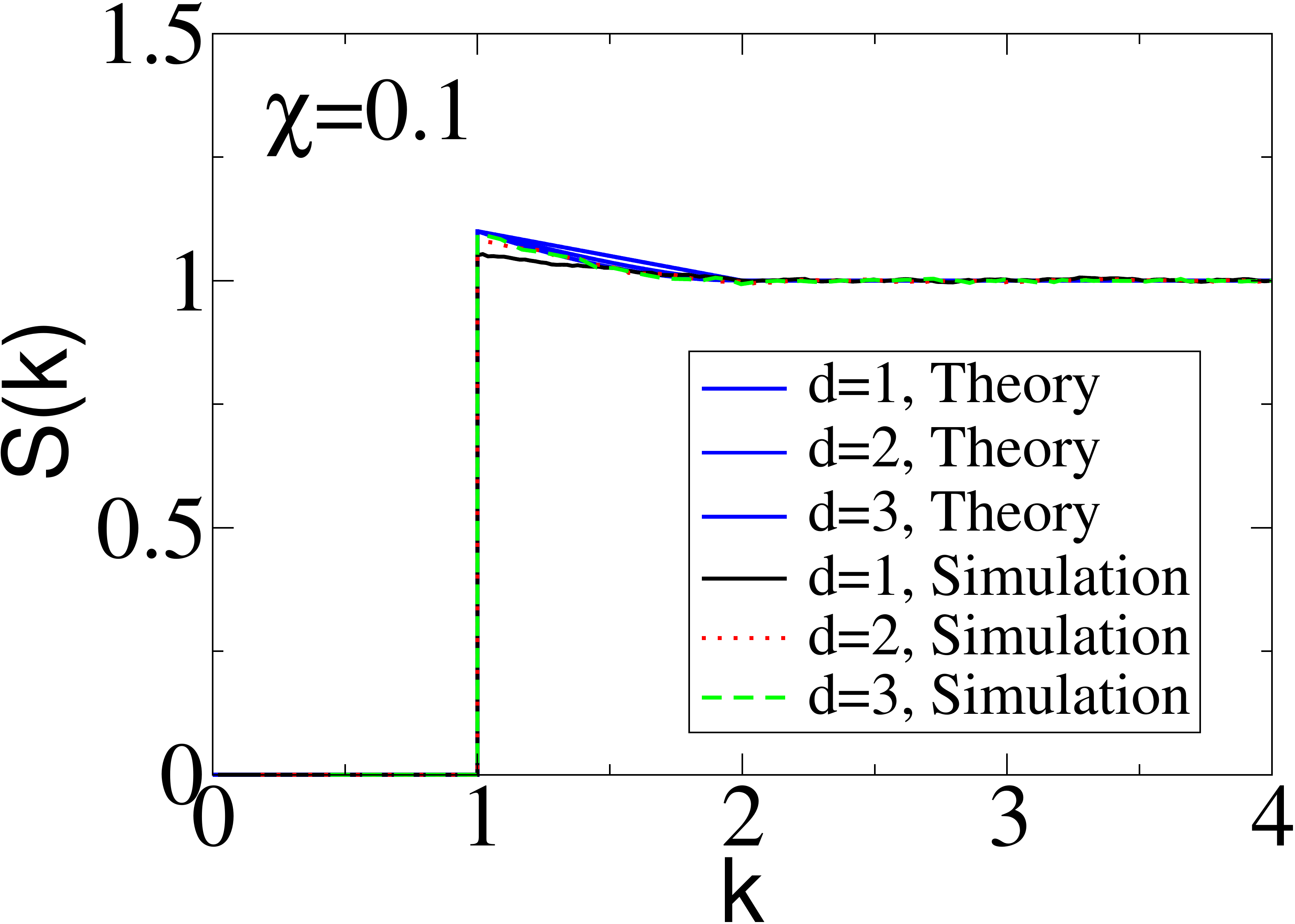}
\includegraphics[width=0.3\textwidth]{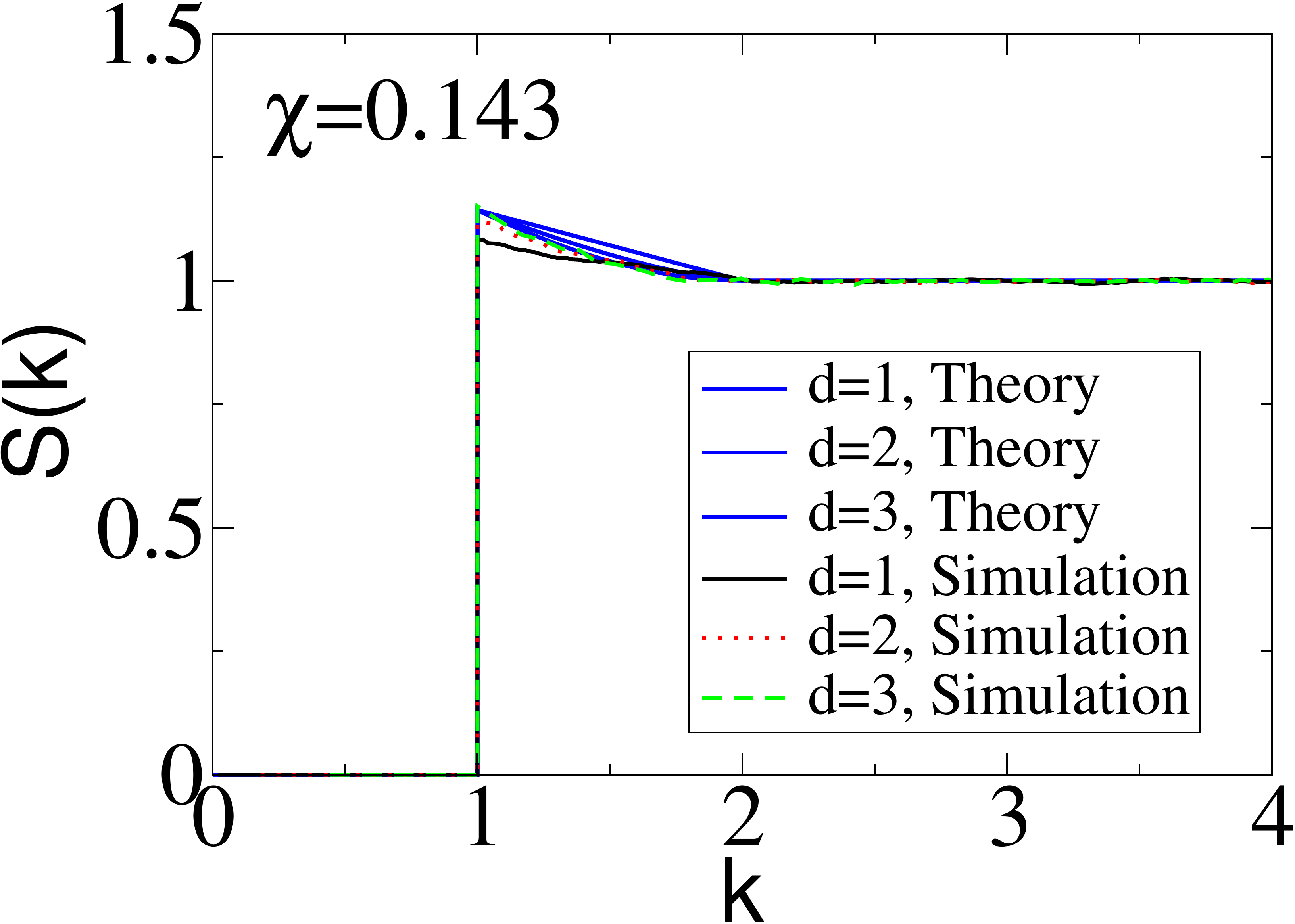}

\includegraphics[width=0.3\textwidth]{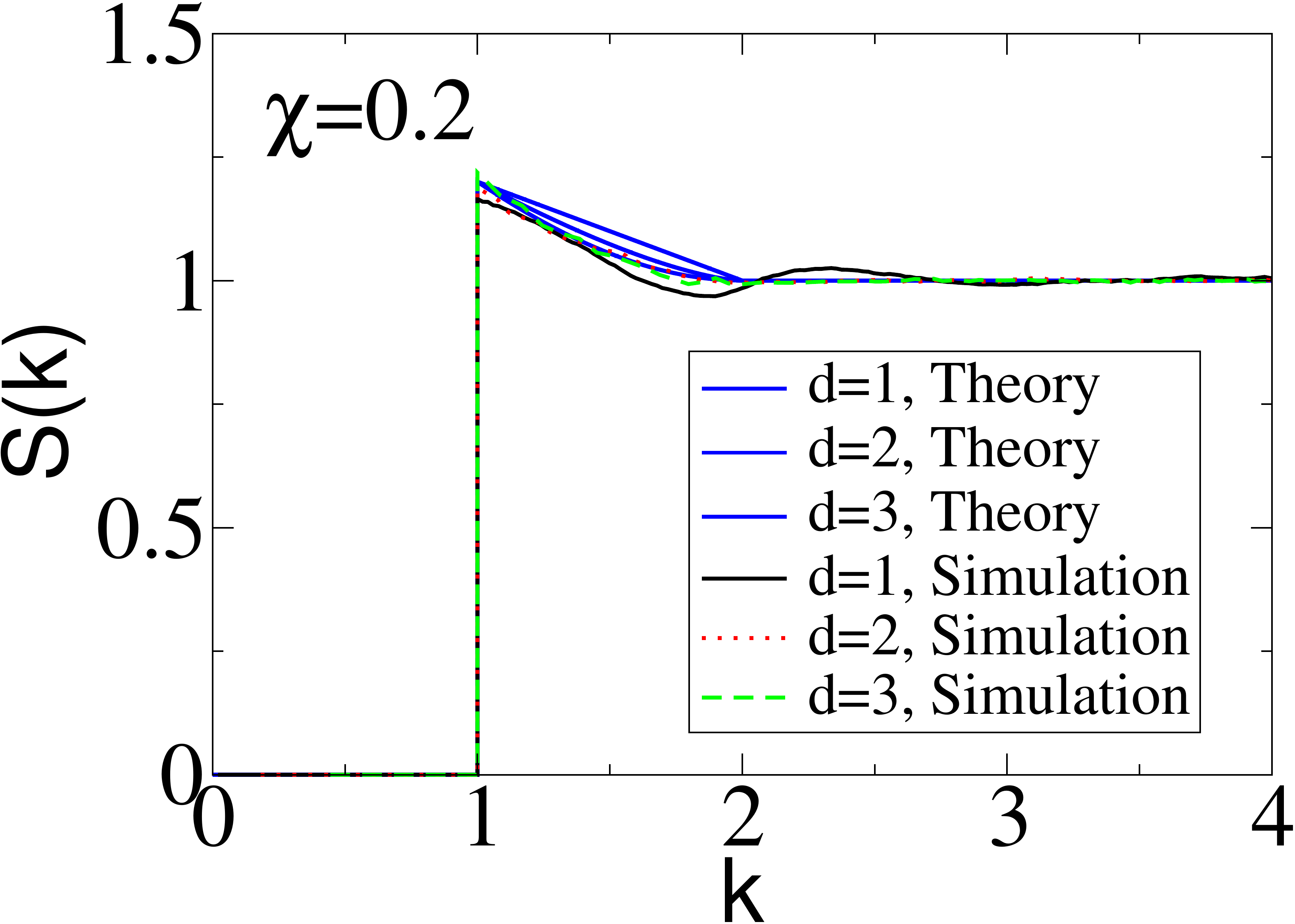}
\includegraphics[width=0.3\textwidth]{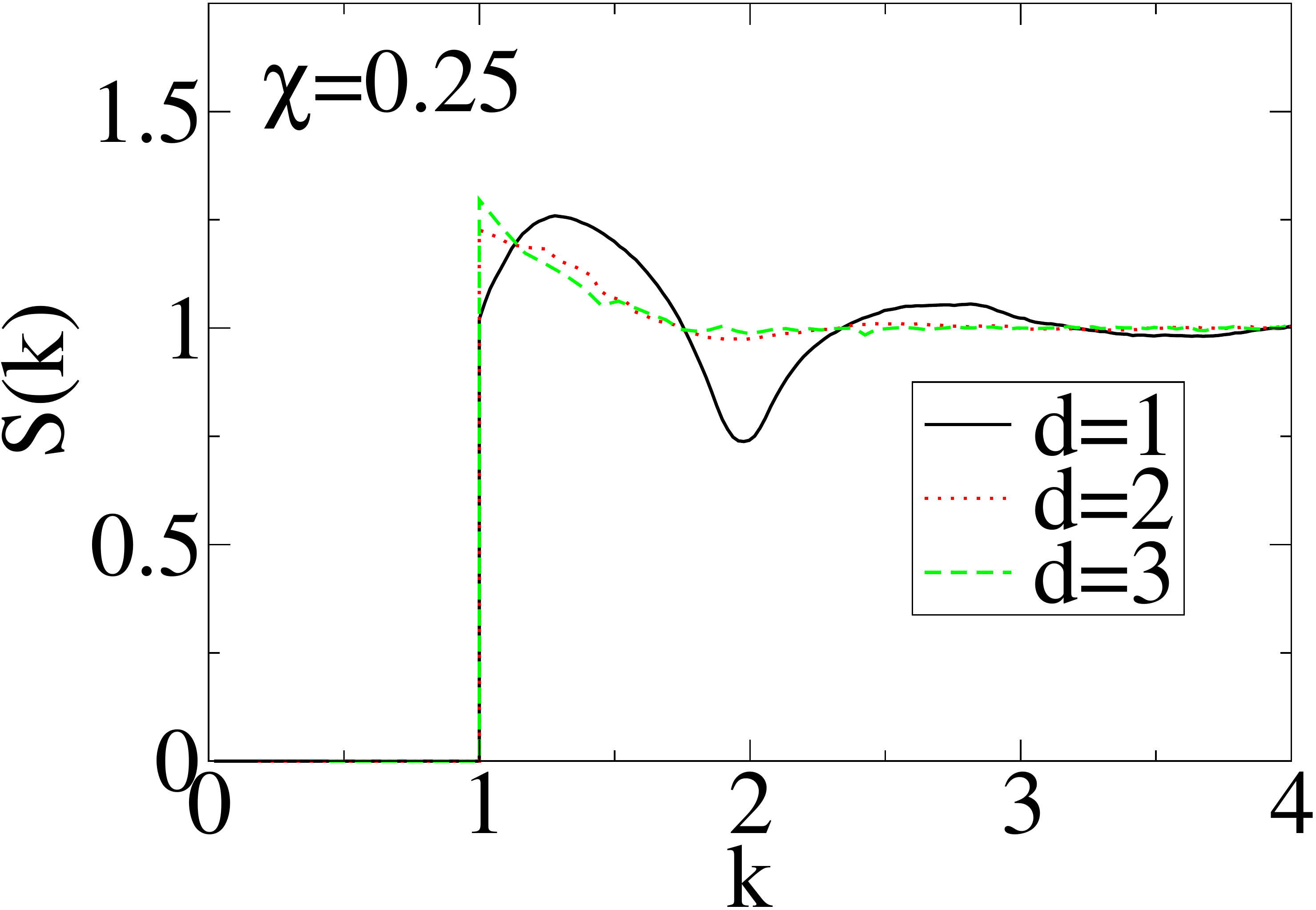}
\includegraphics[width=0.3\textwidth]{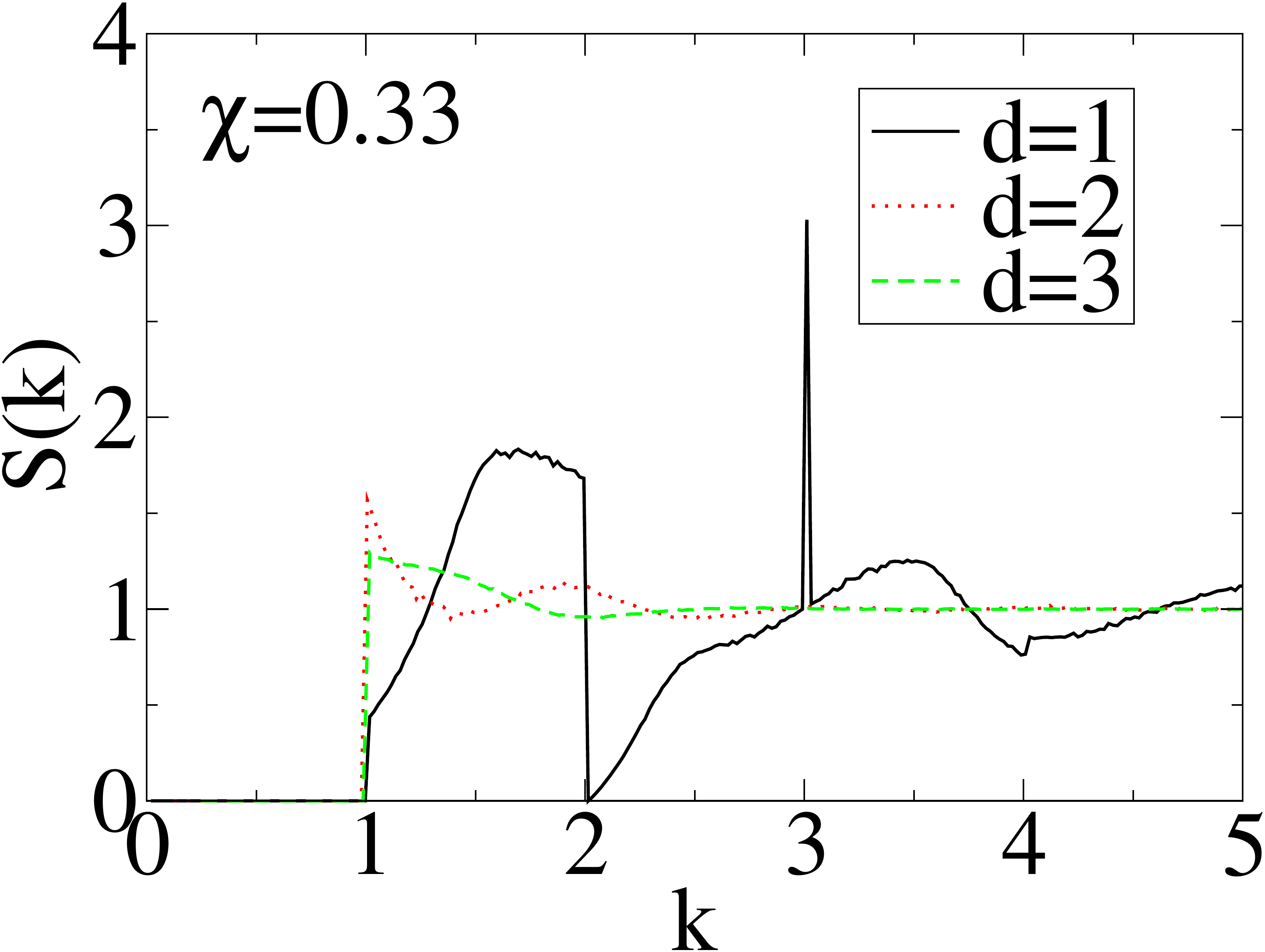}

\end{center}
\caption{(Color online) Structure factors for $1 \le d \le 3$ for $0.05 \le \chi \le 0.33$ from simulations and theory \cite{torquato2015ensemble}. The smaller $\chi$ simulation results are also compared with the theoretical results in the infinite-volume
limit \cite{torquato2015ensemble}. For $\chi \le 0.1$, the theoretical and simulation curves
are almost indistinguishable, and the structure factor is almost independent of the space dimension. However, simulated $S(k)$ in different dimensions become very different at larger $\chi$. Theoretical results for $\chi \ge 0.25$ are not presented because they are not valid in this regime.}
\label{s_dim}
\end{figure}

\begin{figure}[H]
\begin{center}
\includegraphics[width=0.3\textwidth]{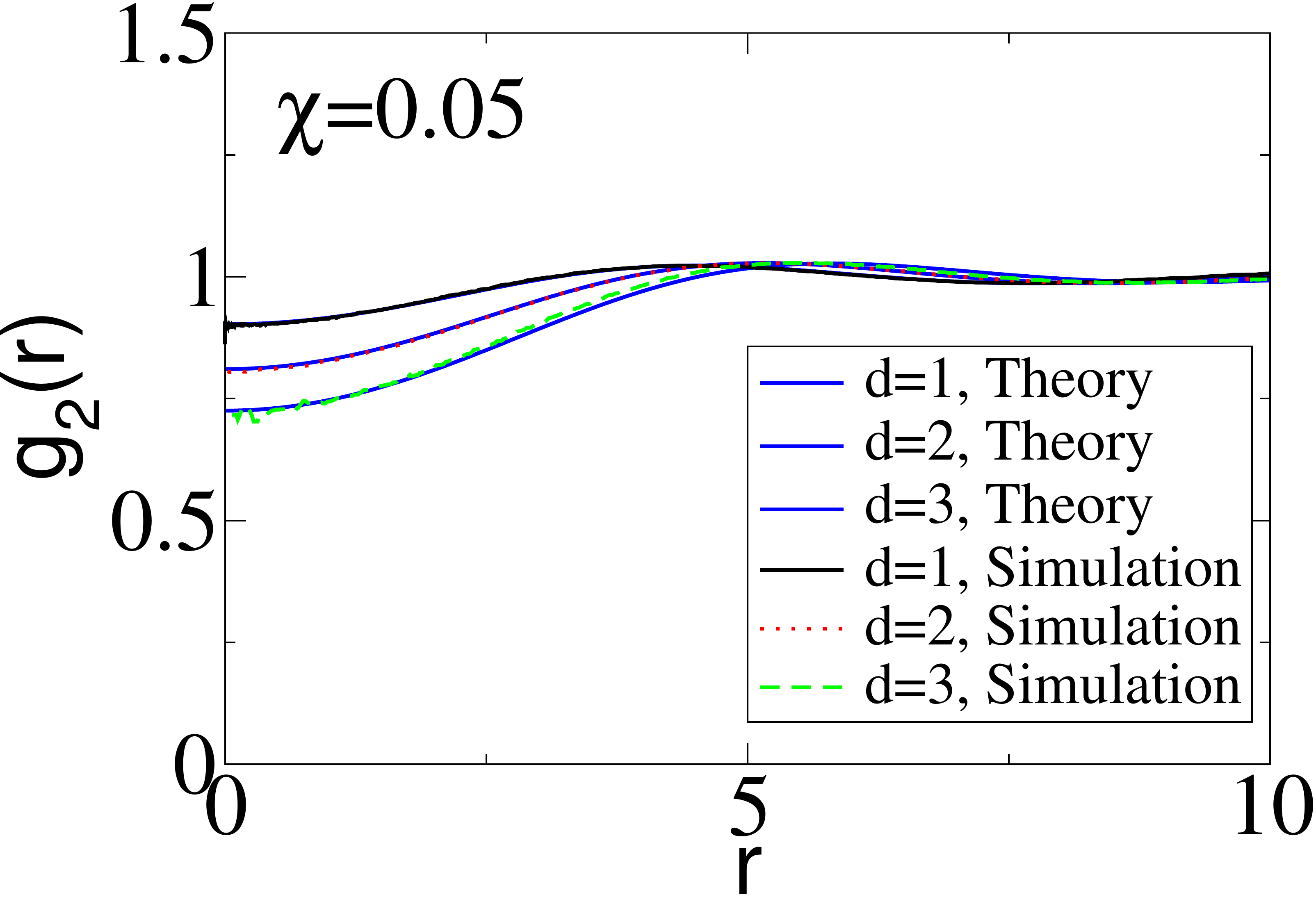}
\includegraphics[width=0.3\textwidth]{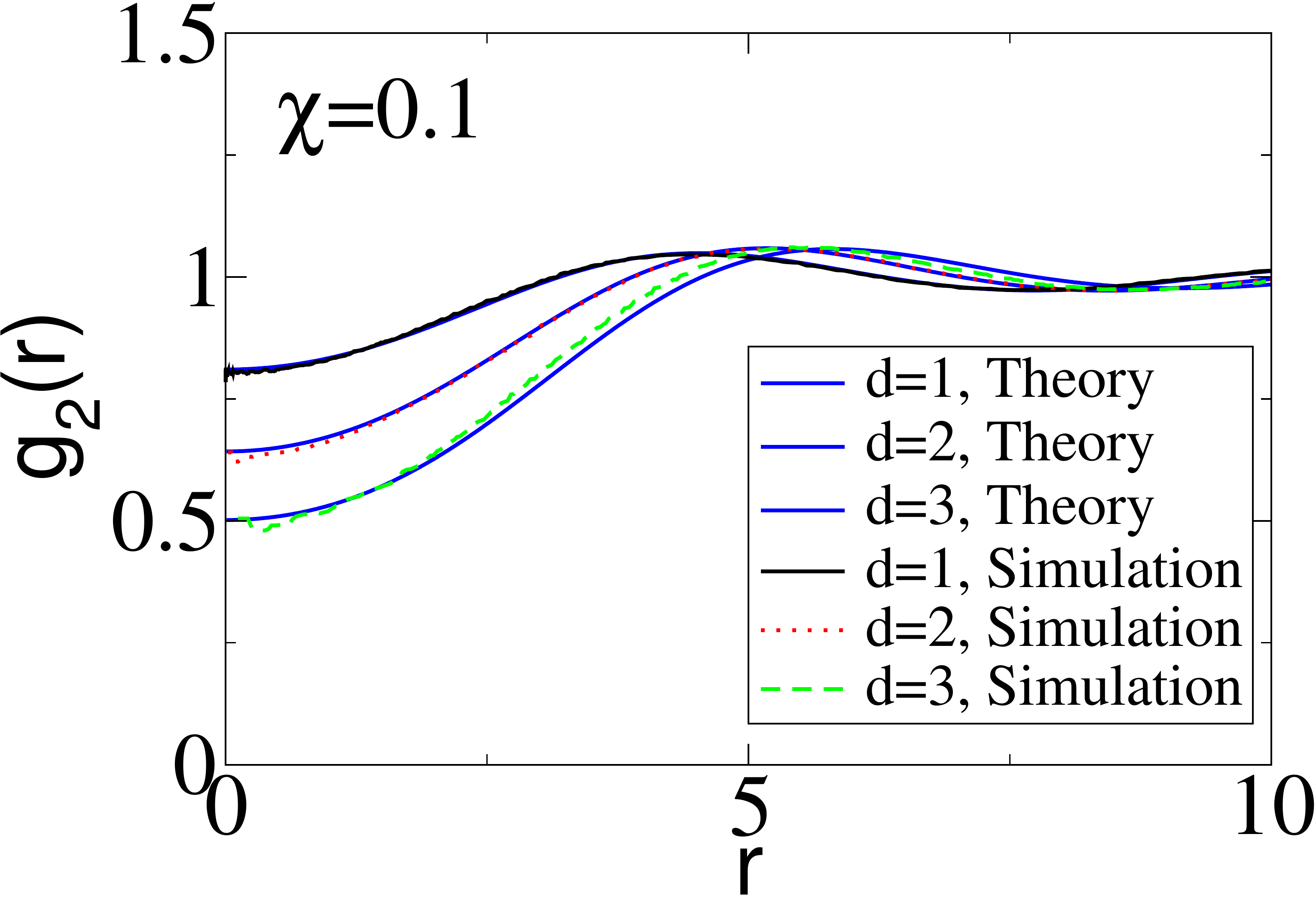}
\includegraphics[width=0.3\textwidth]{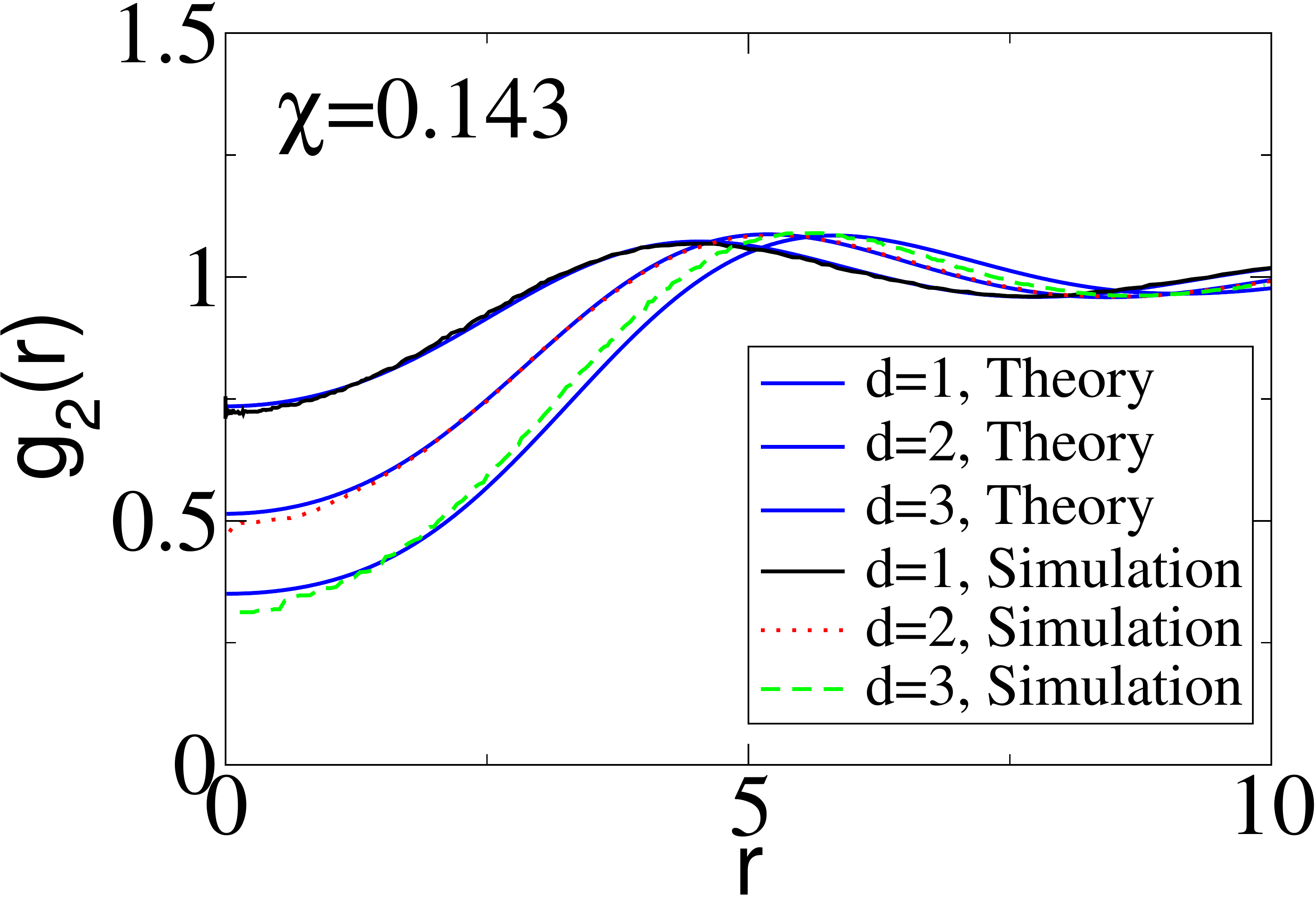}

\includegraphics[width=0.3\textwidth]{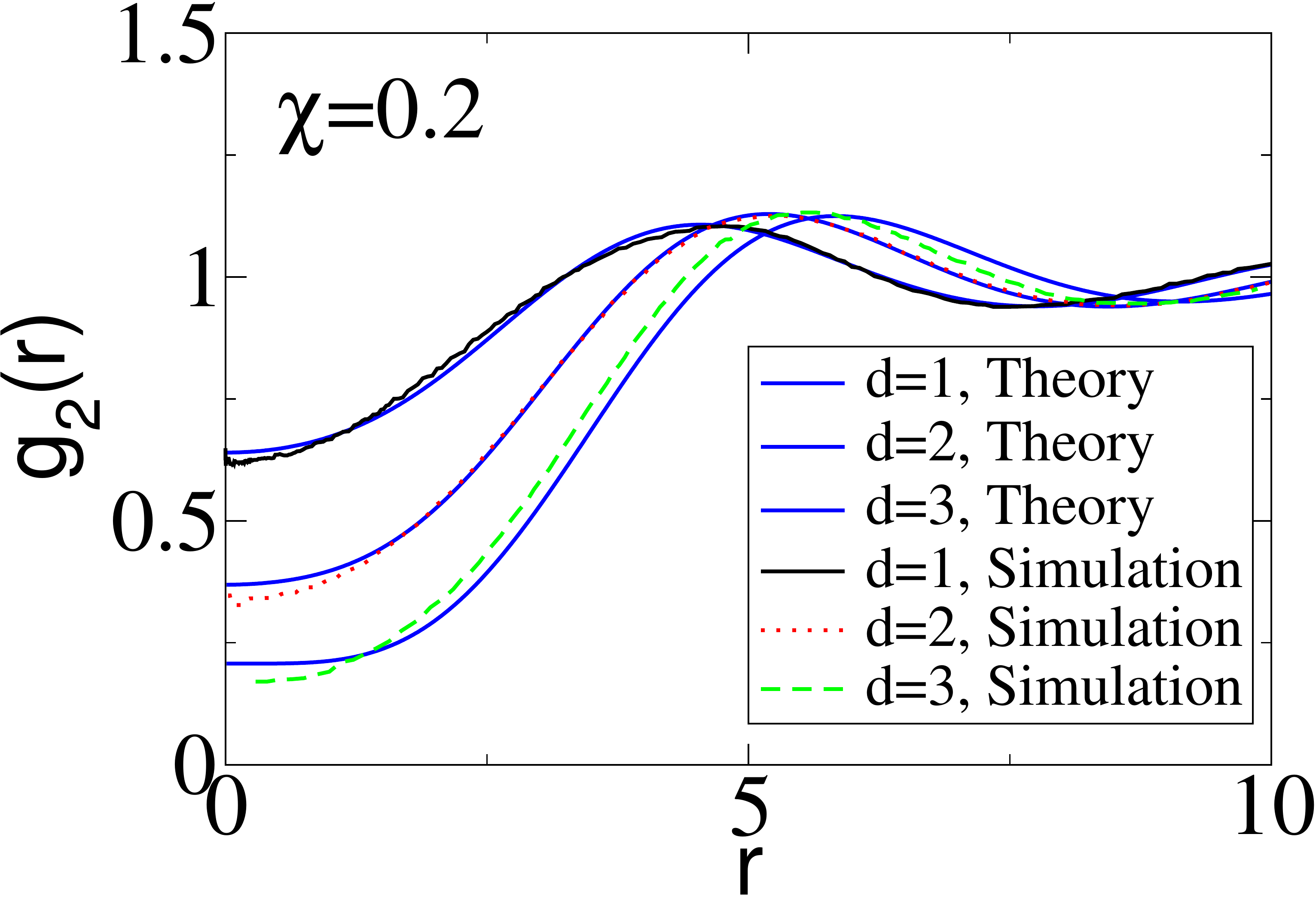}
\includegraphics[width=0.3\textwidth]{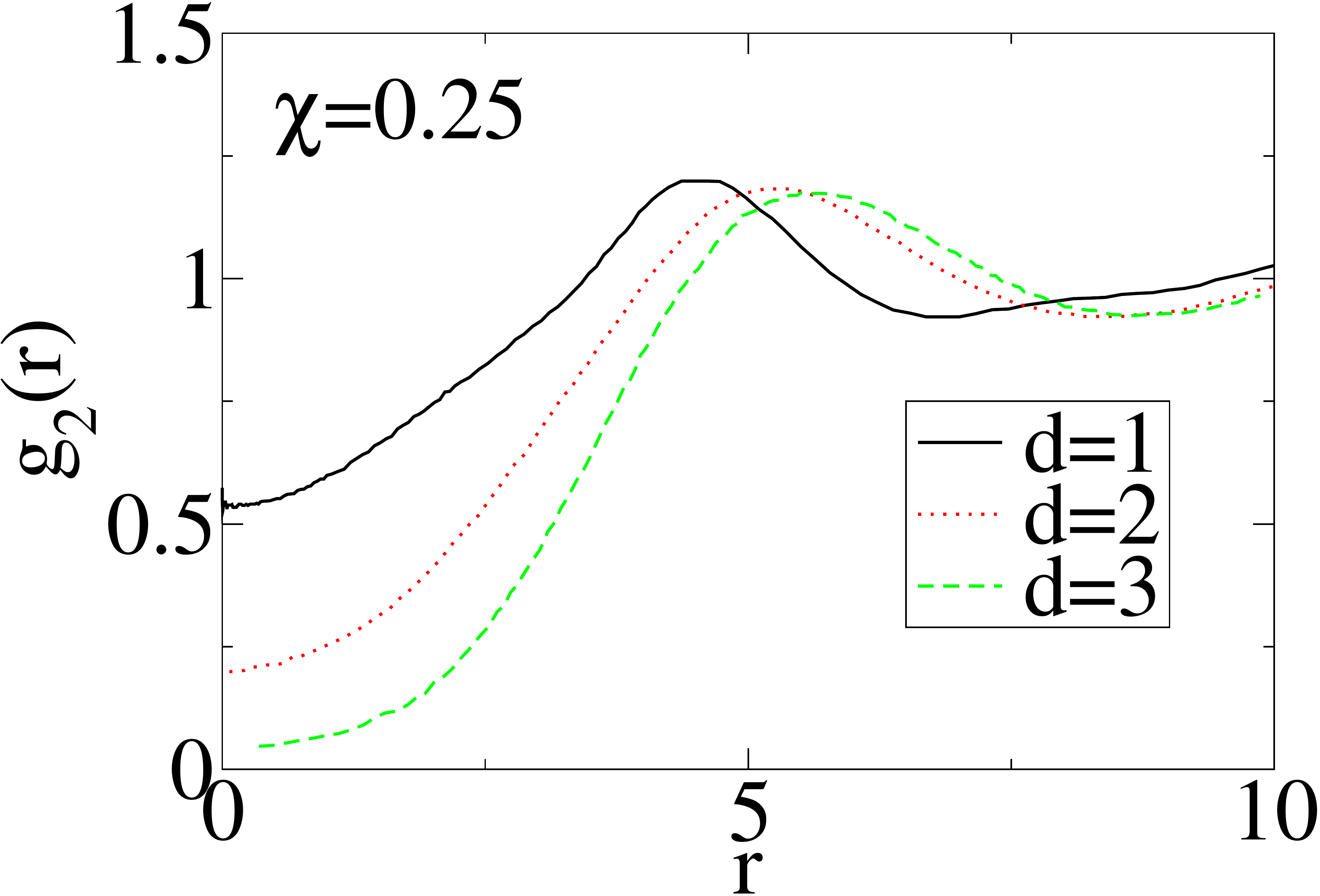}
\includegraphics[width=0.3\textwidth]{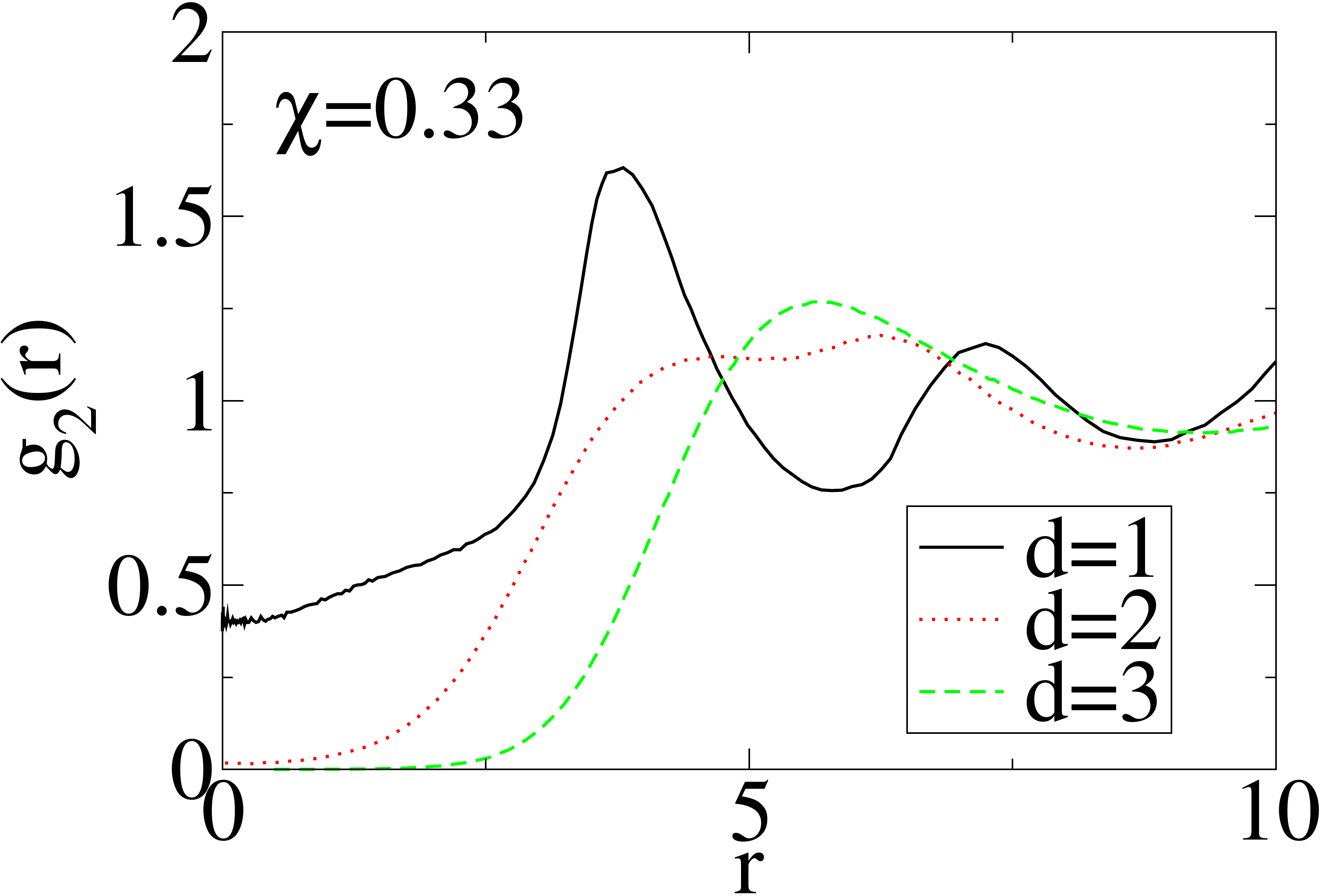}

\end{center}
\caption{(Color online) Pair correlation functions for $1 \le d \le 3$ for $0.05 \le \chi \le 0.33$ from simulations and theory \cite{torquato2015ensemble}. The smaller $\chi$ simulation results are also compared with the theoretical results in the infinite-volume
limit \cite{torquato2015ensemble}. For $\chi \le 0.1$, the theoretical and simulation curves
are almost indistinguishable. Theoretical results for $\chi \ge 0.25$ are not presented because they are not valid in this regime.}
\label{g2_dim}
\end{figure}
\twocolumngrid

We have calculated the pair correlation functions and the structure factors for various $\chi$ values. 
Results for $0.05 \le \chi \le 0.33$ are shown in Figs.~\ref{s_dim}~and~\ref{g2_dim}.
The $\chi < 0.2$ results are in excellent agreement with the ``pseudo-hard-sphere ansatz,'' which states that the structure factor behaves like pseudo equilibrium hard-sphere systems in Fourier space \cite{torquato2015ensemble}.
However, the theory gradually becomes invalid as $\chi$ increases.

The pair correlation functions of the entropically favored stealthy ground states are shown in Fig.~\ref{g2_dim}. 
When $\chi \le 0.2$, since the structure factor is similar to the pair correlation function of the hard-sphere system, inversely the pair correlation function is also similar to the structure factor of the hard-sphere system.
As $\chi$ grows larger, the pseudo hard-sphere ansatz gradually deviates from the simulation result.

We have checked that these pair statistics are consistent with four theoretical integral conditions of the pair statistics in the infinite-volume limit \cite{torquato2015ensemble}.
The first three conditions are Eqs.~(58), (59), and (63) of Ref.~\onlinecite{torquato2015ensemble}, which are 
\begin{equation}
\int_{\mathbb R^d} P(r) d \mathbf r =0,
\end{equation}
\begin{equation}
\int_{\mathbb R^d} P(r) v(r) d \mathbf r =0,
\end{equation}
and
\begin{equation}
g_2(0)=1-2d\chi +2d^2 \chi \int_K^\infty k^{d-1} {\tilde Q}(k) dk,
\end{equation}
where $P(r)$ is the inverse Fourier transform of $\Theta(k-1){\tilde Q}(k) $, $\Theta(x)$ is the Heaviside step function, 
and ${\tilde Q}(k)=S(k)-1$.

The fourth condition is that the pressure calculated from the ``virial equation'' \cite{torquato2015ensemble} has to be either nonconvergent or convergent to the pressure calculated from the energy route \cite{torquato2015ensemble}.
All pair statistics in Figs.~\ref{s_dim} and \ref{g2_dim} were generated using the step-function potential [the $V(k)=1$ case of Eq.~\eqref{stealthy}], but this potential does not lead to a convergent virial pressure.
However, as we have shown earlier, the stealthy ground states that we generated here are also the ground states of other stealthy functional forms ${\tilde v}({\mathbf k})$.
In one dimension, to test our simulation procedure,
we used the potential form $V(k)=(1-k)$
to calculate the pressure from both the virial equation (Eq.~(43) of Ref.~\onlinecite{torquato2015ensemble}) and the energy equation (Eq.~(41) of Ref.~\onlinecite{torquato2015ensemble}).
The pressure from the virial equation converges and agrees with the exact pressure from the energy equation, thus confirming the accuracy of our numerical results.
These checks involve integrals of $g_2(r)$ and $S(k)$ that are only slowly converging. Therefore, passing them demonstrates that our results have very high precision.

\begin{figure}[H]
\begin{center}
\includegraphics[width=0.45\textwidth]{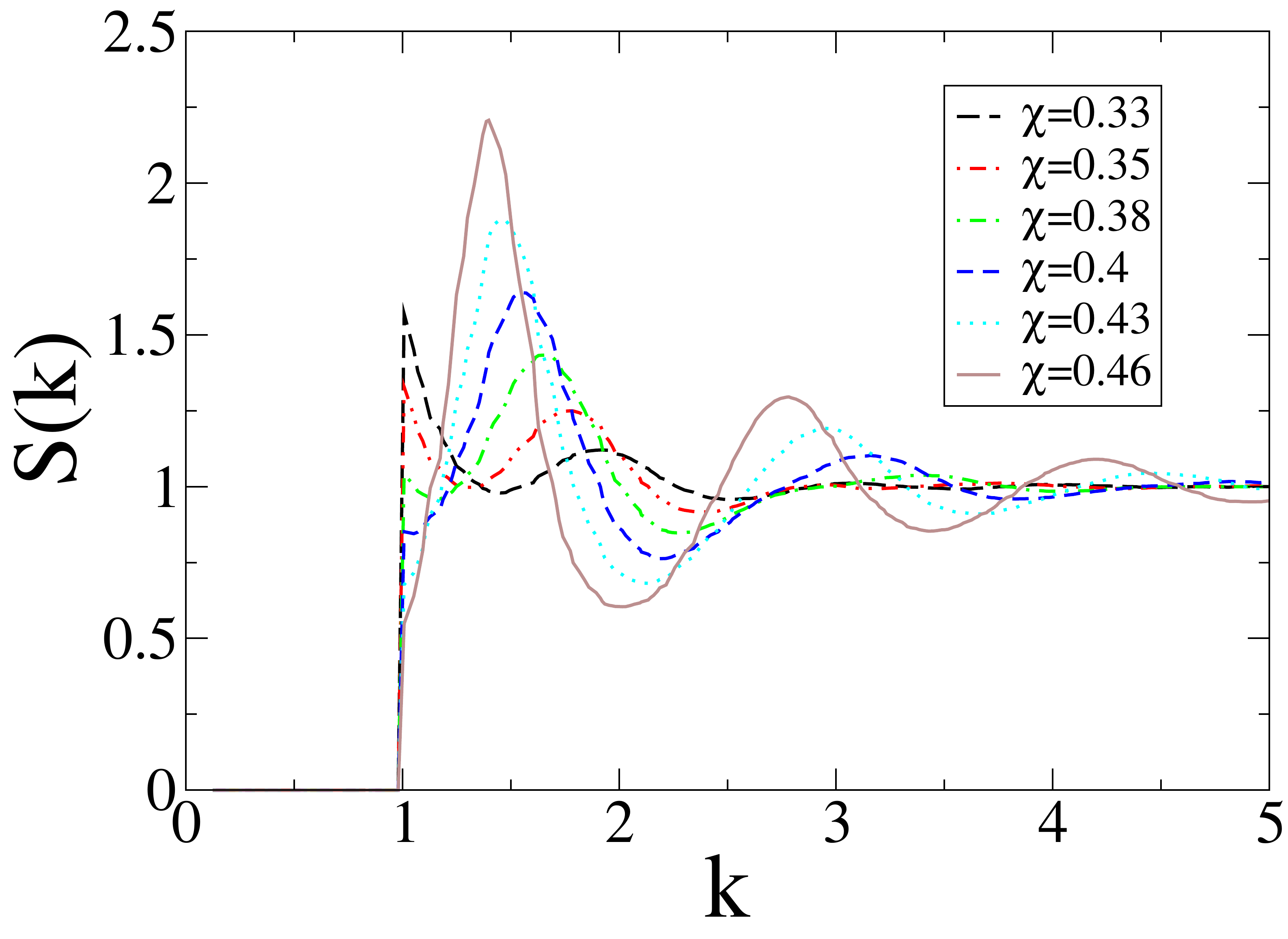}
\includegraphics[width=0.45\textwidth]{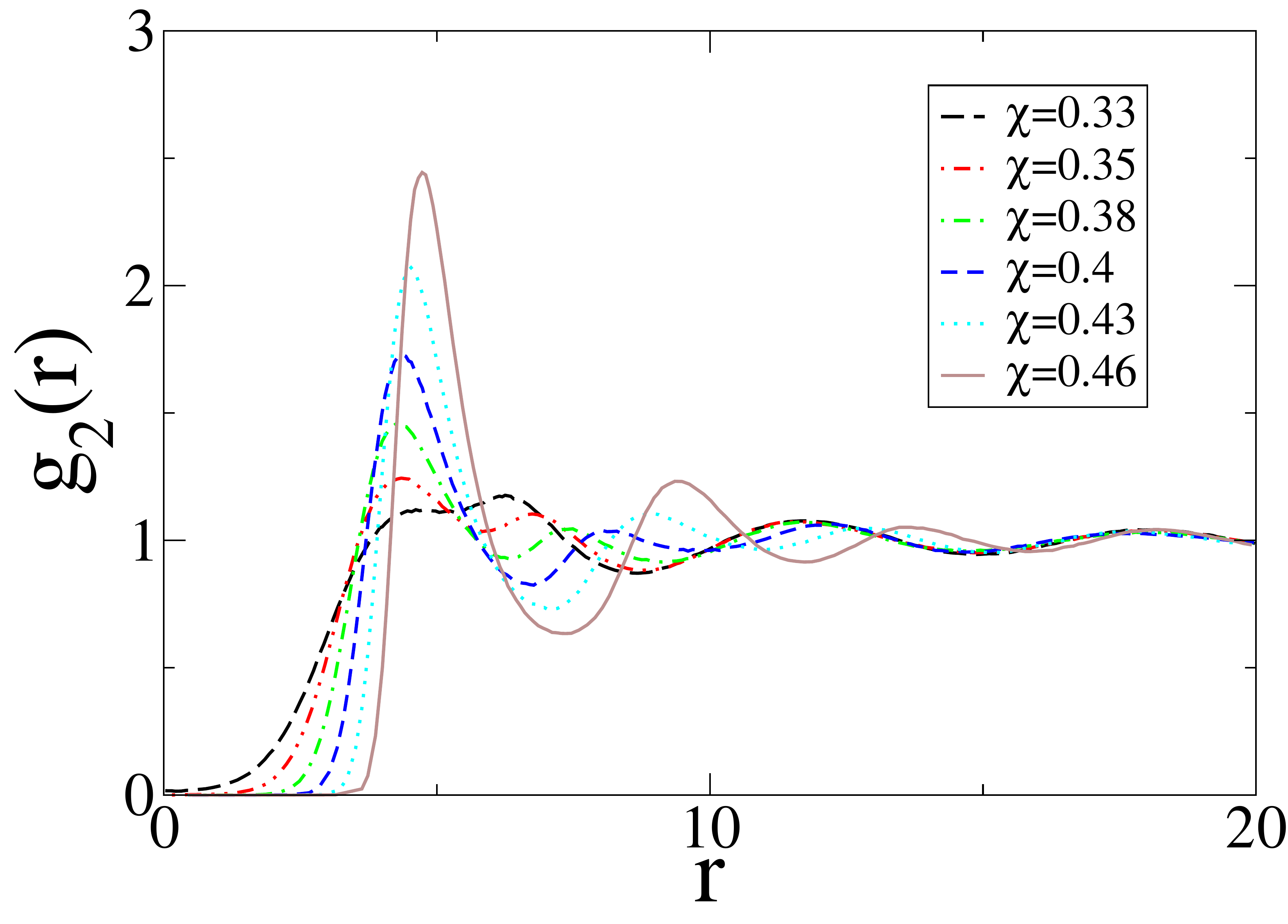}
\end{center}
\caption{(Color online) Structure factor and pair correlation function for $d=2$ for $0.33 \le \chi \le 0.46$, as obtained from simulations.}
\label{s_peak}
\end{figure}

For smaller $\chi$ values, the maximum of the structure factor is at the constraint cutoff $k=K^+$. However, for higher $\chi$ values, the maximum of $S(k)$ is no longer at $k=1^+$.
To probe this transition we have calculated the structure factor in two dimensions for $0.33 \le \chi \le 0.46$. The results are shown in Fig.~\ref{s_peak}.
As $\chi$ increases, the peak at $k=1^+$ gradually decreases its height, while the subsequent peak gradually grows and engulfs the first peak.

\begin{figure}[H]
\begin{center}
\includegraphics[width=0.3\textwidth]{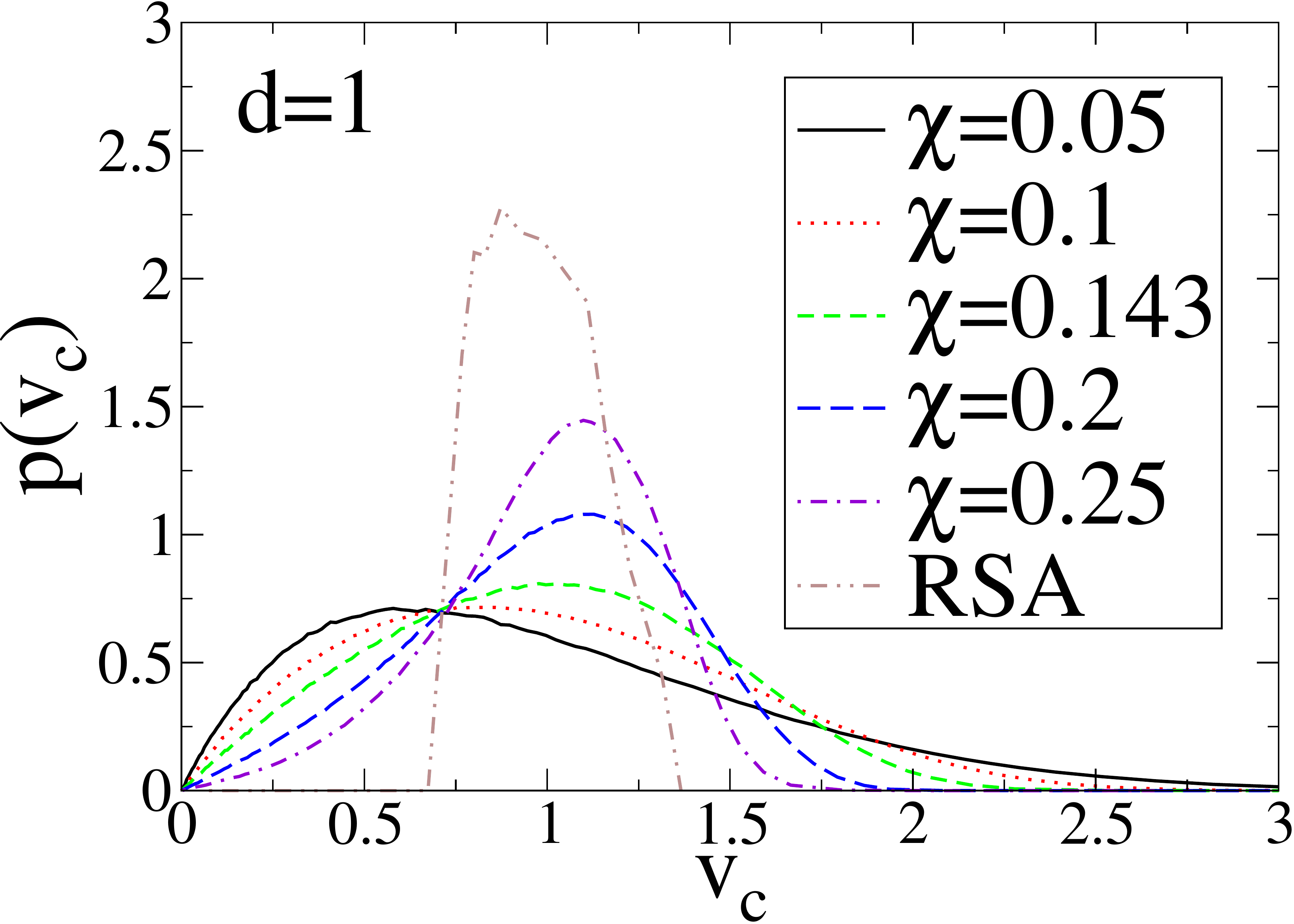}
\includegraphics[width=0.3\textwidth]{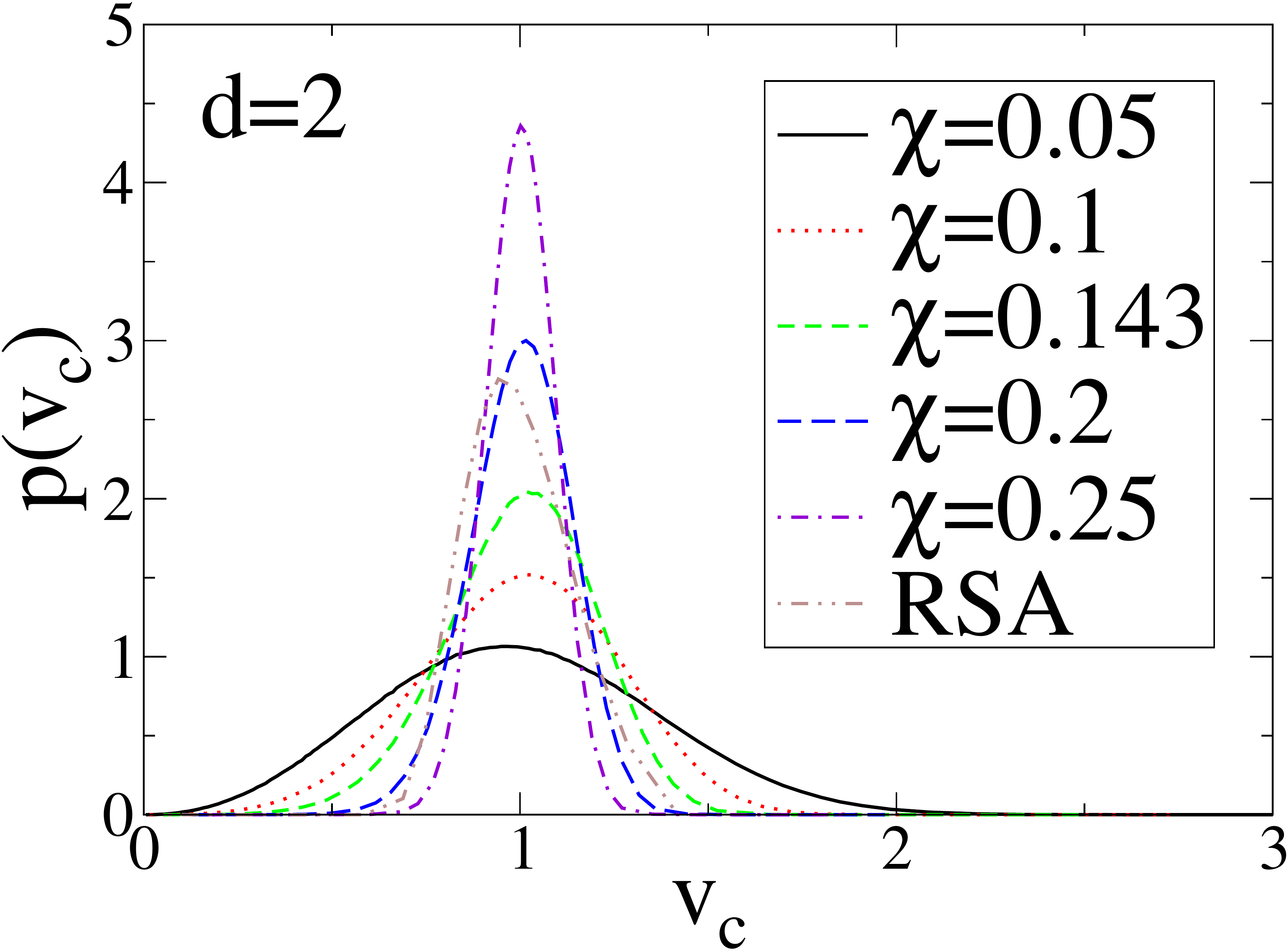}
\includegraphics[width=0.3\textwidth]{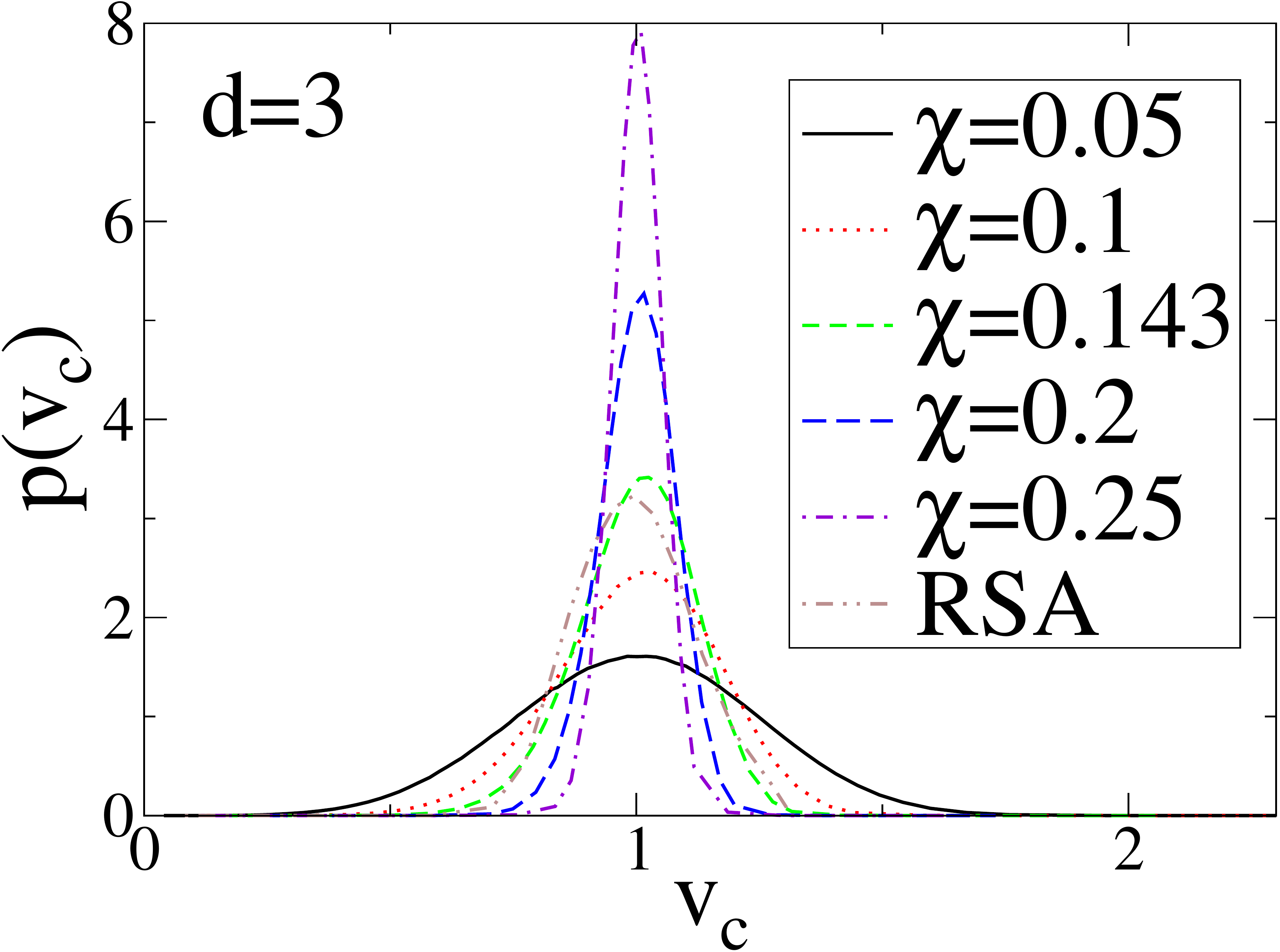}

\end{center}
\caption{(Color online) Voronoi cell-volume distribution for $1 \le d \le 3$ for $0.05 \le \chi \le 0.25$. For the same dimension, the Voronoi cell-volume distribution becomes narrower when $\chi$ increases. For the same $\chi$, the Voronoi cell-volume distribution also becomes narrower when dimension increases. We also present Voronoi cell-volume distributions of RSA packings at saturation here.}
\label{v_dim}
\end{figure}

Besides pair statistics, other widely used characterization of point patterns include certain statistics of the Voronoi cells \cite{uche2004constraints, gabrielli2004voronoi, okabe2009spatial, klatt2014characterization}. A Voronoi cell is the region consisting of all of the points closer to a specific particle than to any other.
We have computed the Voronoi tessellation of the entropically favored stealthy ground states using the {dD} Convex Hulls and {Delaunay} Triangulations package \cite{cgal:hs-chdt3-15a} of the {C}omputational {G}eometry {A}lgorithms {L}ibrary \cite{cgal}.
Since the number density of the stealthy ground states depends on the dimension and $\chi$, we rescaled each configuration to unity density for comparison of the Voronoi cell volumes.
The probability distribution function $p(v_c)$ of the Voronoi cell volumes (where $v_c$ is the volume of a Voronoi cell) are shown in Fig.~\ref{v_dim}. 
In the same dimension, as $\chi$ increases, the distribution of Voronoi cell volumes narrows.
This is expected because the system becomes more ordered as $\chi$ increases.
For the same $\chi$, the distribution also narrows as the dimension increases, consistent with theoretical results that at fixed $\chi$, the nearest-neighbor distance distribution narrows as dimension increases \cite{torquato2015ensemble}.
In Fig.~\ref{v_dim}, we additionally show the Voronoi cell-volume distribution of saturated random sequential addition (RSA) \cite{zhang2013precise, torquato_2006_RSA, widom_1966_RSA} packings, the sphere packings generated by randomly and sequentially placing spheres into a large volume subject to the nonoverlap constraint until no additional spheres can be placed.
Saturated RSA packings are neither stealthy nor hyperuniform \cite{zhang2013precise, torquato_2006_RSA}.
However, the Voronoi cell-volume distributions of saturated RSA packings look similar to that of the entropically favored stealthy ground states. 
This is not unexpected because Voronoi cell statistics are {\it local} characteristics, and hence are not sensitive to the stealthiness, which is a large-scale property.

%\onecolumngrid

%\twocolumngrid

\begin{figure}[H]
\begin{center}
\includegraphics[width=0.3\textwidth]{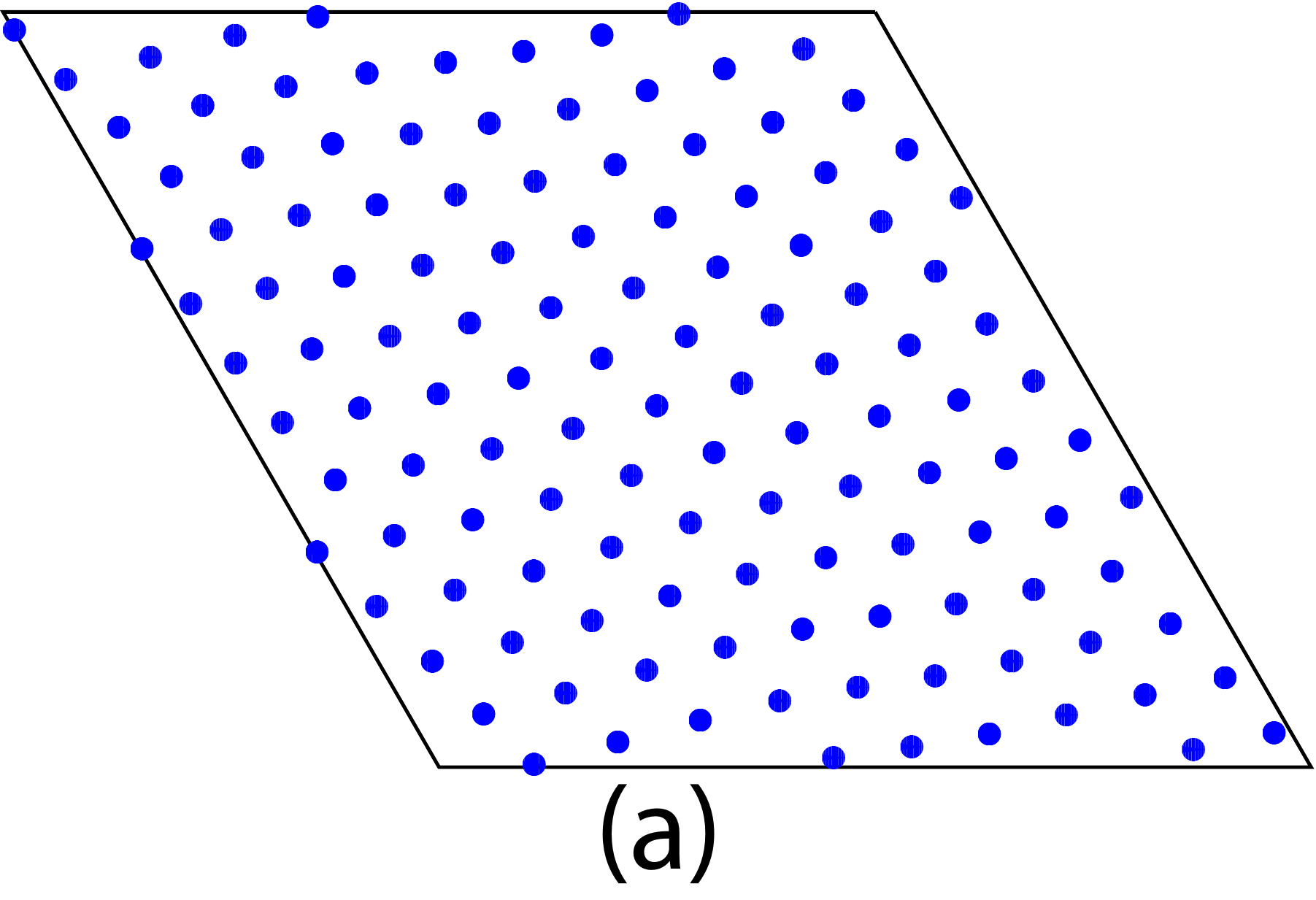}\vspace{0.2in}
\includegraphics[width=0.3\textwidth]{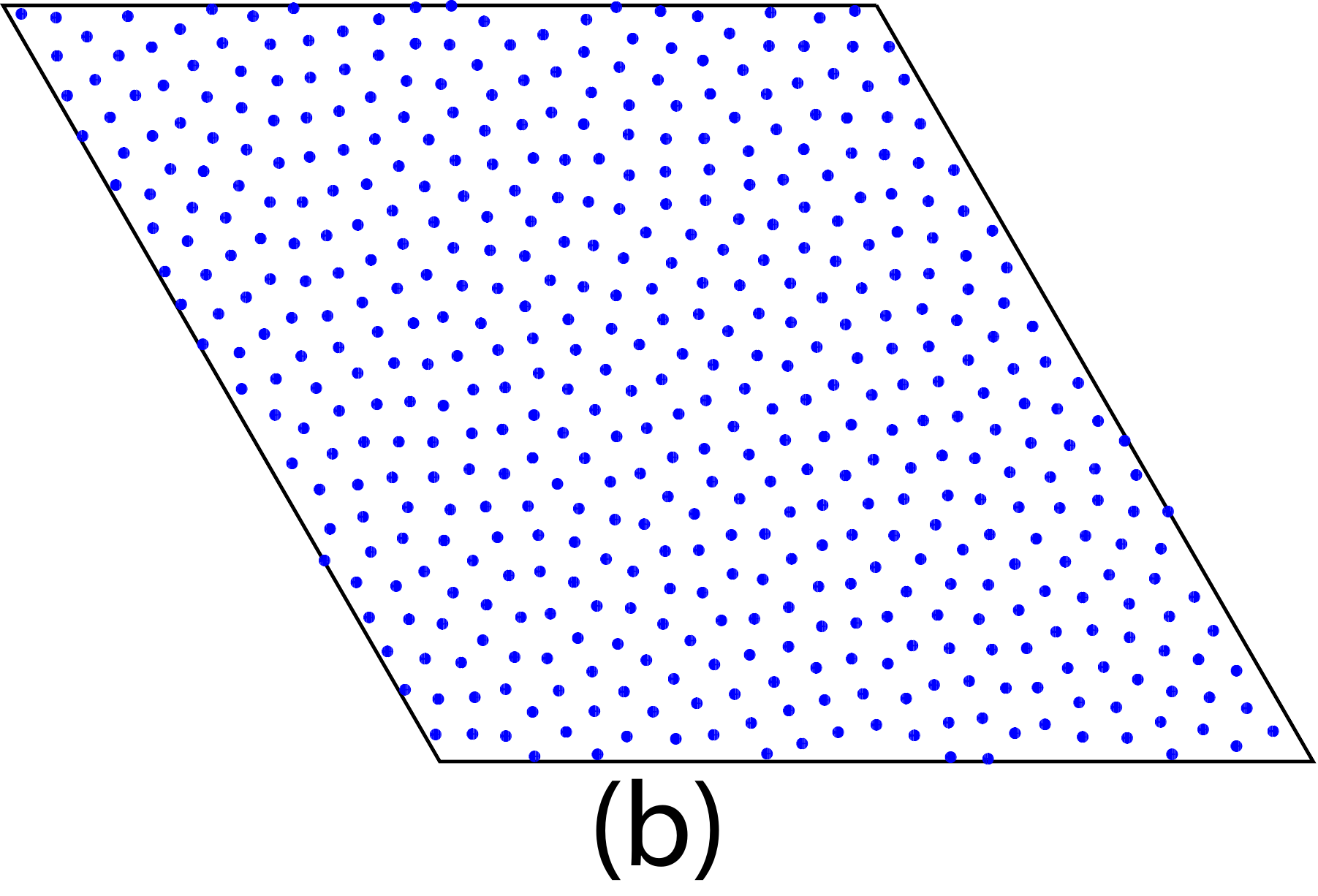}
\end{center}
\caption{(Color online) (a) Low-temperature MD snapshot of a 126-particle system at $\chi=0.48$; the ground-state configuration is crystalline. (b) MD snapshot of a 504-particle system at the same $\T$ and $\chi$; the system does not crystallize and is indeed disordered
without any Bragg peaks.}
\label{finite_crystal}
\end{figure}

One interesting phenomenon is that as $\chi$ increases and approaches 1/2, systems that are not sufficiently large can become crystalline.
In Fig.~\ref{finite_crystal}, we show two snapshots of MD simulations at $\chi=0.48$.
The smaller configuration is crystalline.
However, systems that are 4 times larger remain disordered at the same $\chi$ and temperature. Therefore, this strongly indicates that crystallization is a finite-size effect for $\chi$ tending to 1/2 from below.

\subsection{$\chi \ge 0.5$ region}

\begin{figure}[H]
\begin{center}
\includegraphics[width=0.3\textwidth]{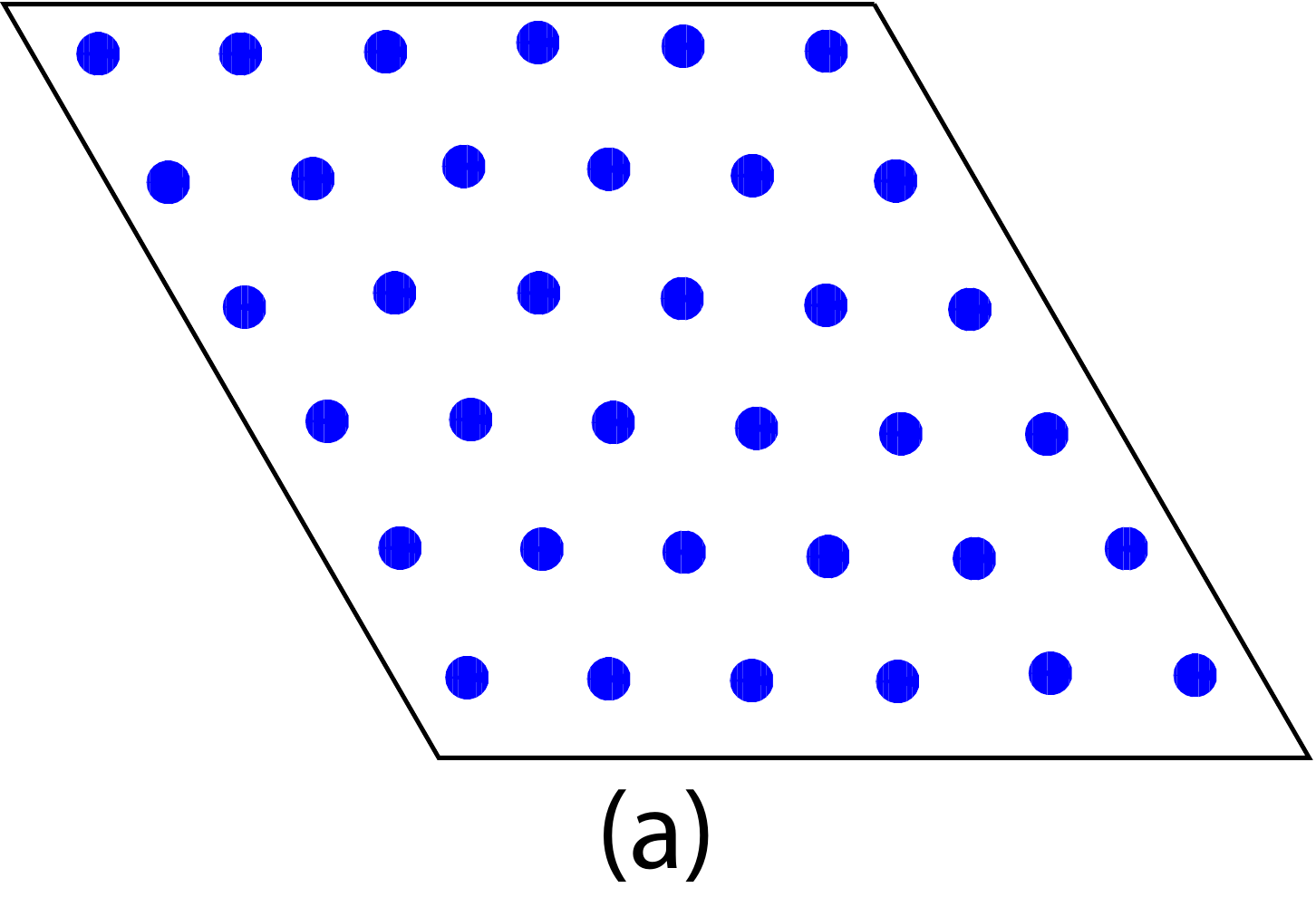}
\includegraphics[width=0.3\textwidth]{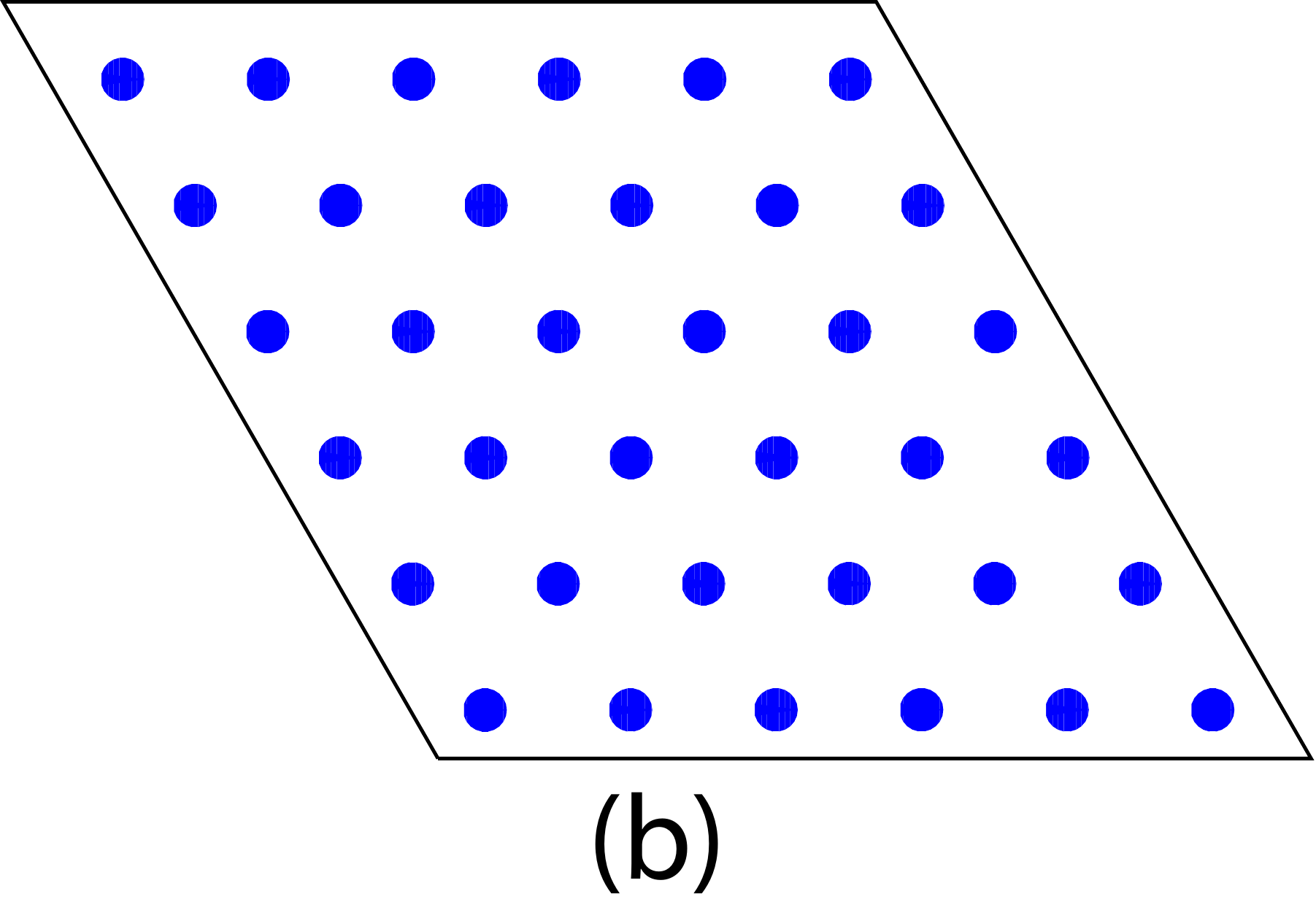}
\includegraphics[width=0.3\textwidth]{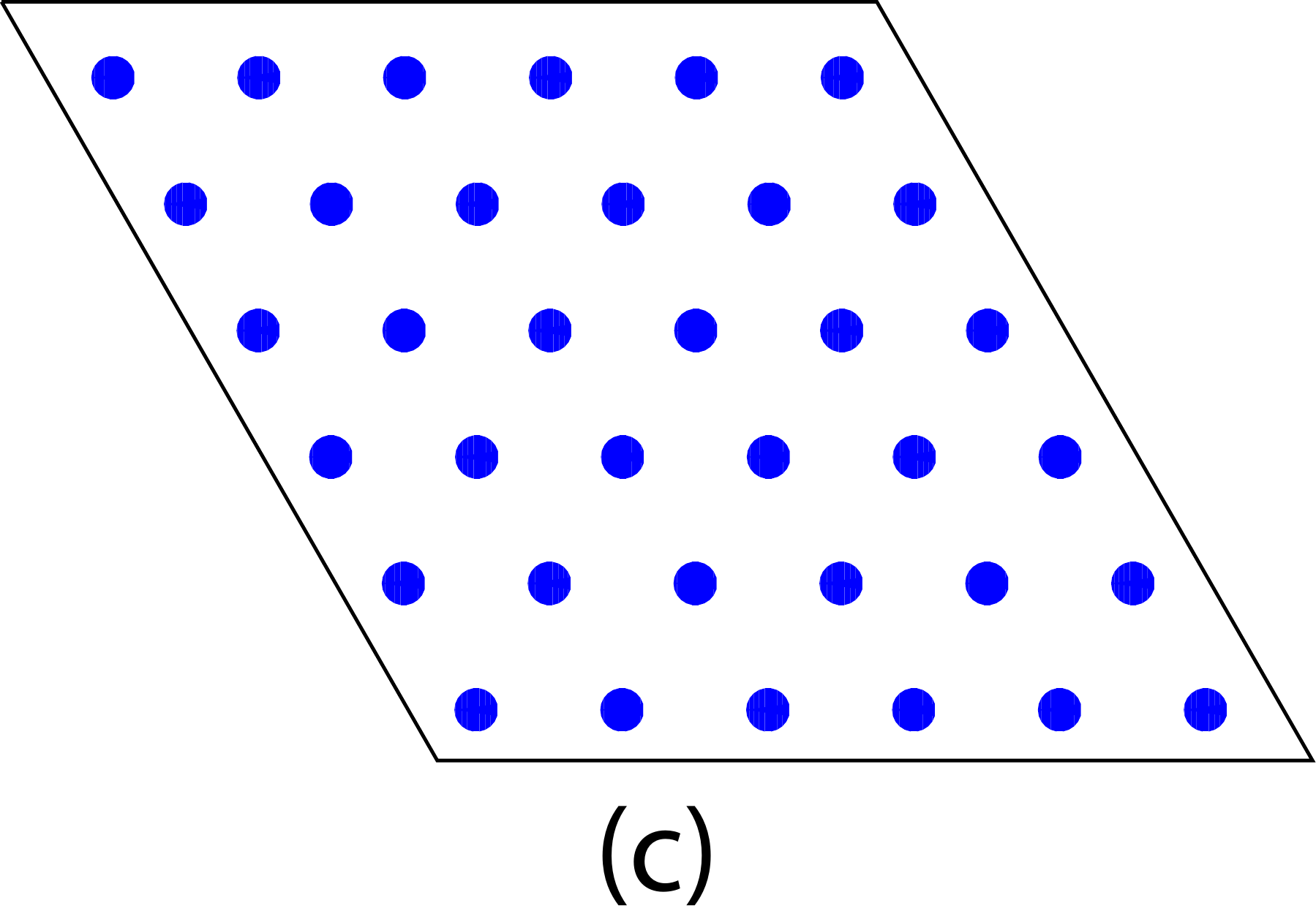}
\end{center}
\caption{(Color online) MD-based simulated annealing result at (a) $\chi=0.55$, (b) $\chi=0.73$, and (c) $\chi=0.81$. The ending configuration is triangular lattice except for small deformations in the $\chi=0.55$ case.}
\label{crystal}
\end{figure}

As explained in Sec.~\ref{detail}, we perform MD-based simulated annealing with Monte Carlo moves of the simulation box for $\chi > 0.5$, since this method works better with rough potential energy surface and can mitigate the finite-size effect.
We performed this simulation at $\chi=0.55$, $\chi=0.73$, and $\chi=0.81$ in two dimensions. The results are shown in Fig.~\ref{crystal}.
The resulting configuration is always triangular lattice.
Even though the ground-state manifold in this $\chi$ regime contains aperiodic ``wavy'' phases discovered previously \cite{uche2004constraints} [but which are called ``stacked-slider" phases in the sequel to this paper \cite{zhang2015ground2}, since they are aperiodic configurations with
a high degree of order in which rows (in two dimensions) or planes (in three dimensions) of particles can slide past each other] as well as crystals other than the triangular lattice, the entropically favored ground state is always a triangular lattice.
This means that the triangular lattice has a higher entropy than stacked-slider phases, 
although the latter appear to be more disordered \cite{WavyEntropy}.

Although we will show analytically that crystals are more entropically favored than stacked-slider phases in the upcoming paper of this series, we still need simulation results to determine which crystal structure has the highest entropy. 
The results of MD-based simulated annealing with Monte Carlo moves of the simulation box suggest that triangular lattice has the highest entropy in two dimensions.
It seems natural to apply the same technique to three dimensions to determine the entropically favored crystal structure. 
However, we were unable to crystallize the system in three dimensions. Even the longest cooling schedule that we tried resulted in stacked-slider phases.

%\onecolumngrid

%\twocolumngrid

\begin{figure}[H]
\begin{center}
\includegraphics[width=0.22\textwidth]{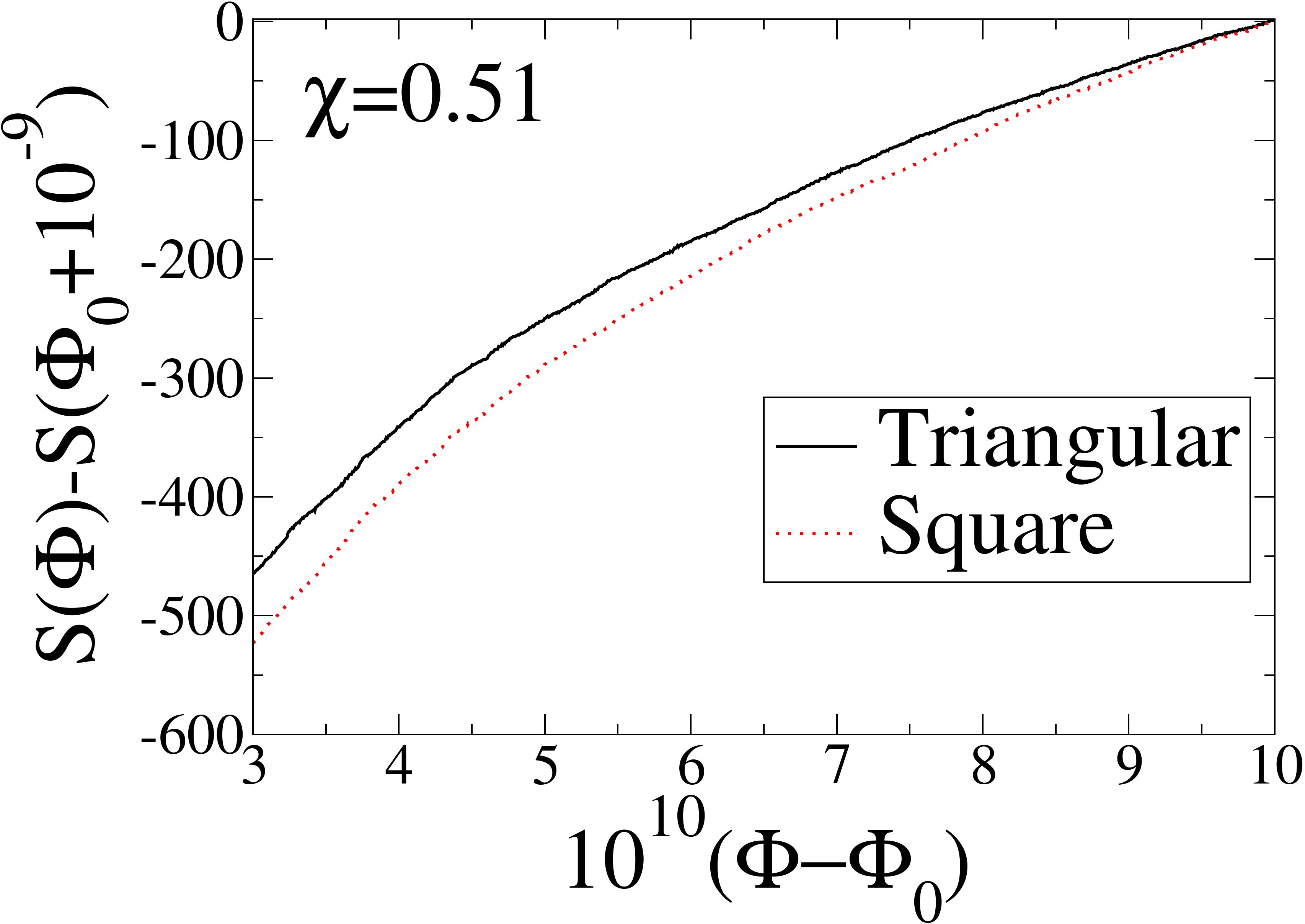}
\includegraphics[width=0.22\textwidth]{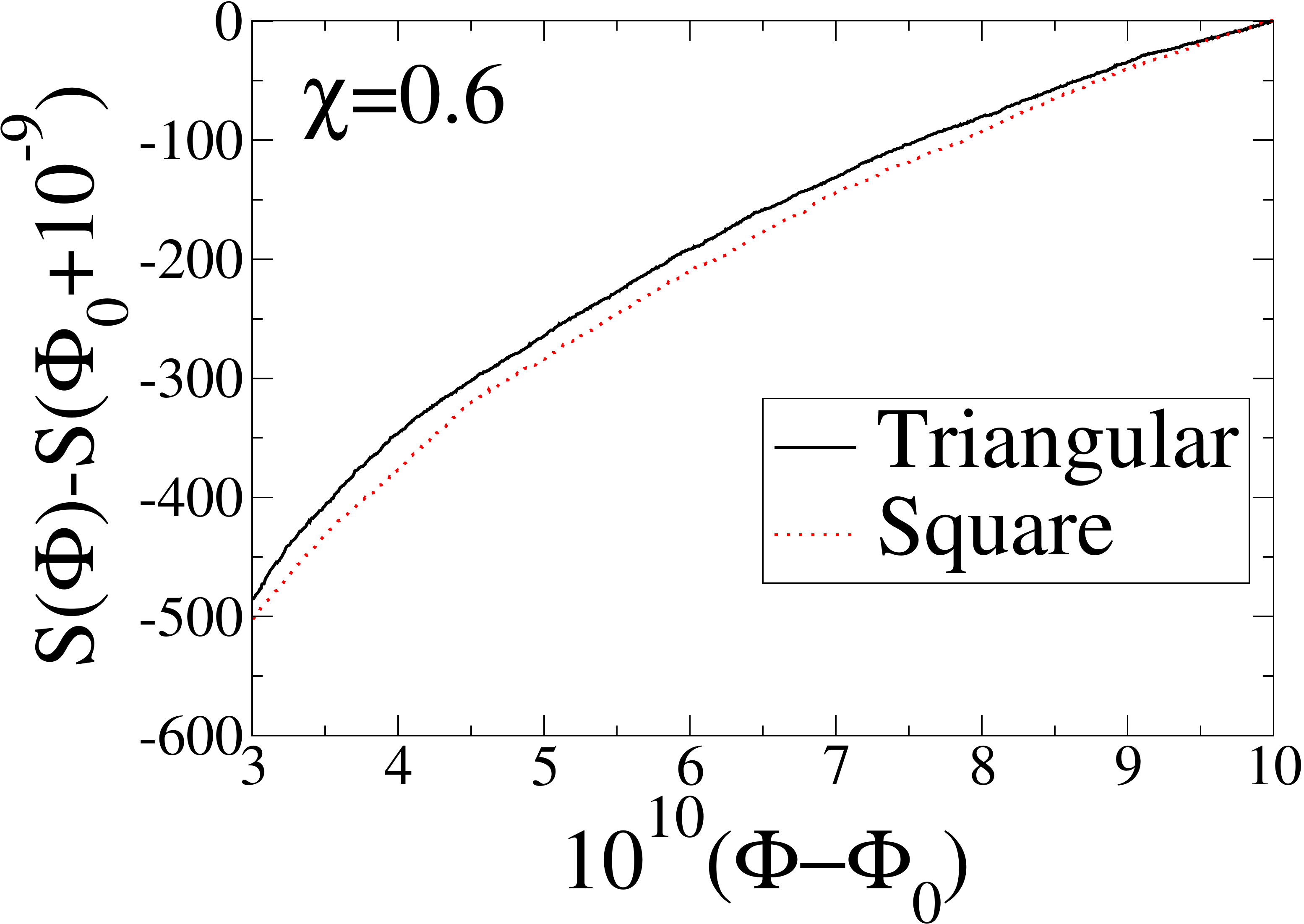}
\includegraphics[width=0.22\textwidth]{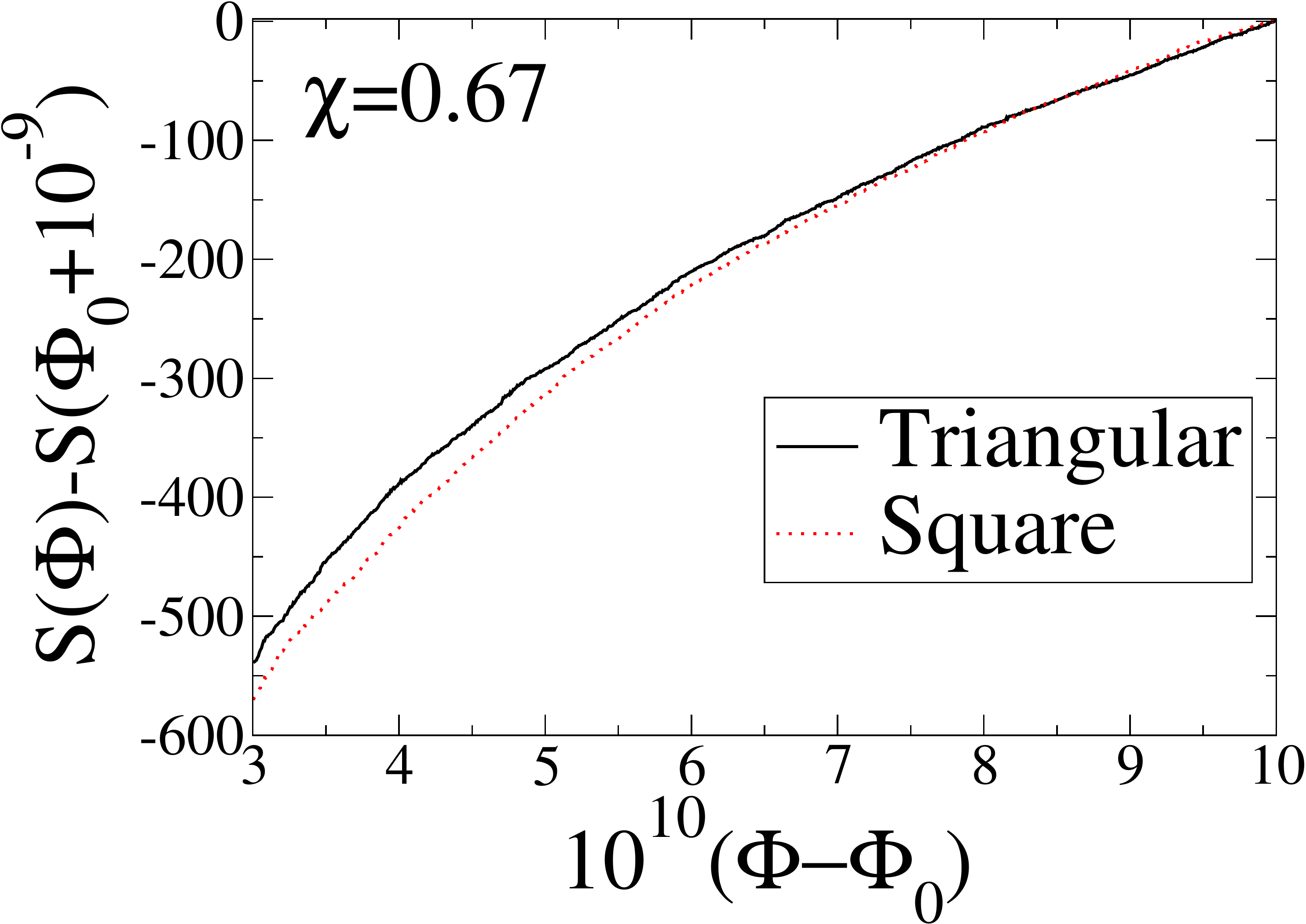}
\includegraphics[width=0.22\textwidth]{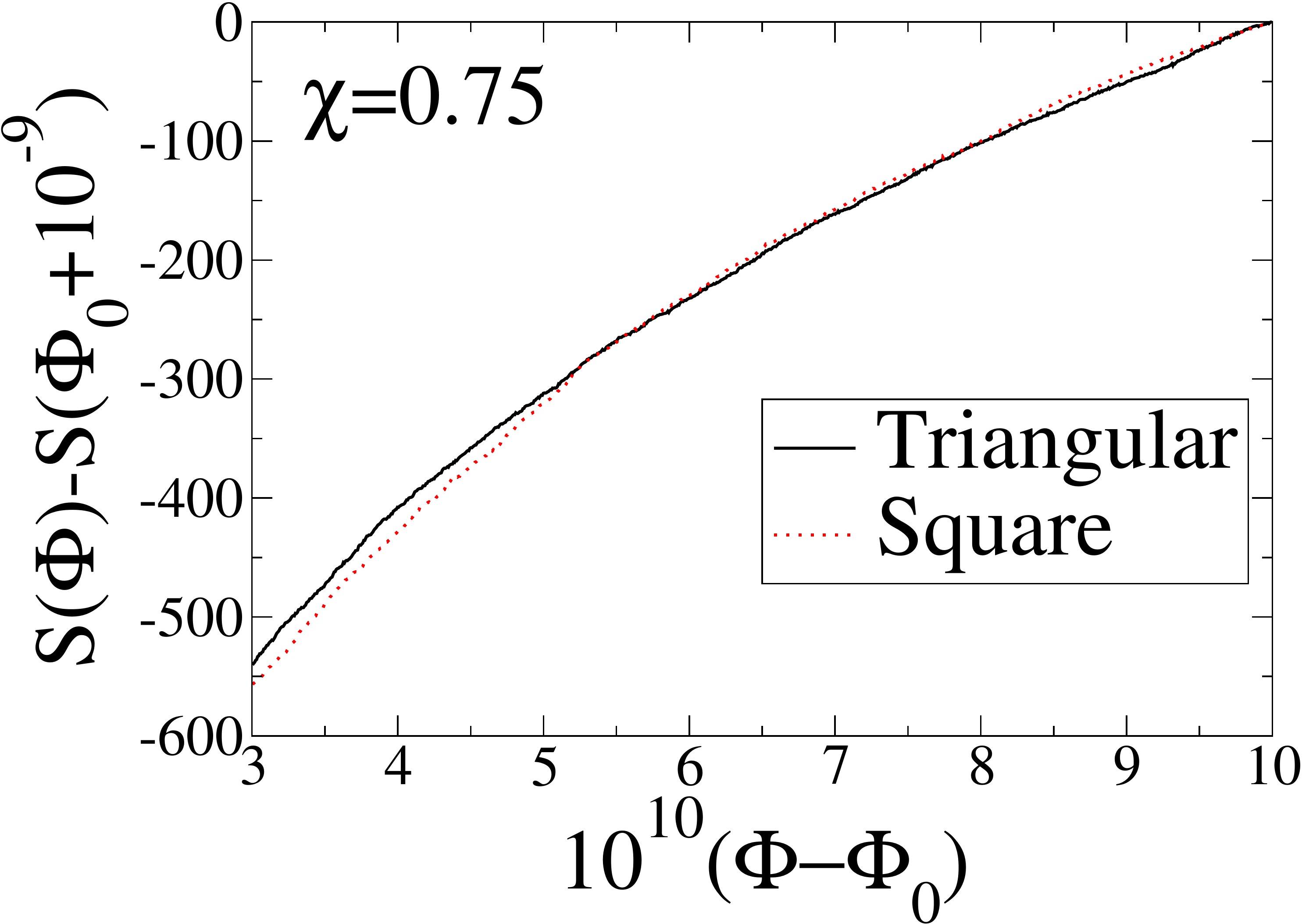}
\end{center}
\caption{(Color online) Microcanonical entropy as a function of energy $\mathscr S(\Phi)$ calculated from Wang-Landau Monte Carlo of triangular lattice and square lattice at various $\chi$'s. Here $\Phi_0$ denotes the ground-state energy.}
\label{wlmc_2d}
\end{figure}

\begin{figure}[H]
\begin{center}
\includegraphics[width=0.22\textwidth]{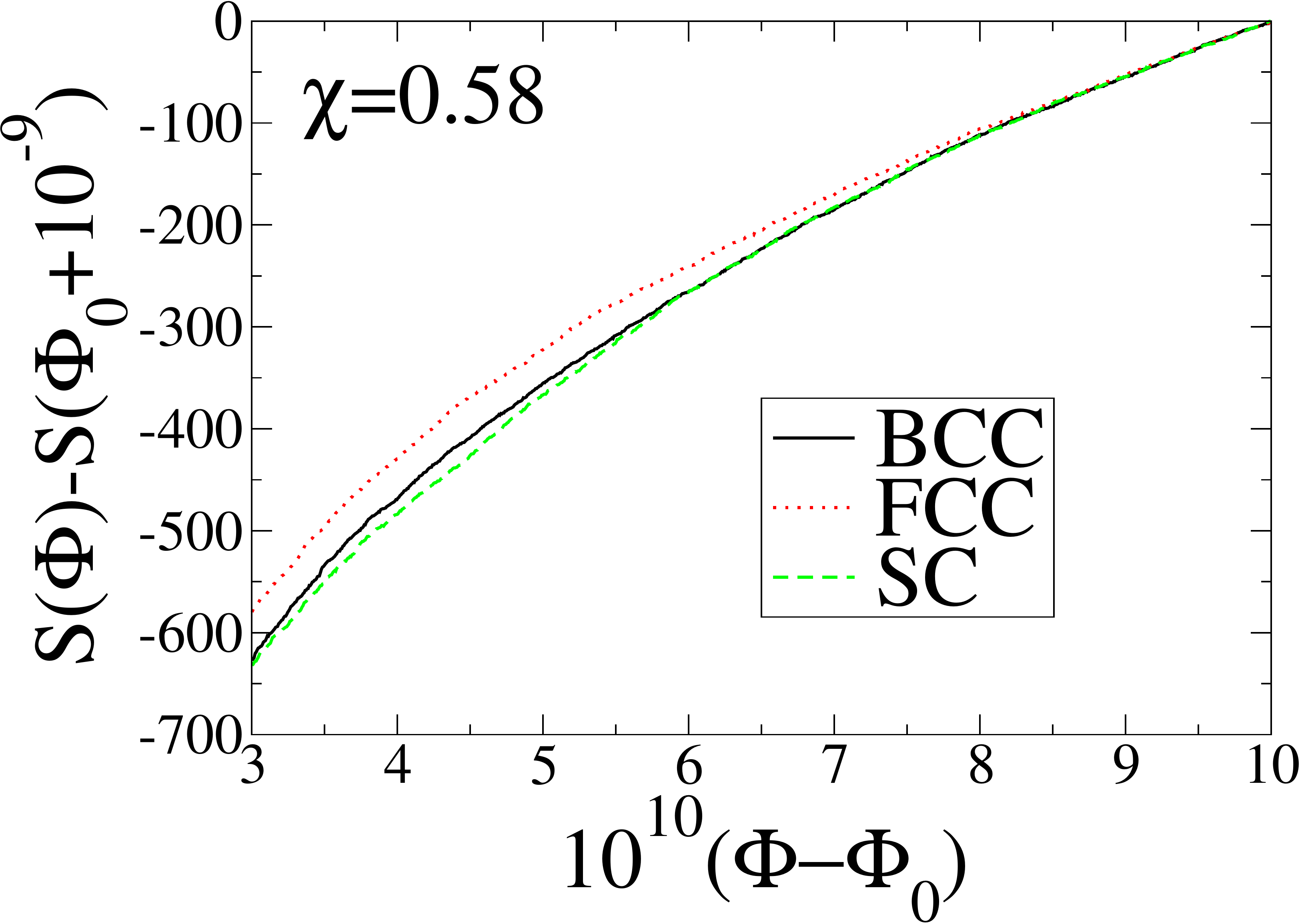}
\includegraphics[width=0.22\textwidth]{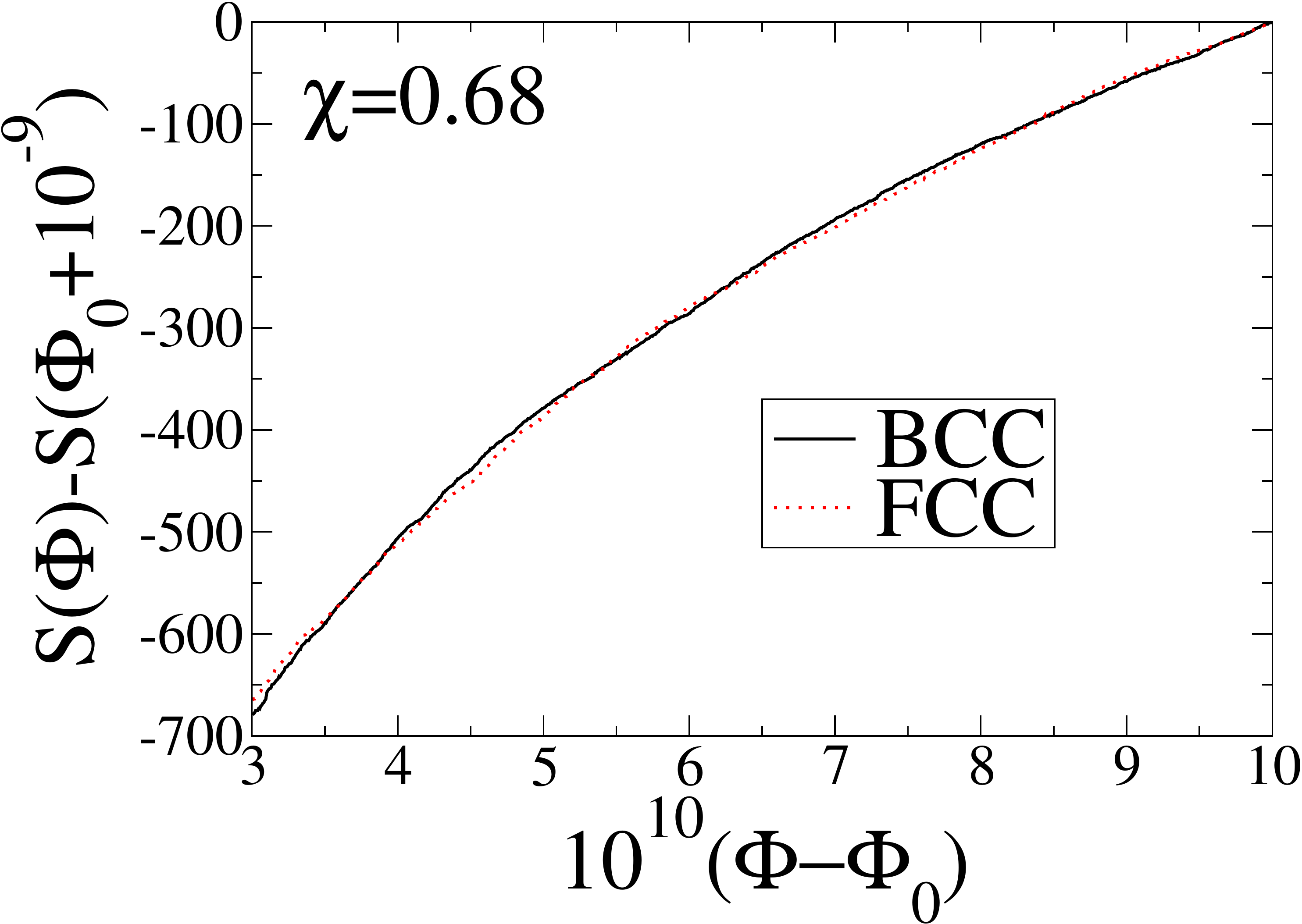}
\includegraphics[width=0.22\textwidth]{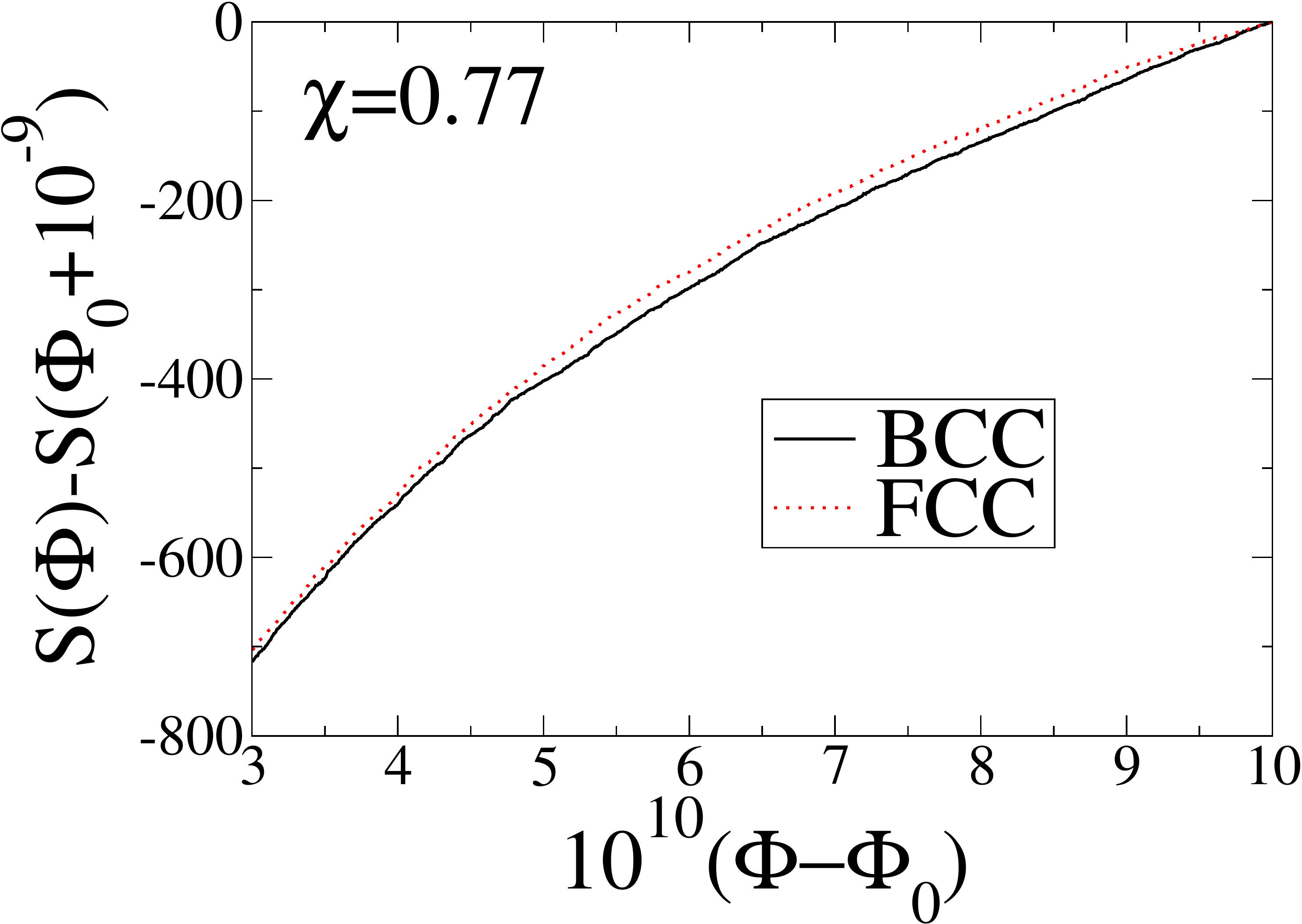}
\includegraphics[width=0.22\textwidth]{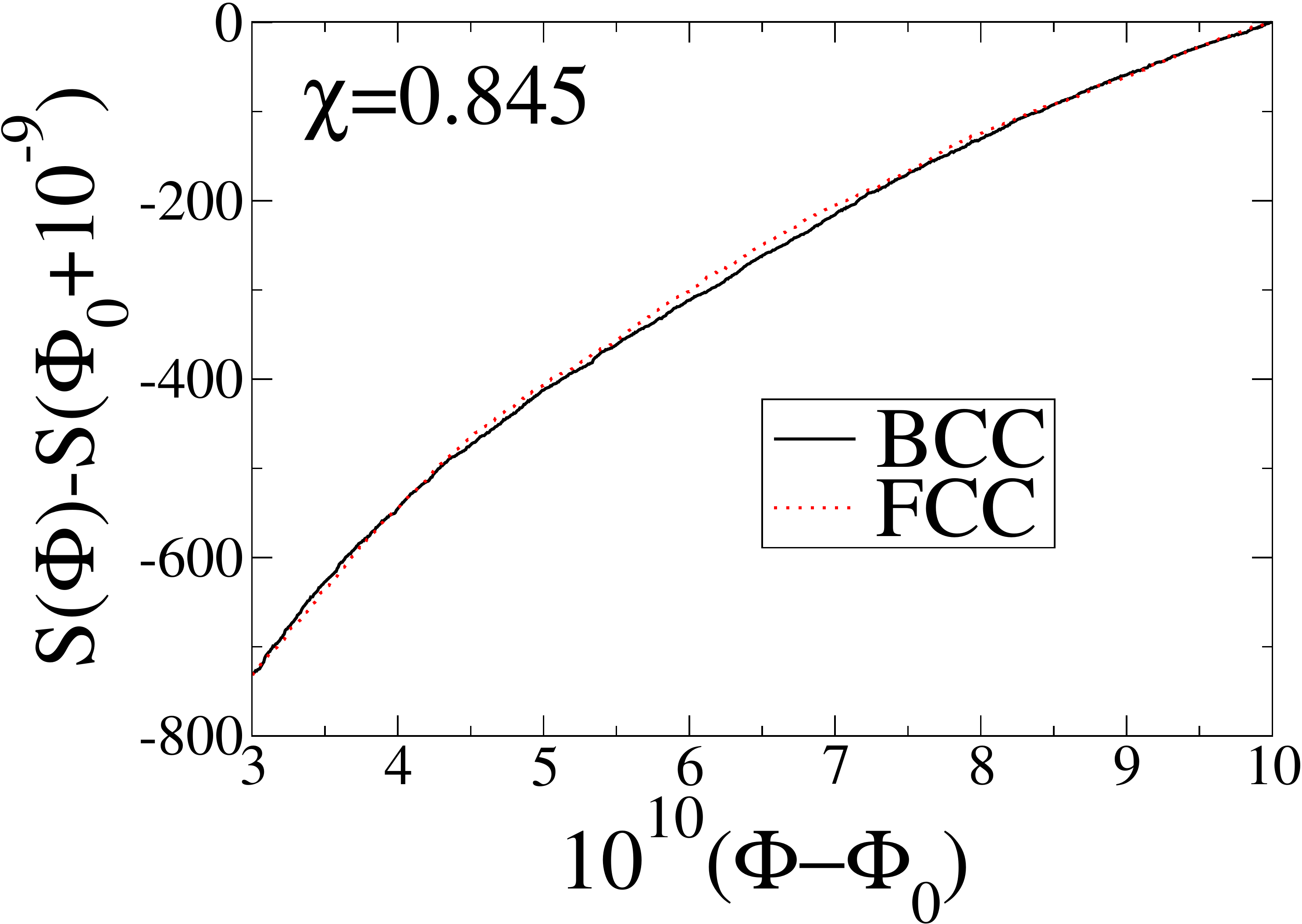}
\end{center}
\caption{(Color online) Microcanonical entropy as a function of energy $\mathscr S(\Phi)$ calculated from Wang-Landau Monte Carlo of BCC lattice, FCC lattice, and SC lattice at various $\chi$'s. A curve for SC lattice is not presented for $\chi \ge 0.68$ because the latter is not a ground state at such high $\chi$ values. Here $\Phi_0$ denotes the ground-state energy.}
\label{wlmc_3d}
\end{figure}

Another way to find the entropically favored crystal is to use Wang-Landau Monte Carlo to directly calculate the entropy of different crystal structures as a function of the potential energy.
We have performed this simulation on two-dimensional triangular lattice, square lattice, and three-dimensional body-centered cubic (BCC) lattice, face-centered cubic (FCC) lattice, and simple cubic (SC) lattice. The results are shown in Figs.~\ref{wlmc_2d}~and~\ref{wlmc_3d}. In all cases the entropy decreases as the energy decreases. In two dimensions, the entropy of the square lattice clearly decreases faster than that of the triangular lattice at every $\chi$ value, confirming that the triangular lattice is entropically favored over the square lattice in the zero-temperature limit. In three dimensions at $\chi=0.58$, the entropy of the FCC lattice decreases more slowly than that of the BCC and SC lattice, suggesting that the entropically favored ground state in three dimensions at $\chi=0.58$ is the FCC lattice. At higher $\chi$ values, the scaling of the entropy of the FCC lattice and the BCC lattice become very close to each other, preventing us from determining the entropically favored ground state at these $\chi$ values.

%\onecolumngrid

%\twocolumngrid

\section{Conclusions and discussion}
\label{conclusion}

The uncountably infinitely degenerate classical ground states of the stealthy potentials have been sampled previously using energy minimizations. We demonstrate here that this way of sampling the ground states to produce ensembles of configurations introduces dependencies on the energy minimization algorithm and the initial configuration. Such artificial dependencies are avoided in studying the canonical ensemble in the $T \to 0$ limit. We sample this ensemble by performing MD simulations at sufficiently low temperatures, periodically taking snapshots, and minimizing the energy of the snapshots. 

The configurations in this ensemble become more ordered as $\chi$ increases and obey certain theoretical conditions on their pair statistics \cite{torquato2015ensemble}, similarly to previous energy minimization results.
However, other properties of this ensemble are unique. First, our numerical results demonstrate that the pair statistics of this ensemble displays no ``clustering effect'' [divergence of $g_2(r)$ as $r \to 0$] for any $\chi$ value, and is independent of the functional form of the stealthy potential. Second, we numerically verify the theoretical ansatz \cite{torquato2015ensemble} that for
sufficiently small $\chi$ stealthy disordered ground states behave like ``pseudo" disordered equilibrium hard-sphere systems in Fourier space, i.e., $S(k)$ has the same functional form as
the pair correlation function for equilibrium hard spheres for sufficiently
small densities. Third, when $\chi$ is above the critical value of 0.5, our results strongly indicate that
crystal structures are entropically
favored in both two and three dimensions in the infinite-volume limit. Our numerical evidence suggests that the entropically favored crystal in two dimensions is the triangular lattice. However, we could not determine the entropically favored crystal structure in three dimensions. For finite systems, the disordered-to-crystal phase transition can happen at a slightly lower $\chi$. A theoretical explanation of this phenomenon remains an open problem. 

Besides ground states of stealthy potentials, other disordered degenerate ground states of many-particle systems have been studied using energy minimizations. Specifically, previous researchers have constrained the structure factor to have some targeted functional form other than zero 
for prescribed wave vectors \cite{uche2006collective, batten2008classical, zachary2011anomalous}. Finding the configurations corresponding to such targeted structure factors amounts to 
finding the ground states of two-, three- and four-body potentials,
in contrast to the two-body stealthy
potential studied in the present paper.
This situation is the most general application
of the collective-coordinate approach. It will be interesting
to study the resulting pair statistics of the ground states for these more general
interactions in the zero-temperature limit of the canonical ensemble.

The collective-coordinate approach is an independent and fruitful addition to the basic statistical mechanics problem of connecting local interactions to macroscopic observables. One important feature of collective-coordinate interactions is that it has uncountably infinitely degenerate classical ground states \cite{torquato2015ensemble}. In the case of isotropic pair interactions,
the only other system that we know with this feature is the hard-sphere system. However, there are two important differences between hard-sphere systems and collective-coordinate ground states. 
First, while the dimensionality of the configuration
space of equilibrium hard-sphere systems consisting of $N$ particles within
a periodic box is fixed [simply determined by the nontrivial number of degrees of freedom, $d(N-1)$], the
dimensionality of the collective-coordinate ground-state configuration space decreases as $\chi$ increases and, on a per particle basis,
eventually vanishes \cite{torquato2015ensemble}. The decreased dimensionality of the ground-state configuration space creates challenges for accurate sampling 
of the entropically favored ground states
using numerical simulations
and hence the development of
better sampling methods is a fertile
ground for future research.

\begin{figure}[H]
\begin{center}
\includegraphics[width=0.45\textwidth]{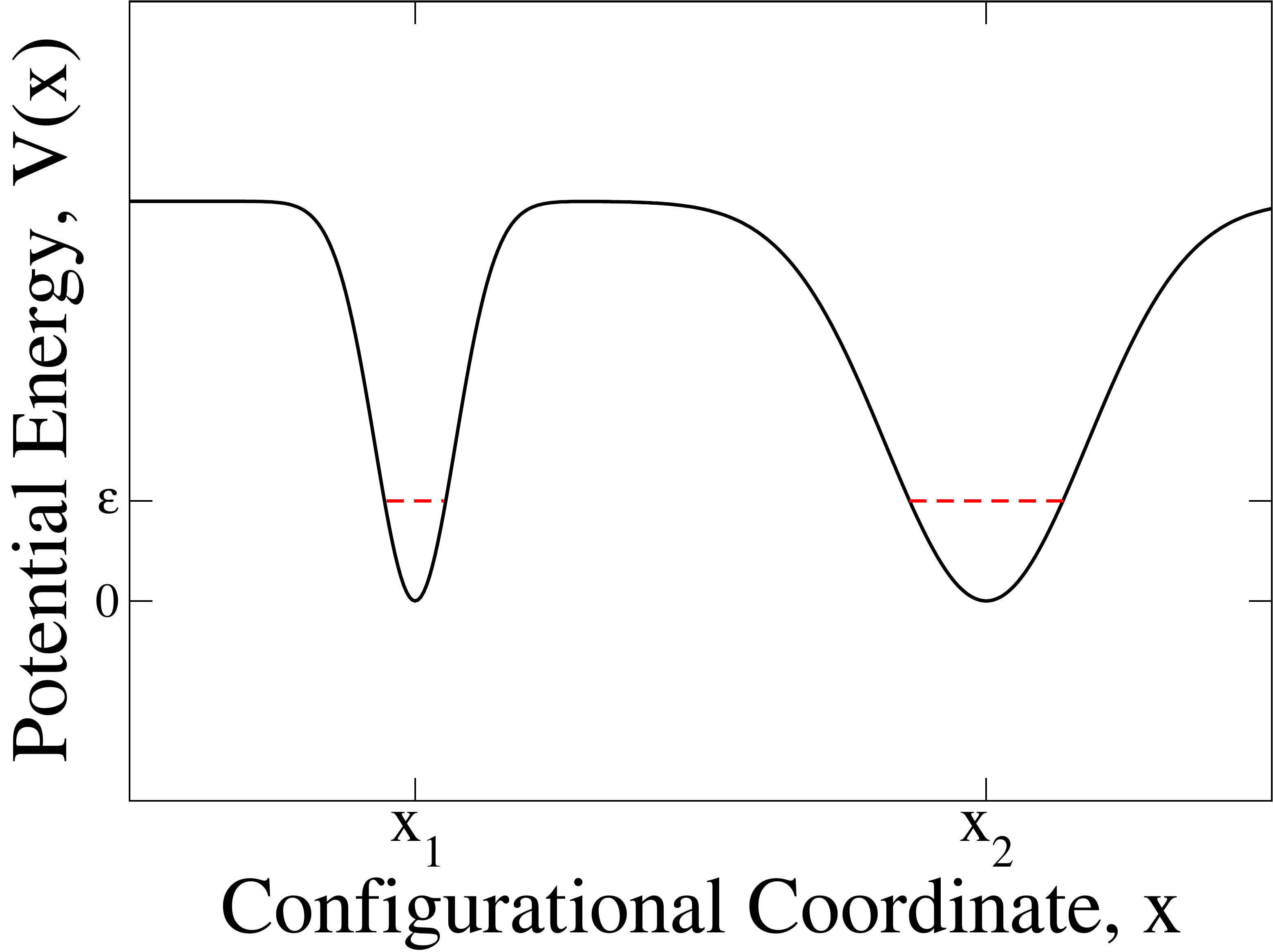}
\end{center}
\caption{(Color online) A model one-dimensional energy landscape with two wells located at $x_1$ and $x_2$ of the same depth but different curvatures. The ``feasible regions,'' i.e., regions where $V(x) < \varepsilon$, is marked by red dashed lines.}
\label{fig17}
\end{figure}

Second, while the probability measure of the equilibrium hard-sphere system is uniform over its entire ground-state manifold, that of the stealthy ground states is not uniform. To illustrate this point, imagine a one-dimensional energy landscape that has a double-well potential behavior in a portion of the configuration space, as shown in Fig.~\ref{fig17}. Each minimum represents a degenerate ground state (as we find with stealthy potentials) and therefore the well depths of the minima are the same. Let us now consider harmonic approximations of the two wells in the vicinity of $x_1$ and $x_2$, respectively,
$$
V_1(x)=a_1 (x-x_1)^2 ,
$$
and 
$$
V_2(x)=a_2 (x-x_2)^2,
$$
where $x$ is the configurational coordinate. At very low temperature, to a good approximation, the system can only visit the part of the configuration space with energy less than $\varepsilon$, and $\varepsilon \to 0$ as $T \to 0$. Solving $V_i(x)<\varepsilon$, where $i=1, 2$, one finds the feasible region of configuration space associated with both wells:
$$
x_1-\sqrt{\varepsilon/a_1}<x<x_1+\sqrt{\varepsilon/a_1} ,
$$and
$$
x_2-\sqrt{\varepsilon/a_2}<x<x_2+\sqrt{\varepsilon/a_2} .
$$
When $a_1 \neq a_2$, we see that the feasible regions associated with the two potential wells have different ranges. Therefore, the weights associated with the two minima, i.e., the relative probabilities for finding the system in the vicinity of those minima, will also differ. Similarly, in the stealthy multidimensional configuration space that we are studying, the magnitude of the eigenvalues of the Hessian matrix will determine the relative weights. Therefore, the probability measure of the stealthy ground states is not uniform over the ground-state manifold, unlike the degenerate ground states of classical hard spheres. Our low-temperature MD simulations sample ground states with this nonuniform probability measure.
It would be useful to devise theories to estimate the weights of different portions of the ground-state manifold. However, a feature that complicates the problem is that the Hessian matrix has zero eigenvalues. In the associated directions of the eigenvectors of the configuration space, the energy scales more slowly than quadratically (harmonically) but we do not know the specific form. 

This paper, which investigates the entropically favored ground states, is the first of a two-paper series. In the second paper, we will study aspects of the ground-state
manifold with an emphasis on configurations that are not 
entropically favored for
$\chi$ above 1/2 (the ordered
regime). In particular, we will 
more fully investigate the nature
of so-called ``wavy'' crystals or ``stacked-slider" phases, discovered in Ref.~\onlinecite{uche2004constraints}. Using an analytical description of such states, we will demonstrate that they are part of the ground state but are not entropically favored. Our analytical model will also demonstrate that stacked-slider phases exist in three and higher dimensions.

\begin{acknowledgments}
G. Z. thanks Steven Atkinson for his careful reading of some parts of the manuscript.
This research was supported by the U.S. Department of Energy, Office of Basic Energy Sciences, Division of Materials Sciences and Engineering under Award No. DE-FG02-04-ER46108.
\end{acknowledgments}

\appendix
\section{Real-space potential in finite systems}

\begin{figure*}
\begin{center}
\newcommand{\inc}[1]{\includegraphics[width=0.18\textwidth]{#1-eps-converted-to.pdf}}
\begin{tabular}{c c c}
\inc{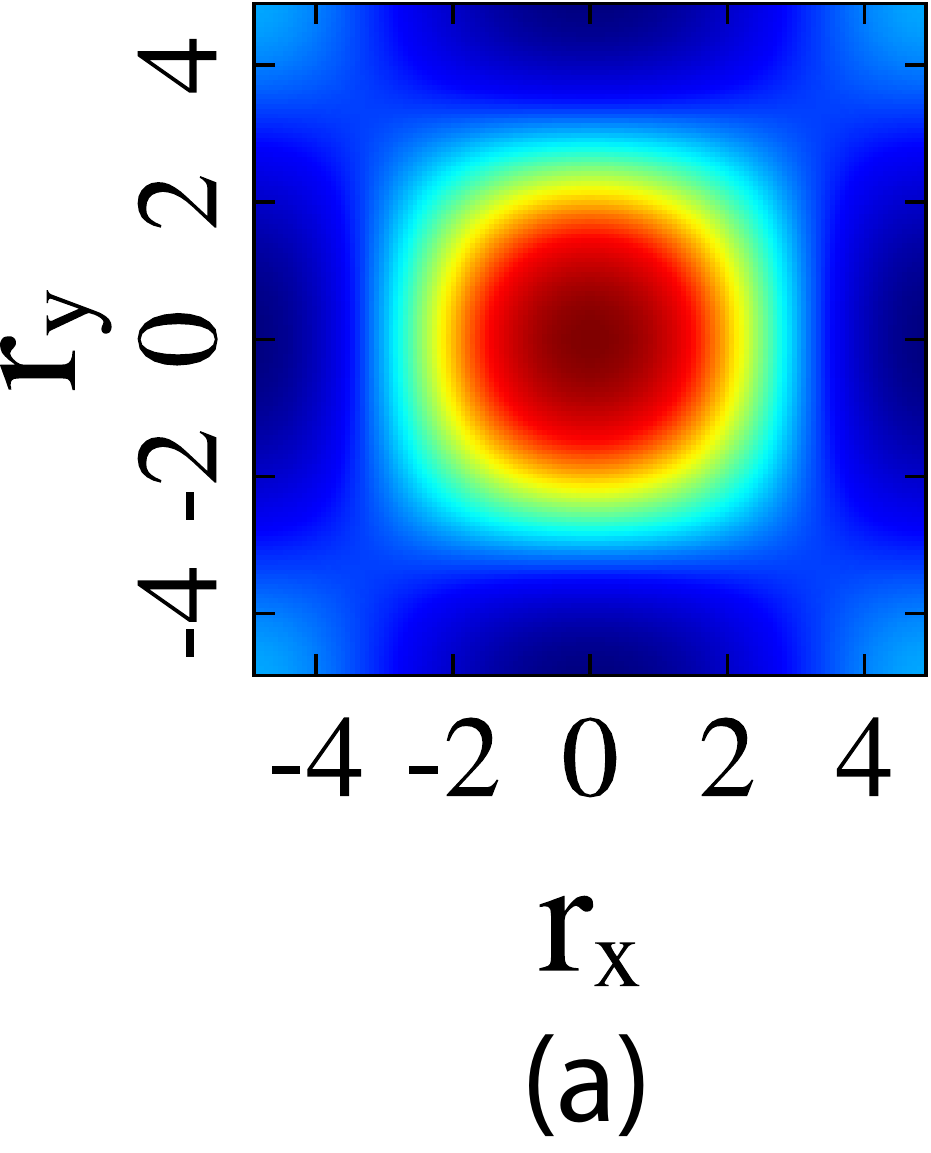} & \inc{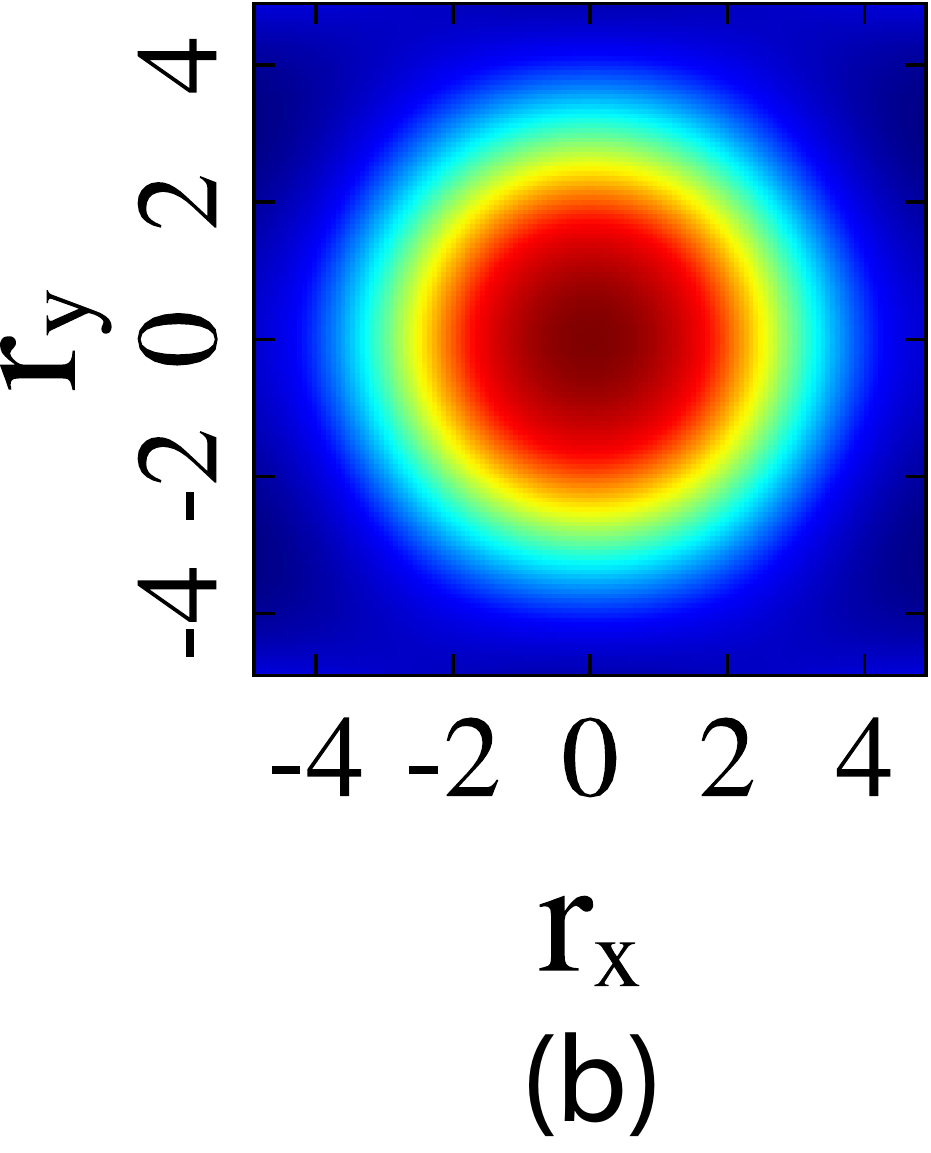} & 
\multirow{4}{*}{\includegraphics[width=0.1\textwidth, trim=-50 0 0 0, clip]{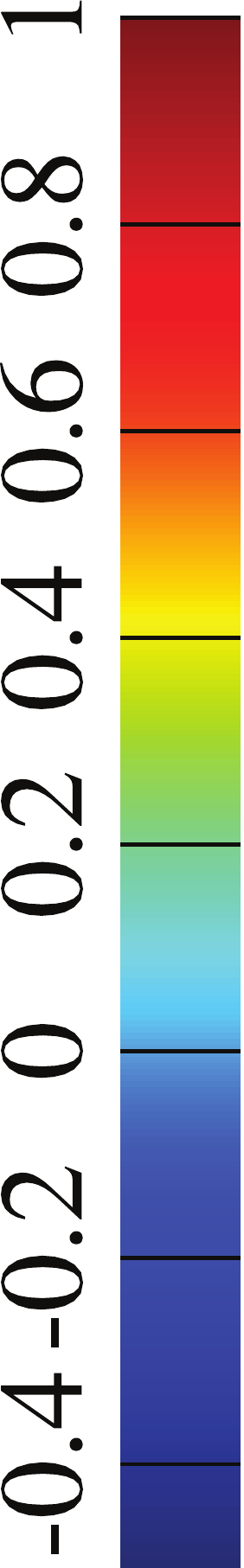}} \\
\inc{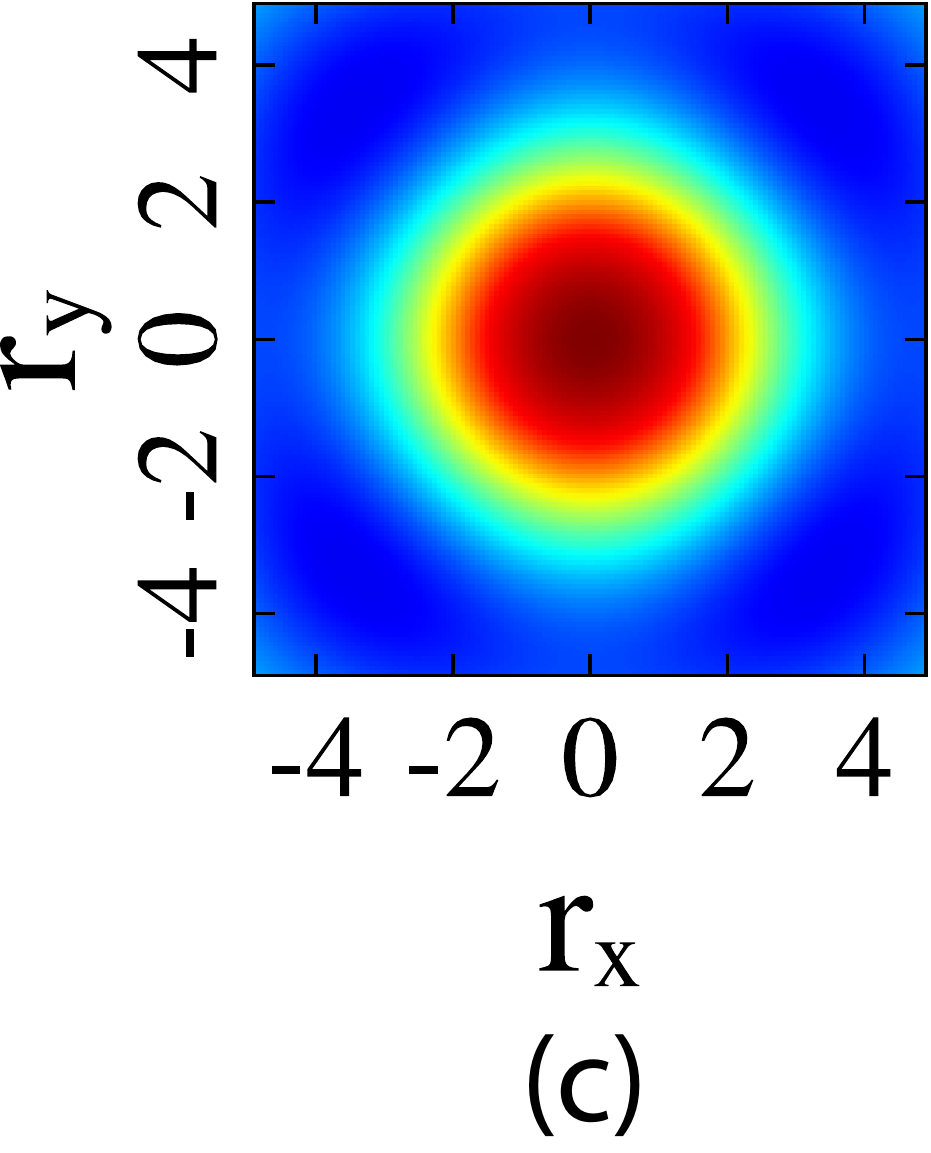} & \inc{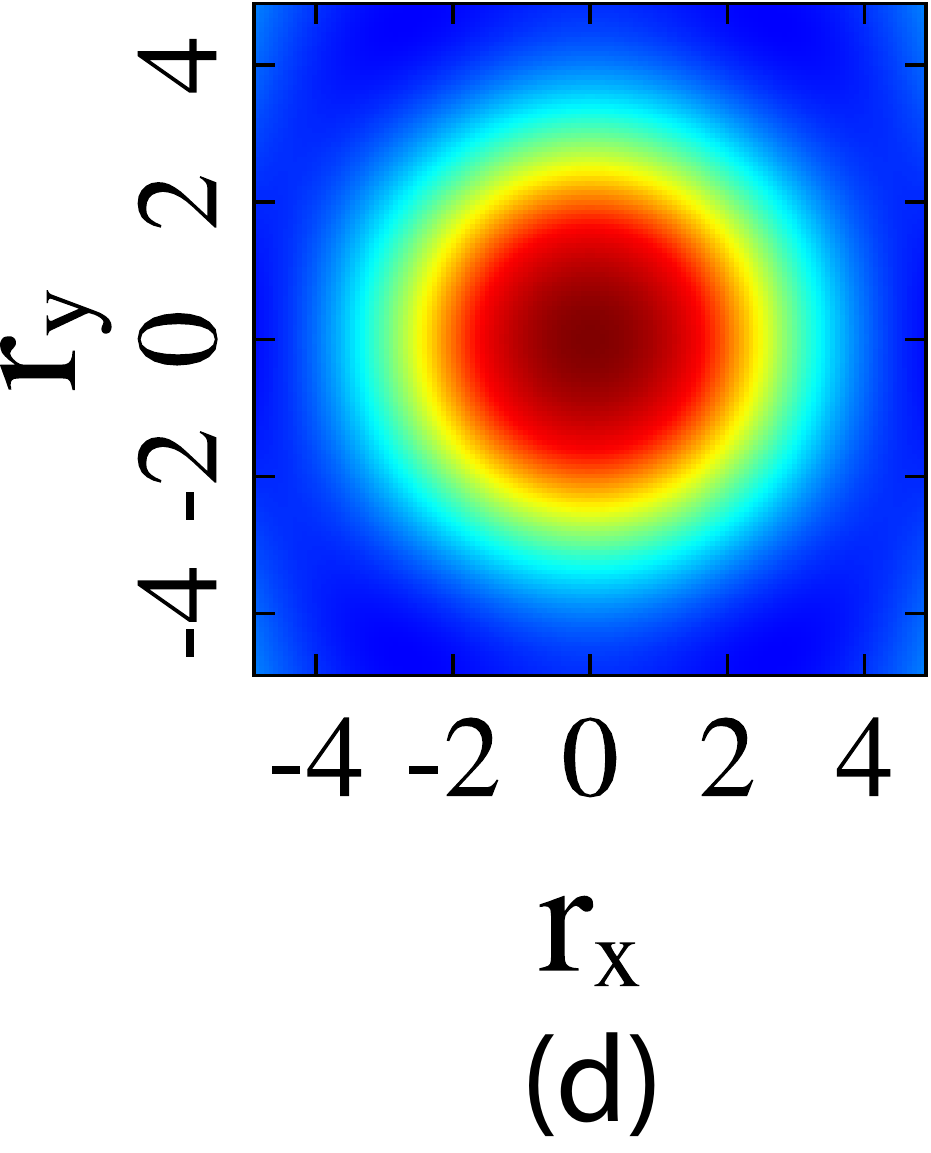} \\
\inc{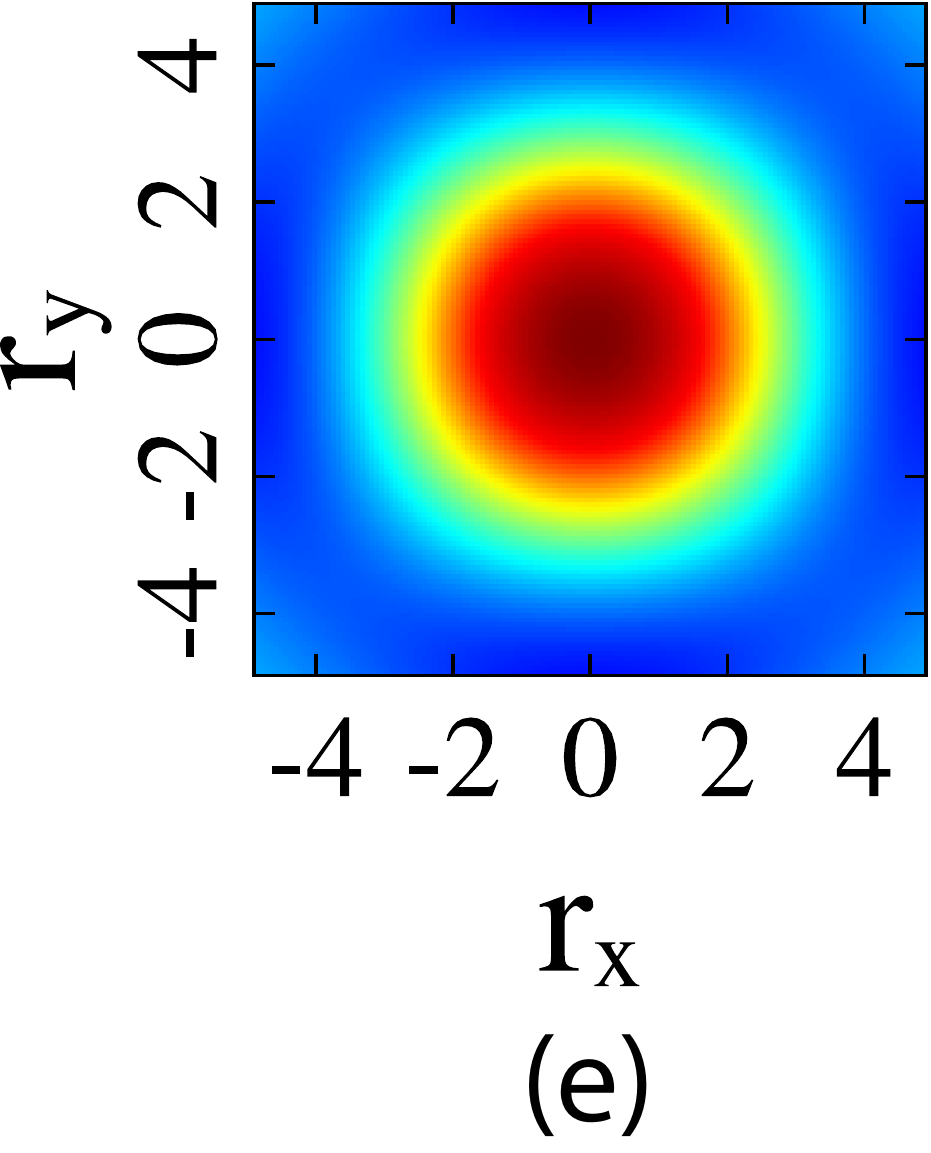} & \inc{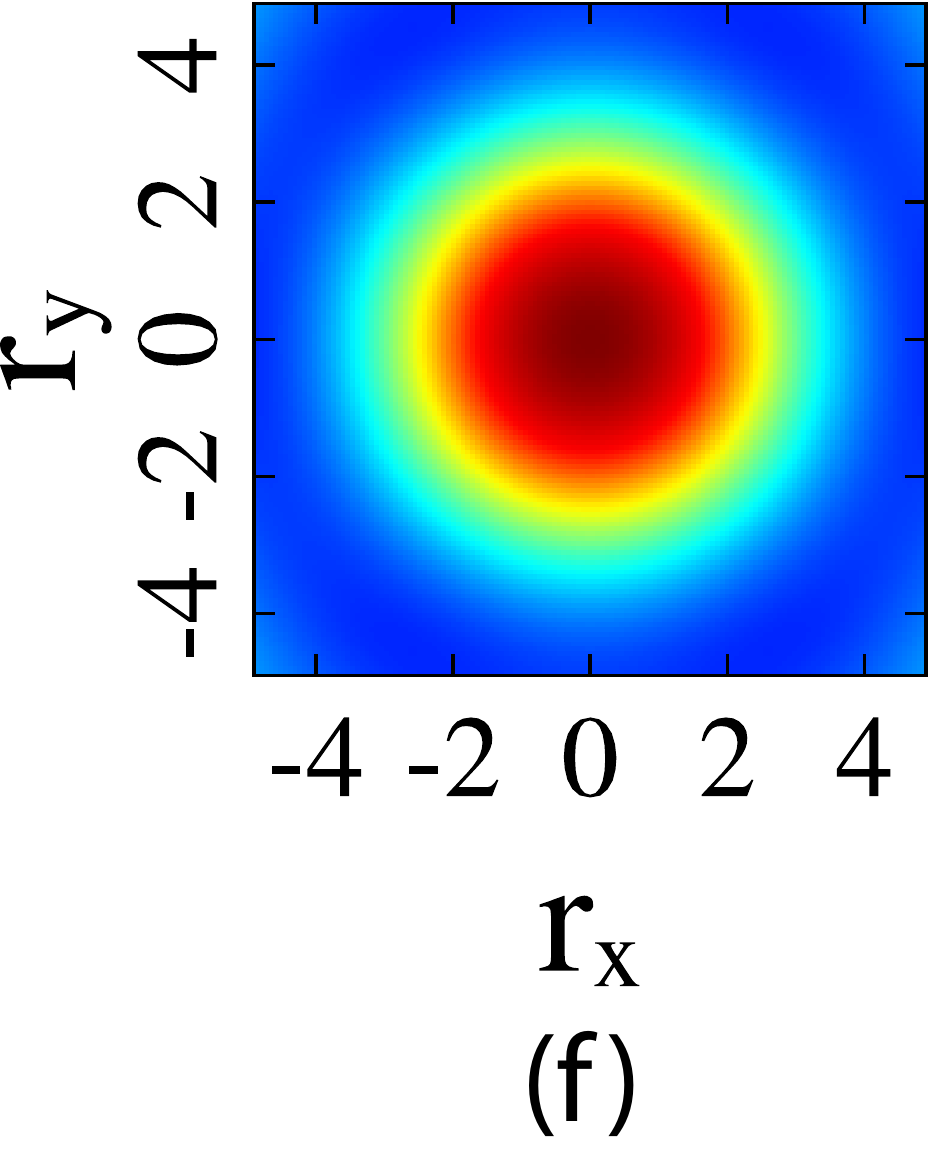} \\
\multicolumn{2}{c}{\inc{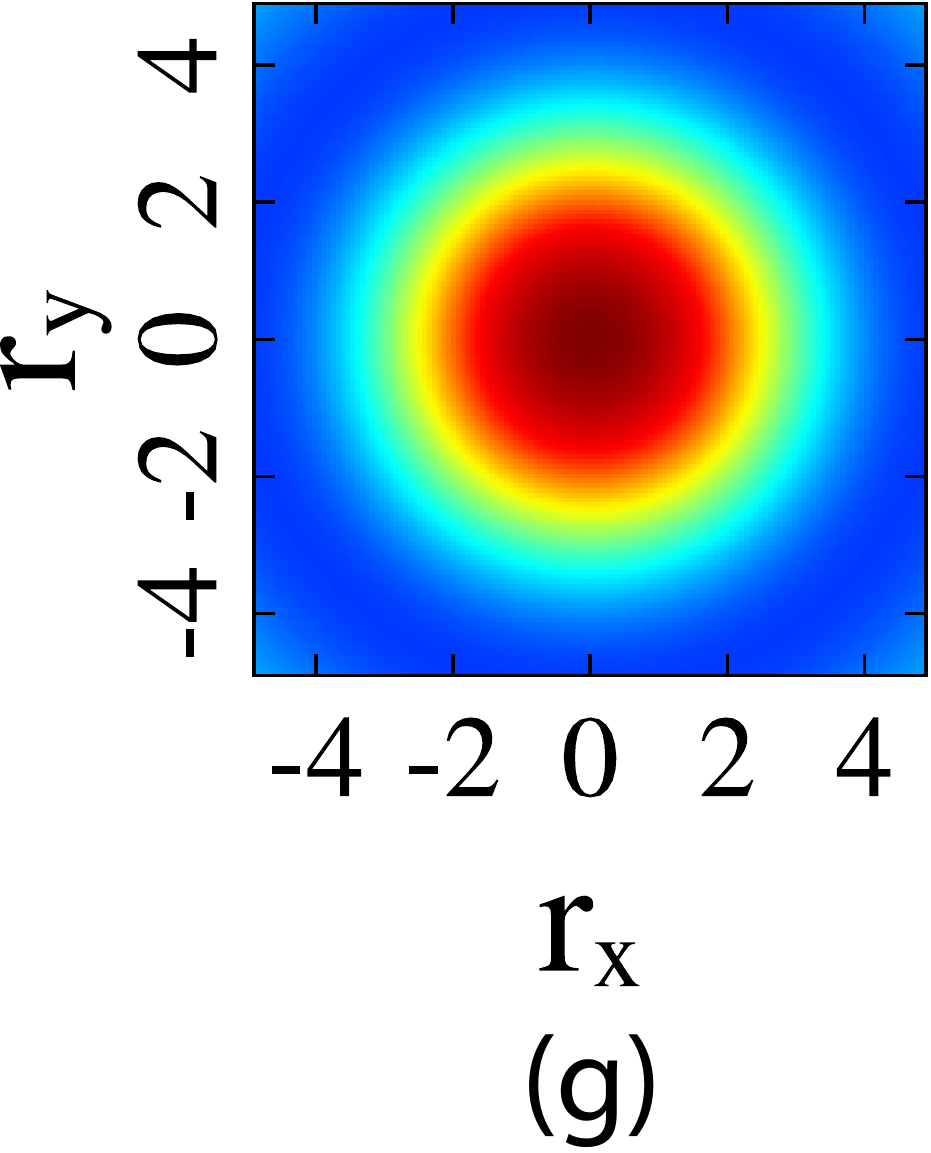}} \\
\end{tabular}

\end{center}
\caption{(Color online)
A portion of the real-space potential $v(\mathbf r)$ around the origin for the stealthy potential (\ref{stealthy}) with $K=1$ and $V(k)=1$.
(a)-(f) Real-space potential in a periodic simulation box that is [(a), (c), and (e)] square or [(b), (d), and (f)] rhombic in shape; the latter has a $60^\circ$ interior angle.
The volumes of the simulation boxes, $v_F$, are [(a) and (b)] 100, [(c) and (d)] 400, and [(e) and (f)] 1385.
Panels (a)-(d) use unrealistically small simulation boxes and is intended to illustrate finite-size effect only.
(g) The real-space potential in the infinite-system-size limit.
All potentials are normalized by their respective values at the origin since scaling does not affect the ground state.
Note that, starting from the center, the dark (red) region indicates the highest values of the potential, whereas towards the edge of the box, the dark (blue) region indicates the lowest values of the potential.
}
\label{RealSpacePotential}
\end{figure*}

\label{finite}
In the infinite-system-size limit, an isotropic ${\tilde v}({\mathbf k})$ correspond to an isotropic real-space pair potential $v(\bf r)$. 
However, for finite systems, the corresponding $v(\bf r)$ is anisotropic.
To illustrate the finite-size effect, we compare the two-dimensional real-space potential $v(\bf r)$ in the infinite-system-size limit to corresponding potentials associated with finite-sized
fundamental cells of square and rhombic shapes of different volumes in Fig.~\ref{RealSpacePotential}.
The real-space potential in the rhombic simulation box with a $60\degree$ interior angle is appreciably more isotropic than the real-space potential in a square simulation box.
Therefore, in this paper, we will henceforth use rhombic fundamental cells in two dimensions.
Similarly, in three dimensions, we always use a simulation box shaped like a fundamental cell of a body-centered cubic (BCC) lattice since BCC lattice is the unique ground state at $\chi_{\mbox{max}}^*$.

\section{Local Gradient Descent Algorithm}
\label{LocalGradientDescent}
Most optimization algorithms are designed for efficiency. They use complex rules to determine the direction of the next step and take as large steps as possible. These features make their path less obvious. To minimize energy in the path following the gradient vector, we designed a ``local gradient descent algorithm'' with the following steps:
\begin{enumerate}
\item Start from an initial guess, $\mathbf{x}$, and find the function value $f(\mathbf{x})$ and derivative $f'(\mathbf{x})$.
\item Start from a relatively large ($10^{-3}$ times the simulation box side length) step size, $s$, and calculate the vector to the next step $\Delta \mathbf{x} = -s \frac{f'(\bf{x})}{|f'(\bf{x})|}$. Find the function value at the next step $f(\mathbf{x}+\Delta \mathbf{x})$. Calculate the change of function value $\Delta f = f(\mathbf{x}+\Delta \mathbf{x})-f(\mathbf{x})$.
\item If we are following the path of steepest descent accurately, the change of the function value should be close to $f'(\mathbf{x}) \cdot \Delta \mathbf{x}$. If the difference between $\Delta f$ and $f'(\mathbf{x}) \cdot \Delta \mathbf{x}$ is less than $1\%$, we accept this move. Otherwise, we abort this move and half the step size $s$.

\item Repeat the above steps until a minimum is found with enough precision.
\end{enumerate}

\section{Number of particles of every system in Sec.~\ref{ensemble}}
\label{number}

\begin{table}[h!]
\setlength{\tabcolsep}{12pt}
\caption{The number of particles $N$ of each systems shown in Figs.~\ref{s_dim}~and~\ref{g2_dim}.}
\begin{tabular}{c c c c}
\hline
$\chi$ & $N$ for $d=1$ & $N$ for $d=2$ & $N$ for $d=3$ \\ \hline
0.05 & 1001 & 541 & 261 \\
0.1 & 501 & 270 & 131 \\
0.143 & 351 & 190 & 92 \\
0.2 & 251 & 136 & 66 \\
0.25 & 201 & 109 & 53 \\
0.33 & 151 & 181 & 191 \\
\hline
\end{tabular}
\label{Np1}
\end{table}

\begin{table}[h!]
\setlength{\tabcolsep}{12pt}
\caption{The number of particles $N$ of each systems shown in Fig.~\ref{s_peak}.}
\begin{tabular}{c c c c}
\hline
$\chi$ & $N$ \\ \hline
0.33 & 181 \\
0.35 & 171 \\
0.38 & 161 \\
0.4 & 151 \\
0.43 & 141 \\
0.46 & 131 \\
\hline
\end{tabular}
\label{Np2}
\end{table}

\begin{table}[h!]
\setlength{\tabcolsep}{12pt}
\caption{The number of particles $N$ of each systems shown in Fig.~\ref{v_dim}.}
\begin{tabular}{c c c c}
\hline
$\chi$ & $N$ for $d=1$ & $N$ for $d=2$ & $N$ for $d=3$ \\ \hline
0.05 & 1001 & 541 & 261 \\
0.1 & 501 & 270 & 131 \\
0.143 & 351 & 190 & 92 \\
0.2 & 251 & 136 & 66 \\
0.25 & 201 & 109 & 53 \\
\hline
\end{tabular}
\label{Np3}
\end{table}
In this appendix we report the number of particles $N$ in each system in Sec.~\ref{ensemble}. 
Both configurations in Fig.~\ref{conf_1d} consist of 51 particles. Configurations (a) and (b) in Fig.~\ref{conf_2d} consist of 271 and 151 particles, respectively. Those in Fig.~\ref{conf_3d} consist of 131 and 161 particles, respectively.

The number of particles of each system in Figs.~\ref{s_dim}, \ref{g2_dim}, \ref{s_peak}, and \ref{v_dim} are shown in Tables~\ref{Np1}, \ref{Np2}, and \ref{Np3}, respectively. 
Each configuration in Figs.~\ref{crystal}, ~\ref{wlmc_2d}, and~\ref{wlmc_3d} consist of 36, 400, and 343 particles, respectively.


\begin{thebibliography}{64}
\bibitem{unique} The ground state is unique if all ground state configurations can be transformed to each other using translation, rotation, inversion, particle permutation, or a combination of these operations.
\bibitem{factor} Compared to Eq.~\eqref{pot_f}, a factor of 2 in the denominator of Eq.~\eqref{pot_f2} is removed because each independent constraint corresponds to two constrained $\mathbf k$ points.
\bibitem{WavyEntropy} In Fig.~\ref{finite_crystal}, we show that crystallization for $\chi < 0.5$ is a finite-size effect. So is crystallization for $\chi \ge 0.5$ also a finite-size effect?
In the upcoming paper of this series, we will present an analytical model of ``stacked-slider" phases for $\chi \ge 0.5$, which are relatively ordered
but aperiodic configurations. Using the model, we will show that the entropy of stacked-slider phases is smaller than that of the crystalline structures.
Therefore, crystallization for $\chi \ge 0.5$ is not a finite-size effect.
\end{thebibliography}
\end{document}